\documentclass[]{emulateapj}
\usepackage{natbib}
\usepackage{amsmath}
\usepackage{float}
\usepackage[caption=false]{subfig}
\usepackage{xcolor}
\usepackage{rotating}

\newcommand{\msol}{\,\textrm{M}_\sun}                

\graphicspath{{pics/}{pics/rotationcurve/}{pics/rotationcurve/kineticmodel/}{pics/slitmap_2_ps/kineticmodel/}{pics/slitmap_2_ps/hamap/}{pics/slitmap_2_ps/N2map/}{pics/metallicity_gradients/}}

\shorttitle{Diversity of Metal Gradients}
\shortauthors{Leethochawalit et al.}
\submitted{}

\begin{document}
\title{A Keck Survey of Gravitationally-Lensed Star-Forming Galaxies: High Spatial Resolution Studies of Kinematics and Metallicity Gradients}
\author{Nicha Leethochawalit\altaffilmark{1}, Tucker A. Jones\altaffilmark{2,6}, Richard S. Ellis\altaffilmark{1}, Daniel P. Stark\altaffilmark{3}, Johan Richard\altaffilmark{4}, 
Adi Zitrin\altaffilmark{1,6} and Matthew Auger\altaffilmark{5}}
\altaffiltext{1}{Cahill Center for Astronomy and Astrophysics, California Institute of Technology, MS 249-17, Pasadena CA 91125, USA}
\altaffiltext{2}{Institute for Astronomy, University of Hawaii, 2680 Woodlawn Drive, Honolulu HI 96822, USA}
\altaffiltext{3}{Department of Astronomy, University of Arizona, Tucson, AZ 85721}
\altaffiltext{4}{CRAL, Observatoire de Lyon, F-69561 Saint-Genis-Laval, France}
\altaffiltext{5}{Institute of Astronomy, University of Cambridge, Cambridge CB3 0HA, UK}
\altaffiltext{6}{Hubble Fellow}
\email{Email: nichal@astro.caltech.edu}

\begin{abstract}
We discuss spatially resolved emission line spectroscopy secured for a total sample of 15 gravitationally lensed star-forming galaxies at a mean
redshift of $z\simeq2$ based on Keck laser-assisted adaptive optics observations undertaken with the recently-improved OSIRIS
integral field unit (IFU) spectrograph. By exploiting gravitationally lensed sources drawn primarily from the CASSOWARY survey, we sample these
sub-L$^{\ast}$ galaxies with source-plane resolutions of a few hundred parsecs ensuring well-sampled 2-D velocity data and resolved
variations in the gas-phase metallicity. Such high spatial resolution data offers a critical check on the structural properties of larger samples 
derived with coarser sampling using multiple-IFU instruments. We demonstrate how kinematic complexities essential to understanding
the maturity of an early star-forming galaxy can often only be revealed with better sampled data. Although we include four sources from 
our earlier work, the present study  provides a more representative sample unbiased with respect to emission line strength. 
Contrary to earlier suggestions, our data indicates a more diverse range of kinematic and 
metal gradient behavior inconsistent with a simple picture of well-ordered rotation developing concurrently with established steep metal 
gradients in all but merging systems. Comparing our observations with the predictions of hydrodynamical simulations suggests that gas and
metals have been mixed by outflows or other strong feedback processes, flattening the metal gradients in early star-forming galaxies.

\end{abstract}


\section{Introduction}

Statistical surveys of star forming galaxies at the peak era of activity, corresponding to a redshift interval $z\simeq$1.5--3, have been increasingly complemented over the past few years by spatially resolved spectroscopic observations. Early work focused on characterizing global trends such as the evolution of star formation rate density with redshift \citep{Madau14}, and evolution of the ``main sequence" of star formation rate (SFR) as a function of galaxy stellar mass \citep{Noeske07, Peng10}. Spatially resolved data from integral field unit (IFU) spectrographs has provided complementary information on the ionized gas kinematics \citep{Forster06,Stark08, Law09, Forster09, Jones10a, Wisnioski11, Swinbank12, Wisnioski15}, the size and spatial distribution of giant star forming regions \citep{Genzel08, Jones10b, Livermore12}, and radial metal abundance gradients \citep{Cresci10, Jones10b, Yuan11, Swinbank12, Jones13, Stott14, Troncoso14, Jones15}. 2-D kinematic data and the properties of clumpy star forming regions provide valuable insight into the emergence of primitive disks, the role of instabilities, and the gradual assembly of central bulges possibly formed from inwardly migrating clumps \citep{Genzel11}. The properties of these clumps provides important evidence of gravitational instabilities in the primitive disks whereas correlations between their ages and radial positions insight into the morphological evolution of these early systems \citep{Forster11}.

Metallicity gradients also provide an important opportunity to study the amount of feedback, i.e., the energy returned to the surrounding medium by star formation and nuclear activity. Energy input can arise from various mechanisms including supernovae, radiation pressure, and cosmic rays. These effects combine to drive large-scale outflows of gas and metals. Outflows are ubiquitously observed from systems with high star formation rate densities \citep{Heckman01}, including virtually all star forming galaxies at high redshifts \citep[e.g.][]{Shapley03}. Cosmological simulations confirm that outflow mass loss rates must be comparable and often larger than star formation rates in order to explain measured stellar mass functions and the mass-metallicity relation \citep[e.g.][]{Vogelsberger14,Crain15}. The rates of mass and metal loss are degenerate with rates of subsequent accretion and hence are poorly known. However, this ``galactic fountain" cycle may be constrained via its imprint on metallicity gradients. To first order, high rates of outflow and subsequent accretion will redistribute heavy elements, resulting in a flatter gradient. Furthermore, the redistribution of interstellar gas will affect the radial profile of star formation and hence future metal production and feedback. Evidence for these effects has been recently observed in the form of flattened metal gradients in the extended disks of nearby galaxies \citep[e.g.][]{Bresolin11}. Several groups have now explored how various forms of feedback affect metal gradients in simulations \citep[e.g.][]{YangKrumholz12,Gibson13, Angles-Alcazar14}. Despite a range of methods, there is a clear consensus that stronger feedback (that is a higher rate of energy injection from stellar winds, supernovae, and other sources) results in flatter gradients due to mixing of gas and metals over larger physical scales. Various prescriptions are able to match observations of local galaxies but predict different behavior at early times. We therefore seek to constrain the degree of gas cycling via feedback, by directly measuring gradients at high redshifts.

Initial IFU surveys targeted modest numbers of galaxies beyond $z\simeq2$ \citep{Forster06, Law09, Jones10a}, reflecting the challenges of securing adequate signal to noise per spatial resolution element. The advent of multi-IFU spectrographs (e.g., KMOS on the ESO VLT; \citealt{Sharples13}) has recently led to a significant improvement in survey capability. For example, \citet{Wisnioski15} report resolved kinematic data for over 100 galaxies within the redshift range $2<z<2.7$ using this impressive instrument. However, the multi-IFU approach comes with a major limitation in terms of angular resolution. Typical 0\farcs6 seeing conditions correspond to a physical scale of 5 kpc at $z\simeq2$, whereas the half-light radius of a typical L$^{\ast}$ galaxy at this redshift is only 2 kpc. Clearly only the largest and most massive systems can be adequately probed with seeing-limited data. In fact, even with adaptive optics, the physical resolution is only $\simeq$1 kpc \citep{Genzel06, Law09}. \citet{Jones10a, Jones13} illustrate the difficulties of correctly interpreting velocity fields and abundance gradients from data with so few resolution elements across each galaxy. As an example, early IFU surveys suggested that compact galaxies at $z\simeq2$ were characterized by dispersion-dominated kinematics with little or no rotation \citep{Forster09}. Deeper observations with adaptive optics showed that most of these sources in fact harbor rotating thick disks, with circular velocities commensurate with their lower masses compared to rotation-dominated systems \citep{Newman13}, confirming earlier results from lensed galaxies with superior resolution \citep{Jones10a}.

Studying gravitationally lensed systems with adaptive optics represents a highly valuable route to addressing the physics of galaxy formation, particularly for the less massive and more abundant systems at $z>2$. Lensing magnification enables higher spatial resolution and better sampling than is otherwise possible. In a pioneering study, \citet{Stark08} illustrated the potential of the lensing approach by securing resolved kinematic data for MACSJ2135-0102, a $z$=3.075 star-forming galaxy magnified in angular size by a factor of $\times$8 along its major axis by a foreground galaxy and galaxy cluster. Using the Keck OSIRIS IFU spectrograph with adaptive optics, the source-plane velocity field was sampled with a resolution of 120 pc leading to 20 independent points on its rotation curve. Subsequently \cite{Jones10a} studied a sample of 6 lensed galaxies, demonstrating that a high fraction of such $\lesssim$ L$^{\ast}$ systems display well-ordered velocity fields, in contradiction to the claims of earlier less well-sampled studies of unlensed galaxies of similar luminosities \citep{Law09}. Likewise, \citet{Jones10b} derived a radial metal abundance gradient for a lensed $z=2.00$ galaxy utilizing the spatial variation of [N II]/H$\alpha$ and [O III]/H$\alpha$ on 500 pc scales for a system with a half-light radius of 2.9 kpc. The necessity of mapping weaker lines such as [N II] makes the metal gradient aspect particularly challenging, but four systems were studied in \citet{Jones13} showing a diverse range of gradients, possibly dependent upon the kinematic properties (see also \citet{Cresci10}). \citet{Jones13} introduced a simple model which suggests that metal gradients and their evolution should provide insight into the radial variation in the mass loading factor governing the amount of outflowing gas. More recently, a KMOS study of a larger sample of 21 galaxies at $z\simeq1$ suggests the metal gradient tends to be less steep in more intensely star-forming systems \citep{Stott14}. In view of the diverse results, clearly larger samples are required, particularly at high redshift where only lensed galaxies provide the necessary physical resolution.

This paper presents the results of a significantly enlarged IFU survey of lensed galaxies. Two practical developments have motivated us to extend the original sample presented in \citet{Jones13}. Firstly the number of lensed targets with known spectroscopic redshifts has increased following the CASSOWARY survey of lensed star-forming galaxies located primarily in the Sloan Digital Sky Survey \citep{Stark13}. The second is a substantial improvement in the performance of the OSIRIS IFU spectrograph following its transfer to Keck I. This exploits the more powerful center launch laser which offers a brighter and more compact LGS beacon, ensuring a much improved Strehl. The installation of a new grating also improved the instrumental throughput by a factor of $\times$1.5--2.5 depending on wavelength.

The aim of the present survey is to exploit recent improvements at the Keck Observatory to extend the original campaign, obtaining IFU spectroscopy of a larger and more representative sample of lensed galaxies. The improved throughput has enabled us to discard the selection criterion adopted by \cite{Jones13} whereby targets were initially pre-screened with a long-slit spectrograph (NIRSPEC) to ensure adequately strong emission lines, possibly biasing the sample to unusually active sources. Our goals are two-fold. First we aim to measure the radial metallicity gradient on sub-kpc scales for a representative sample of $z\simeq2$ galaxies and to examine further the origin of the diverse results obtained by various groups. \citet{Jones13} and \citet{Yuan13} argued that much of the discrepancy in metal gradient measurements, e.g. \citet{Queyrel12} might arise from poorly-sampled data, whereas \citet{Stott14} suggested correlations with the specific star formation may be the cause. Secondly we aim to characterize the kinematics and star formation on sub-kpc scales for a larger sample. Our well-sampled kinematics and star formation morphologies allows us to evaluate the utility and conclusions drawn from complementary larger surveys being undertaken, e.g. with KMOS \citep{Stott14,Wisnioski15}, at coarser $\sim$5 kpc resolutions.

A plan of the paper follows. In \S2 we discuss the sample drawn primarily from the CASSOWARY survey and the relevant selection criteria. We present the OSIRIS spectroscopic observations and their initial reduction. \S3 discusses the reductions of the resolved spectroscopic data into the source plane utilizing the available mass models for the foreground lenses. We discuss the kinematic properties in \S4 and the metal gradients in \S5. \S6 discusses the overall results and we summarize our conclusions in \S7.

\section{Data}
\subsection{Sample and Observations}
\label{sec:obs}

\begin{deluxetable*}{lllccccccccc}
\setlength{\tabcolsep}{0.001cm} 
\tablewidth{\linewidth}
\tablecaption{Observation Log}
\tablehead{\colhead{ID} & \colhead{ $z$ } & \colhead{Coordinates} & \colhead{Dates} & \colhead{Filter} & \colhead{Lines} & \colhead{$t_{exp}$} & \colhead{FWHM}& \colhead{FWHM} & \colhead{$\mu$}&\colhead{$\mbox{A}_{\text{H}\alpha}$}&\colhead{SFR}\\
		  \colhead{}    & \colhead{ } 	 & \colhead{} 			& \colhead{} 	     & \colhead{} 		& \colhead{} 	     & \colhead{}         & \colhead{}	      & \colhead{} &  \colhead{}&\colhead{}&\colhead{}\\
		  \colhead{}    & \colhead{ } 	 & \colhead{\scriptsize{RA\hspace{0.7cm} DEC} }& \colhead{\scriptsize{(MM/YY)}}	     & \colhead{} 		& \colhead{} 	     & \colhead{\scriptsize{(ks)}} 			      & \colhead{\scriptsize{(PSF)}}	  & \colhead{\scriptsize{(Source Plane)}} &  \colhead{}&\colhead{\scriptsize{(mag)}}&\colhead{\scriptsize{($\msol yr^{-1}$)}}}
\startdata
\cutinhead{\emph{This paper}}
cswa11  & 1.41& 08:00:12 +08:12:07  &3/13,2/14     & Hn2   & H$\alpha$, N[II] & 16.2	&$0''.21    $& $1.5 \times 1.9 \mbox{ kpc}$ & 1.9   &\nodata             &$99\pm90^b$\\
cswa15  & 2.16& 10:09:01 +19:37:23  &2/14               & Kn2   & H$\alpha$, N[II] & 9            &$0''.53^a$& $2.2 \times 4.4 \mbox{ kpc}$ & 7.6   &$0.52\pm0.10$         &$42\pm5$\\
    	       &         & 	     	     	     	   &2/14               & Hbb   & O[III], H$\beta$  & 3.6         &\nodata     &\nodata                                       &          &                                       &                 \\ 
cswa19  & 2.03& 09:00:03 +22:34:08  &3/13,2/14     & Kn1   & H$\alpha$, N[II] & 14.4       &$0''.14$ 	  &$0.4\times 1.2 \mbox{ kpc}$	  &4.3    &$0.46\pm0.17$         &$59\pm11$ \\
                &         &                                      &3/13               & Hn1   & O[III], H$\beta$  & 3.6          &\nodata    &\nodata                                       &         &                                        &                  \\ 
cswa20  & 1.43& 14:41:49 +14:41:22  &02/14            & Hn2   & H$\alpha$, N[II] & 3.6         &$0''.09$    &$0.1 \times 0.3 \mbox{ kpc}$  &14     & $0.27\pm0.34$         &$6\pm3$\\
                &         &                                      &2/14               & Jn1    & O[III], H$\beta$  & 2.7         &$0''.25$	   &$0.5 \times 1.1 \mbox{ kpc}$  &         &                                         &                     \\
cswa28  & 2.09& 13:43:33 +41:55:13  &2/14              & Kn1 & H$\alpha$, N[II]   & 3.6           &$0''.23$    &$0.3 \times 1.8 \mbox{ kpc}$  & 9.3 &\nodata	            &$12\pm11$\\
cswa31  & 1.49& 09:21:25 +18:10:11 &12/14             & Hn3 & H$\alpha$, N[II]  & 10.8        &$0''.22$ 	   &$0.6 \times 2.1 \mbox{ kpc}$  & 3.3  &\nodata           &$36\pm33$\\
cswa128&2.22& 19:58:35 +59:50:53  &9/13               & Kn2 & H$\alpha$, N[II]  & 5.4           &$0''.15$    &$0.4 \times 0.7 \mbox{ kpc}$  & 10   &$1.96^{+0.23}_{-0.77}$   &$250^{+71}_{142}$\\
                &        &                                       &9/13               & Hbb & O[III], H$\beta$   & 5.4          &\nodata	   &\nodata                                       &         &                                 \\
cswa139&2.54&  08:07:32 +44:10:51  &2/14,12/14  	& Kc5 & H$\alpha$, N[II]   & 10.8  	&$0''.18$ 	   &$0.7 \times 1.2 \mbox{ kpc}$  &9.7   &$0.38^{+0.36}_{-0.52}$            &$33^{+16}_{-24}$\\
                &        &                                       &12/14           & Hbb & O[III], H$\beta$   & 5.4		&$0''.07$ 	   &$0.2 \times 0.4 \mbox{ kpc}$  &          &                                 \\
                &        &                                       &2/14             &Jn3	  & O[II]                       & 1.8	          &\nodata	  &\nodata	                                        &          &                                \\
cswa159&2.30& 22:22:09 +27:45:25  &9/13,12/14  & Kc3 & H$\alpha$, N[II]   & 7.2 	          &$0''.17$ 	  &$1.2\times2.7\mbox{ kpc}$	  &4.6     &$0.44\pm0.95$         &$53\pm47$\\
                &        &                                       &12/14           & Hn3 & O[III], H$\beta$   & 3.6            &\nodata	&\nodata           	                              &           &                               \\
cswa165& 2.13& 01:05:20 +01:44:58  &9/13,12/14  & Kn2 & H$\alpha$, N[II]  & 7.2          &$0''.16$ 	&$0.1 \times 0.2 \mbox{ kpc}$	&42         &$1.03^{+0.2}_{-0.58}$          &$7^{+2}_{-3}$\\
                &         &                                      &9/13,12/14   & Hbb & O[III], H$\beta$   & 5.4          &$0''.20$	&$0.1 \times 0.2 \mbox{ kpc}$	&             &                      	  \\ 
a773       &2.30 & 09:17:57 +51:43:31  &3/13, 2/14,   & Kc3 & H$\alpha$, N[II]  & 12.6        &$0''.39$  &$0.1 \times 1.2 \mbox{ kpc}$	&$20.3$& $2.5\pm0.25$             &$30\pm7$\\   
                 &        &                                       &12/14           &         &                               &                 &                 &                                                     &             &                             \\   
                 &       &                                        &3/13              & Hn3& O[III], H$\beta$   & 7.2          &$0''.41$	&$0.1 \times 1.2 \mbox{ kpc}$   &             &                            \\  
\cutinhead{\emph{Objects published in \citet{Jones13}}}
J0744       &2.21 & 07:44:48 +39:27:26 & 1/08            & Kn2 & H$\alpha$, N[II] & 9              &$0''.11$ 	& $0.3\times0.8 \mbox{ kpc}$	&16	       &$0.53\pm1.2$        &$5.4^{+4.9}_{-1.8}$    \\
                   &        &                                      &2/11            & Hbb & O[III], H$\beta$  & 3.6           &$0''.08$   & $0.3\times0.7 \mbox{ kpc}$	&	       &                                      &                           \\    
J1038       & 2.20& 10:38:42 +48:49:19 &2/11,3/11   & Kn2 & H$\alpha$, N[II] & 9               &$0''.14$   & $0.4\times1.6 \mbox{ kpc}$	&8.4	       &  $0.67\pm1.30$     & $38^{+37}_{-15}$      \\
                   &        &                                      &2/11            & Hbb & O[III], H$\beta$  &  3.6 	         &$0''.14$   & $0.3\times1.7 \mbox{ kpc}$	& 	       &                                      &                    	   \\ 
J1148        & 2.38& 11:48:33 +19:29:59 &2/11           & Kn2 & H$\alpha$, N[II] & 9               &$0''.11$  & $0.6\times0.9 \mbox{ kpc}$	&10.3      &$2.94\pm1.05$       & $210\pm167$         \\
                   &        &                                       &2/11,3/11 & Hbb & O[III], H$\beta$  & 3.6             &$0''.08$  & $0.6\times0.9 \mbox{ kpc}$	&	       &                                      &                    	    \\ 
J1206        & 2.00& 12:06:02 +51:42:30 &5/10           & Kn2 & H$\alpha$, N[II] & 9                &$0''.18$  & $0.5\times3.0 \mbox{ kpc}$	&13.1      &$1.22\pm0.47$        &$68^{+44}_{-24}$     	 \\
                   &        &                                       &5/10           & Hbb & O[III], H$\beta$  & 3.6 	           &$0''.33$ &$0.6\times3.5 \mbox{ kpc}$	&	       &                                       &                    	   \\ 
\enddata
\tablecomments{The 2-D magnification $\mu$ is defined as the ratio between image plane flux and source plane flux. UV magnitudes and H$\alpha$ fluxes are of the source planes. $^a$ tip/tilt star is a galaxy. $^b$ For galaxies with no observations in the H$\beta$ band, their star formation rates are obtained with the weighted mean extinction A$_{\text{H}\alpha}=0.8\pm1.0$ in the sample. \label{tab:observationlog}}
\end{deluxetable*}

During several observing runs in 2013--2014, we observed 11 gravitationally-lensed sources at a mean redshift $z\simeq2$ using the near-infrared integral field spectrograph OSIRIS \citep{Larkin06} on the Keck 1 telescopes with the laser guide star adaptive optics (LGSAO) system \citep{Wizinowich06}. Prior to 2013, we observed four similar sources using OSIRIS with the Keck II AO system and presented the results in \citet{Jones13}. Because of the lower system throughput at that time, the early four sources were pre-screened with the long-slit spectrograph NIRSPEC to have suitably bright [N II] and H$\alpha$ emission lines so as to ensure a reasonable signal-to-noise with OSIRIS in practical integration times. After the transfer of OSIRIS to the Keck II telescope and the installation of a new grating in 2012 December, the AO Strehl was improved and the OSIRIS throughput increased by a factor of $\simeq\times$2 \citep{Mieda14}. Accordingly, to avoid any bias in selecting targets, we abandoned the earlier spectroscopic pre-screening method.  The current sample therefore comprises 15 sources in total. Ten new sources were selected from the CASSOWARY catalog of star-forming lensed galaxies as presented in \citet{Stark13} based on their availability during the scheduled observation period, their rest-frame UV spectroscopic redshifts in the range $z=1.5\--2.5$, and the presence of suitably bright proximate tip-tilt guide stars. We selected an additional cluster-lensed galaxy (Abell 773) from \citet{Belli13} with similar criteria as the CASSOWARY sample. The eleven new sources extend the UV luminosity and SFR range in \citet{Jones13} with UV absolute magnitudes in the range M$_{\textrm{UV}}\sim-22$ to $-18$ and SFR in the range 1 to 80 M$_\odot\textrm{yr}^{-1}$ (uncorrected for dust). Most of the sources are in the sub-L$^{\ast}$ regime while a few extend slightly beyond the L$^{\ast}$ magnitude limit (M$_{\textrm{UV}}<-21$). The total sample is presented in Table \ref{tab:observationlog}. 

Using reddening-corrected star formation rates (SFRs) derives from both H$\alpha$ and ultraviolet continuum measures (using the methods described below), we can compare the properties of our present sample with those in previous studies. Although there is a wide range overall (from $\simeq$5 to 250 $M_{\odot}$ yr$^{-1}$), the bulk of our sample have values in the range 10 to 100, with a median of $\simeq$40$M_{\odot}$ yr$^{-1}$. This is comparable to the rates observed in the unlensed surveys\citep{Law09, Forster09, Wisnioski11, Wisnioski15}. Near-infrared photometry, essential for deriving accurate stellar masses, are available for about half the objects in our sample (and hence are not listed in Table \ref{tab:observationlog}). However, they lie in the range log $M_{\ast}\simeq$9.0-9.6, which is significantly less than for those in the unlensed surveys which generally probe systems from 10$^{10}$ to $10^{11} \,M_{\odot}$. Our survey therefore samples star-forming galaxies somewhat above the main sequence as
defined by \citet{Behroozi13}.

We closely followed the observing technique described in \citet{Jones10b}. Observations were undertaken in one or more near-infrared passbands (J,H, or K) to secure spatially-resolved data on the H$\alpha$ and [N II] emission lines. Where practical, we continued to observe H$\beta$, and the [O III] emission lines within a shorter wavelength band. The average spectral resolution is $R\simeq 3600$ which corresponds to $\simeq 3-6$ \AA \  across the J, H, K bands. We used the 100 mas pixel scale which gives a field of view of at least $1.6\times6.4$ arcsec. We took short exposures of each tip-tilt star to center the position before moving to the lensed galaxy. Each exposure comprised a number of 15-minute sub-exposures with a ABAB dither pattern ABAB of increment $\sim2-3$ arcsec ensuring that the target was present in all frames. Exposure times varied from $1\--4$ hours. The seeing during the observations varied between $0.4''$ to $1.5''$. The median AO-corrected seeing was $0.17''$.    

\subsection{Data Reduction}
We used the latest OSIRIS Data Reduction Pipeline \citep{Larkin06} \footnote{Data Reduction Pipeline Version 3.2} to perform dark subtraction and cosmic ray rejection, followed by a direct or scaled sky subtraction prior to spectral extraction, wavelength calibration and telluric correction using faint standard stars. Adjacent exposures were used as sky reference frames. The data cubes from each exposure were finally combined using a 
$\sigma$-clipped mean.

\subsection{Emission Line Fitting}
\label{sec:Emission Line Fitting}
We fit a Gaussian to the various emission lines to determine the line flux, velocity, and velocity dispersion. In all cases, H$\alpha$ is the most prominent line and a key indicator of the velocity field and star-formation rate distribution. A four-parameter Gaussian curve is fit to the H$\alpha$ line for each spatial pixel using a weighted $\chi^2$ minimization procedure. In each data cube, we select a region devoid of emission lines and calculate the weight for each wavelength from a variance over this region, $w(\lambda)=V^{-1}(\lambda)$, for use in the Gaussian fitting. We require that each H$\alpha$ detection must be above 5$\sigma$ and, where necessary, we spatially smooth the data cube with a Gaussian kernel of FWHM 3 pixels to increase the signal-to-noise. Given a H$\alpha$ detection, we fit a Gaussian of identical width and velocity to the [N II] emission. Where observed, we fit the [O III] and H$\beta$ emission lines independently from the H$\alpha$ and [N II] emission lines but with a similar procedure. The velocities and line widths obtained from the [O III] emission lines are consistent with those obtained from H$\alpha$ to within the $1\sigma$ uncertainties. The intrinsic velocity dispersion is calculated by subtracting the instrumental resolution measured from OH sky lines $\sigma_{inst}\simeq 50\ $km s$^{-1}$ in quadrature from the best-fit line width.

\subsection{Flux Calibration, Extinction and Star Formation Rate}
\label{sec:sfr}
For each galaxy observed in the H$\alpha$ observation band, we use the tip/tilt reference star to calibrate the absolute flux. We fit a PSF to the image of an integrated flux over wavelength of the star and obtain a 1D star spectrum from the derived PSF. The flux calibration is then calculated at H$\alpha$ wavelength of the galaxy. The only exception is CSWA15 whose tip/tilt reference star is an extended galaxy and its flux calibration is measured from a UKIRT infrared standard star FS26 observed in the same night. We crossed check the all H$\alpha$ flux calibrations with the flux calibrations derived from standard stars (FS9, FS11, FS19, or FS26) taken at the end of the same nights. Flux calibration derived from these standard stars generally agree to within 25\%.

Ideally we would do the same process to the H$\beta$ observation band observations. However, there are only three galaxies with tip/tilt reference stars taken in the H$\beta$ bands (CSWA139, CSWA159 and Abell773). Hence, we use the flux calibration from the infrared standard stars to calibrate the H$\beta$ fluxes when the information from tip/tilt stars are not available. The flux calibrations derived from the three tip/tilt reference stars are in agreement with those from standard stars within 15\%.

We follow \citet{Jones13} in the calculation of H$\alpha$ extinction and star formation rate. In short, we used the Balmer line ratios H$\alpha$/H$\beta$ with \citet{Calzetti00} reddening curve to calculate dust extinction $E(B-V)$. We then calculate the extinction of H$\alpha$, A$_{\text{H}\alpha}=3.33E(B-V)$. The star formation rate is computed from total H$\alpha$ flux corrected for extinction and lensing magnification with the H$\alpha$ SFR relation from \citet{Kennicutt98}. For the three galaxies with no observations in H$\beta$ bands (CSWA11,CSWA28, and CSWA31), we use the weighted mean and the standard deviation of A$_{\text{H}\alpha}$ of the other 12 galaxies in our sample, $\overline{\mbox{A}_{\text{H}\alpha}}=0.8\pm1.0$, to calculate the star formation rates. The resulting A$_{\text{H}\alpha}$ and SFR are listed in Table 1.

\section{Source Plane Reconstruction}
\label{sec:sourceplane}

By observing gravitationally-lensed systems we can secure much higher spatial resolution and sampling for our targets than would otherwise be the case. Nonetheless, this gain in resolution is only possible by using accurate mass models for the lens system via which our observations in the image plane can be transferred into the (unlensed) source plane. The key to developing appropriate mass models is the correct identification of multiply-imaged systems ideally with spectroscopic redshifts. Given the variety of lenses, from SDSS galaxies to Abell clusters, surveyed in this study, it has not been possible to adopt a uniform approach to constructing these mass models across our sample.

When Hubble Space Telescope(HST) images or Gemini Science Archive (GSA) images were not available, we took photometric BRI images in good seeing with the Echellette Spectrograph and Imager (ESI) on the Keck II telescopes. In some cases, e.g. CSWA159, mass models were already available in the literature \citep{Dahle09}. Depending on the circumstances, three distinct methods were used in constructing new mass models as detailed below. The methods and imaging data employed to develop the mass models are summarized in Table \ref{tab:massmodels}.

We used the Light-Traces-Mass (LTM) method \citep{Zitrin14} to develop mass models for CSWA11, CSWA15, CSWA19, CSWA28, CSWA31, CSWA139, and CSWA159. Generally these lenses are dominated by a single galaxy with a few associated companions as they represent lenses with angular Einstein diameters in excess of the size of a SDSS fiber ($>3$ arc sec). The method assumes that the lensing mass distribution of lensing galaxies is described by an elliptical power-law, with a gaussian core for the brightest central galaxies, scaled according to their luminosities. The member galaxies are identified spectroscopically and/or via a color cut. To represent the dark matter, the mass distribution of the galaxies is smoothed by a 2D Gaussian kernel. Multiple images are then identified from their similar colors and/or spectroscopic redshifts, if available, in conjunction with a preliminary guess for the mass model. In many cases, spectroscopic redshifts are available in the literature\citep{Stark13, Dahle09, Brewer11} and from our OSIRIS observations. In a few cases where redshifts are not available for identified multiple images, we leave the redshift as a free parameter. In constructing the final best-fit mass model, we employ a Monte-Carlo Markov Chain (MCMC) to minimize the image position $\chi^2$, i.e. the total distance between the predicted and actual position of the multiple images. The model consists of six basic parameters: the exponent of the power law, the smoothing Gaussian width, the relative weight between the galaxy component and the dark matter component, the overall normalization of the mass distribution, and the two external shear parameters. The mass-to-light ratio is fixed for all galaxies except that for the brightest central galaxies that we left as a free parameter in each model.

Some mass models for CASSOWARY sources, CSWA20, CSWA128 and CSWA165 were already available following modeling developed by Matt Auger (MA) as part of that survey \citep{Stark13}. In these cases, $gri$ imaging from SDSS was used to fit each lensing component with a singular isothermal ellipsoid following the procedure discussed in \citet{Auger11}. The same mass model is fitted to each of the filters simultaneously, although the amplitudes of the surface brightness profiles can vary.  

To determine how uncertainties in the lensing models affect our analyses of both the kinematic state of each galaxy and its chemical gradient, we empirically analyzed a marginal case of the lensing models. Since CSWA28 has a relatively high magnification factor and contains only one set of multiple images in constraining the lensing model i.e. the magnification is the least well-constrained, we selected CSWA28 as the appropriate case. We constructed 20 different mass models following the parameters randomly drawn from the MCMC and performed the same kinematic properties analysis (simple disk model fitting) as in Section \ref{sec:kinematics} below. We found that in most of the cases, different lens models primarily affect the position of the source relative to the image but the morphology and size of the galaxy in the source plane is largely the same. From the kinematic model fitting in each source plane, the derived galaxy inclinations, position angles and centers are in agreement given the uncertainties we quote for the best model. The resulting uncertainty in radius at each pixel is $\lesssim10\%$. In a few rare cases, the result from kinematic model fittings yields a radius that is $30\%$ higher. For CSWA28, if we have propagated the expected maximum uncertainty of $10\%$ in radii into the metal gradient calculation, the final uncertainty in the derived N2$_{\text{PP}04}$ metal gradient would have increased by only $2\%$ from $\pm27\%$ to $\pm29\%$. Since CSWA28 is one of the least certain lens models, we expect that the uncertainties due to lensing models are not the dominance source of uncertainty and did not propagate this uncertainty into the subsequent kinetic modeling.

Finally, for the rich cluster, Abell 773, the mass model is available from a detailed study conducted by Johan Richard (JR) \citep{Richard11,Richard10, Limousin12, Livermore12}. In this case, a parametric mass model of the central region of the cluster was developed using the LENSTOOL package \citep{Jullo07}. Briefly, we assumed the cluster mass distribution follows a double Pseudo-Isothermal Elliptical profile \citep{Eliasdottir07} and added one or more central cluster members as smaller scale perturbations to the mass distribution. As above, optimization is conducted by minimizing the predicted and observed positions of the many multiple images in these well-studied clusters. Table 1 summarizes the overall 2-D magnification $\mu$ appropriate for each source although, in the following analysis, full account is taken of the spatial dependence in transforming our data to the source plane.

\begin{deluxetable}{lccc}
\tablewidth{\linewidth}
\tablecaption{Mass Models}
\tablehead{\colhead{ID} & \colhead{Method} & \colhead{Image} & Photometric bands}\
\startdata
cswa11		&LTM 	& Gemini	& $gri$\\
cswa15		&LTM 	& Keck ESI &BRI	\\
cswa19		&LTM	&  HST &GVIH	\\
cswa20		&MA		& SDSS &$gri$	\\
cswa28		&LTM	& HST  &GVI	\\
cswa31		&LTM	& Gemini 	& $gri$\\
cswa128		&MA 	& SDSS &$gri$	\\
cswa139		&LTM	& Keck ESI &VRI \\
cswa159		&LTM	& SDSS &$gri$\\
cswa165		&MA 	& SDSS  &$gri$	\\
a773			&JR   	& HST	&YJH
\enddata
 \label{tab:massmodels}
\end{deluxetable}

\section{Kinematic Properties}
\label{sec:kinematics}

\subsection{Methods and Motivation}

\begin{figure*}
\centering
	\includegraphics[width=4.5cm]{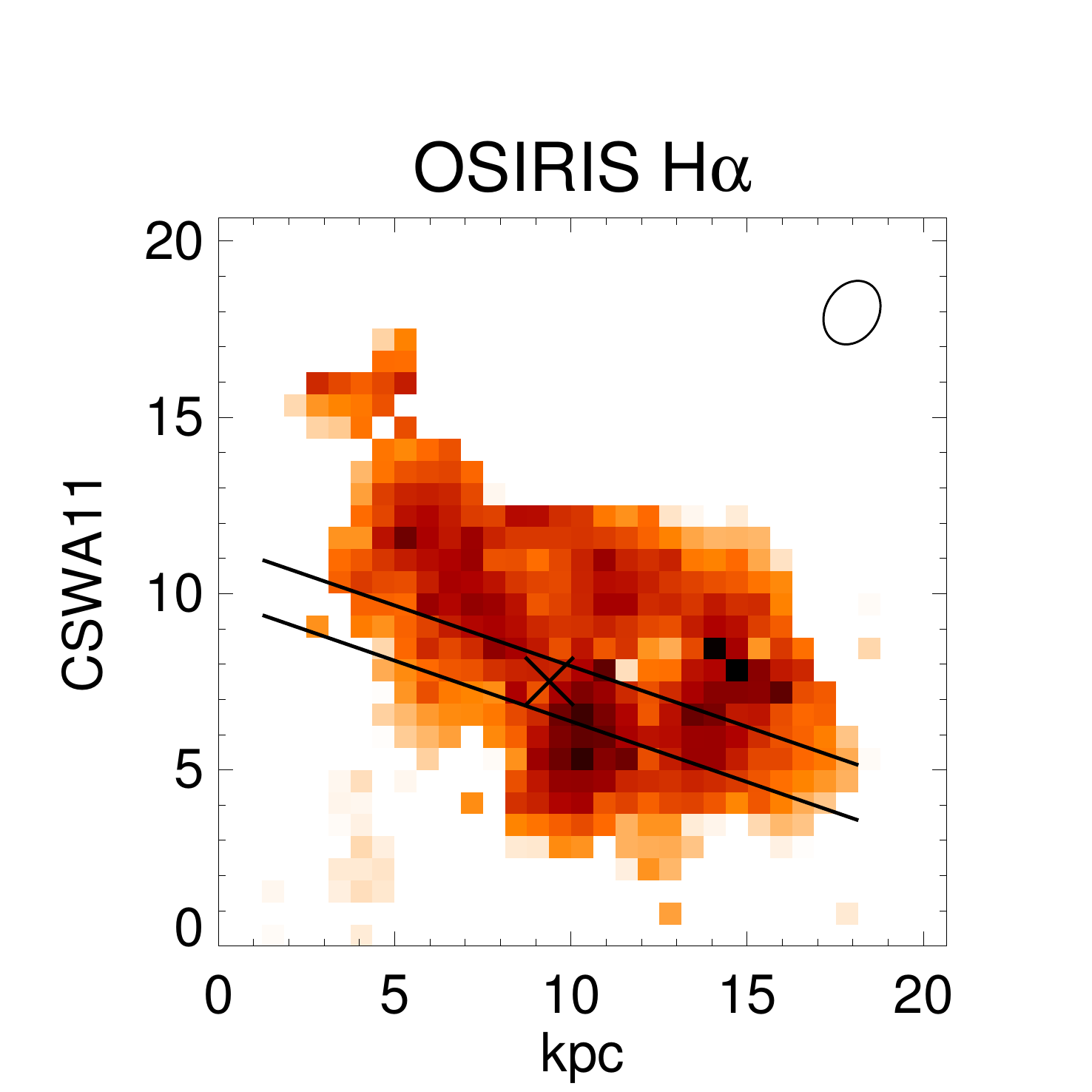}
	\includegraphics[width=4.5cm]{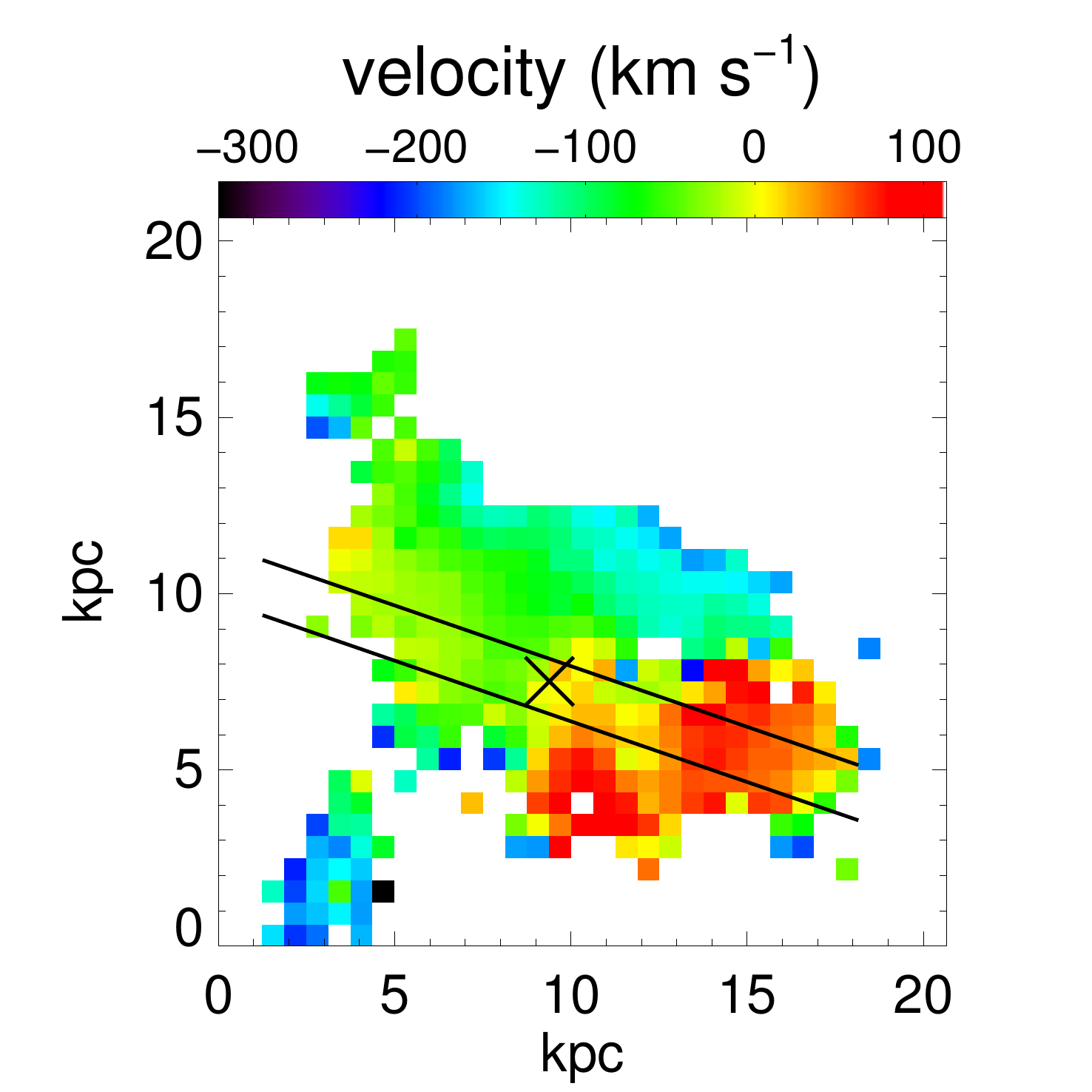}
	\includegraphics[width=5cm]{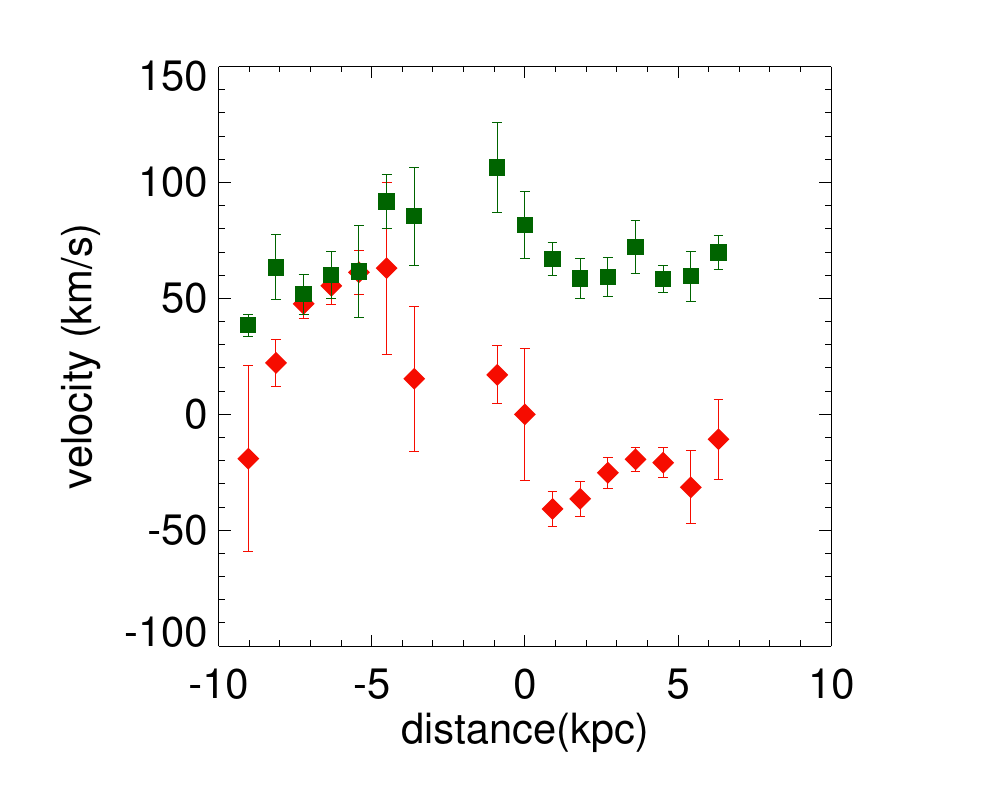}\\
	\includegraphics[width=4.5cm]{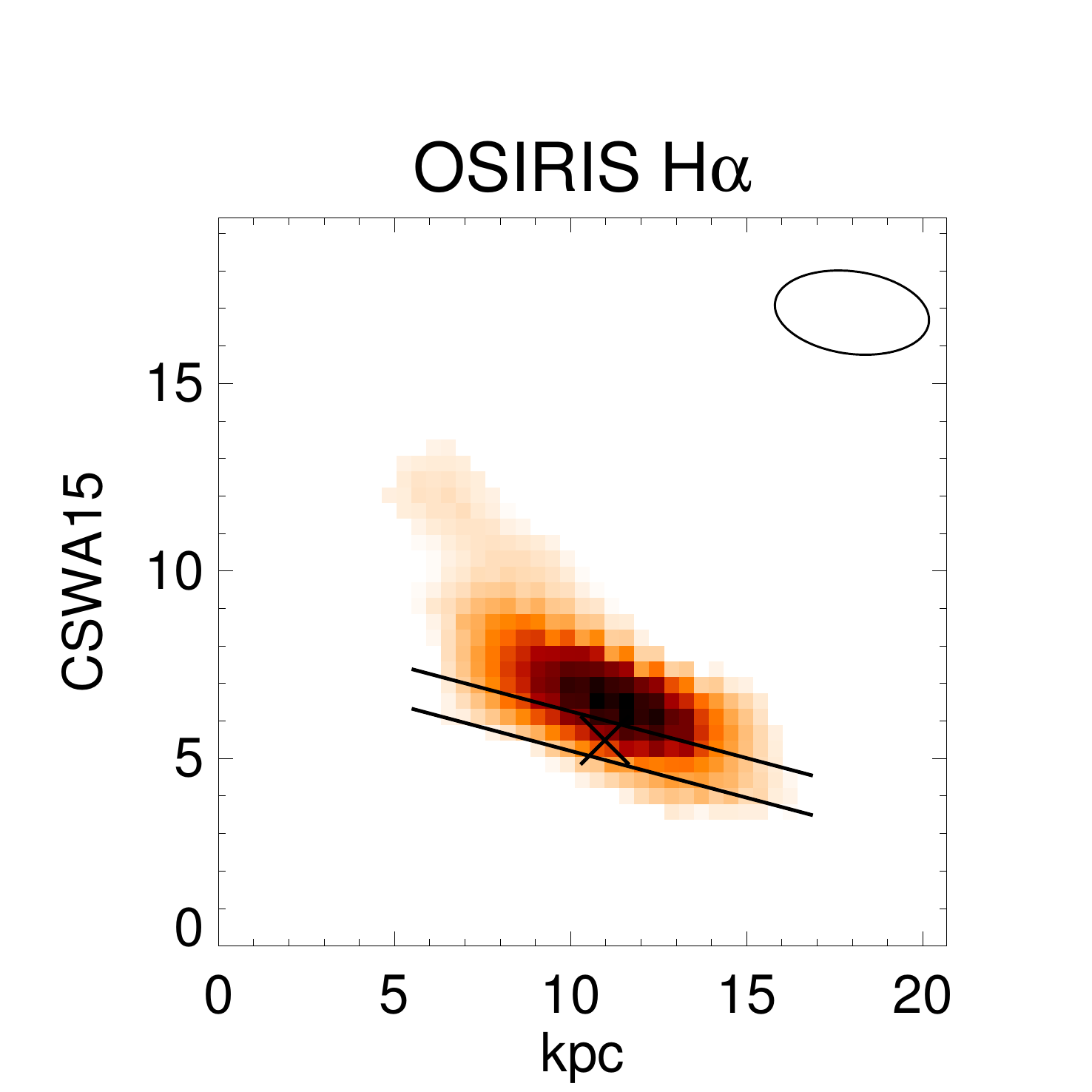}
	\includegraphics[width=4.5cm]{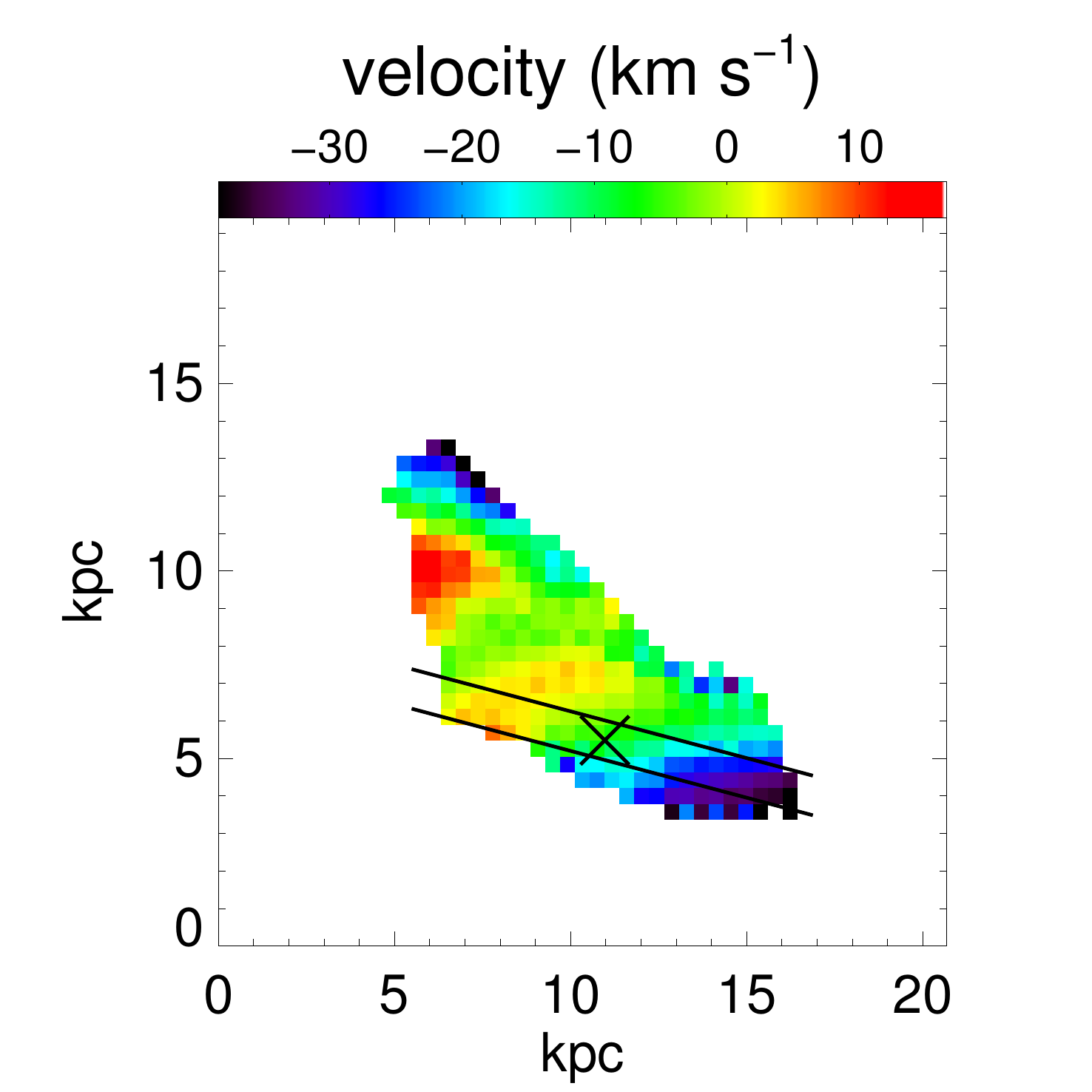}
	\includegraphics[width=5cm]{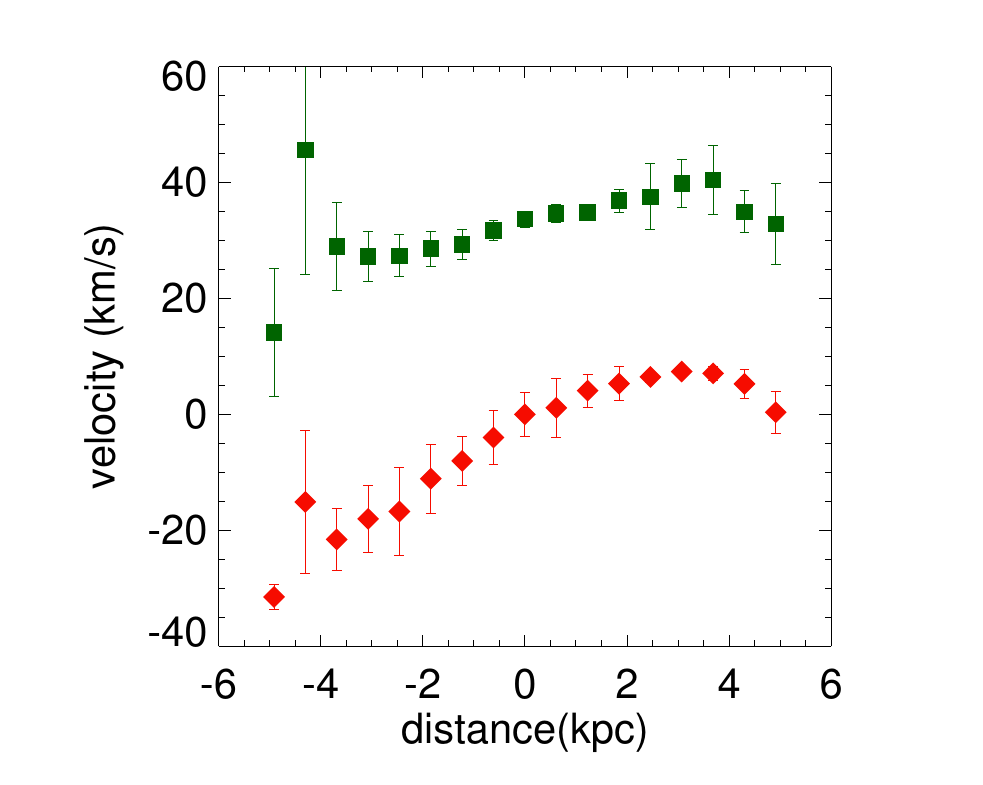}\\
	\includegraphics[width=4.5cm]{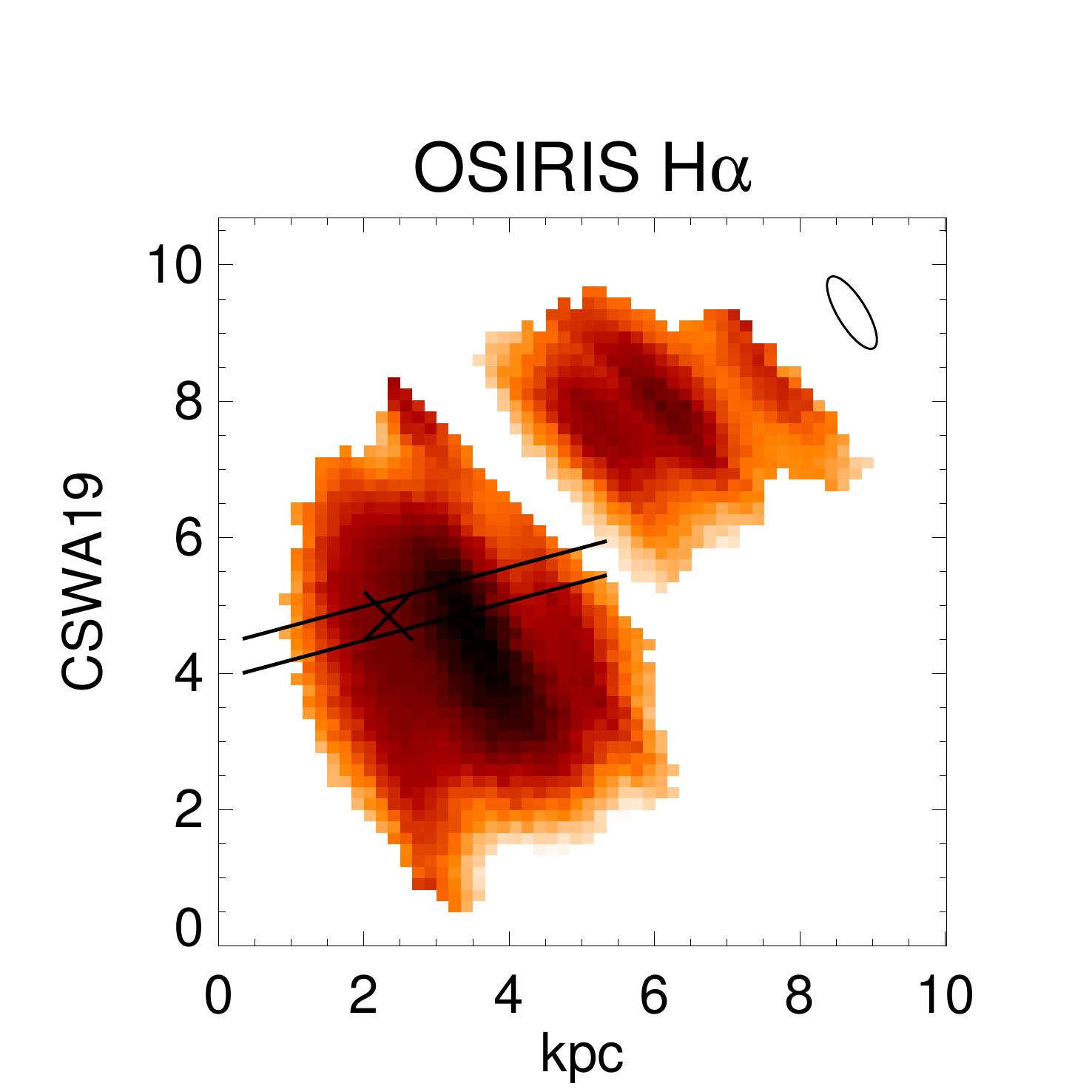}
	\includegraphics[width=4.5cm]{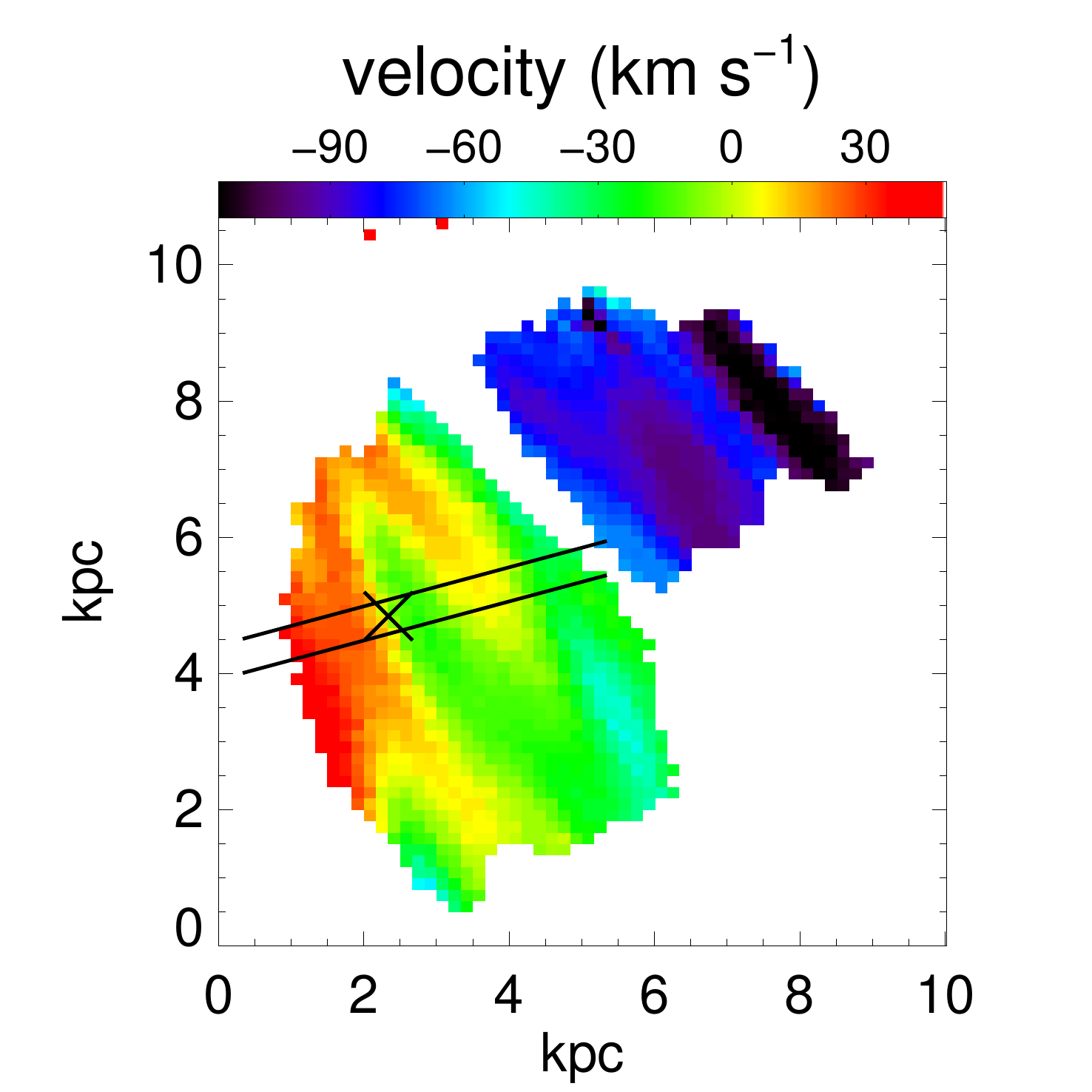}
	\includegraphics[width=5cm]{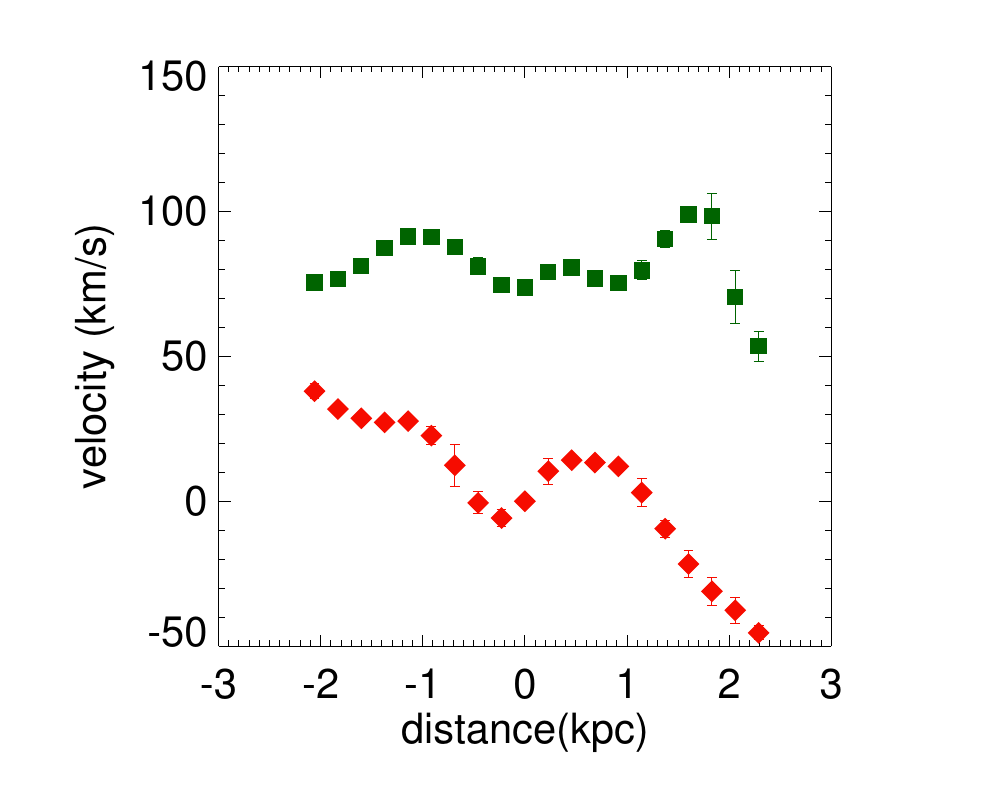}\\
	\includegraphics[width=4.5cm]{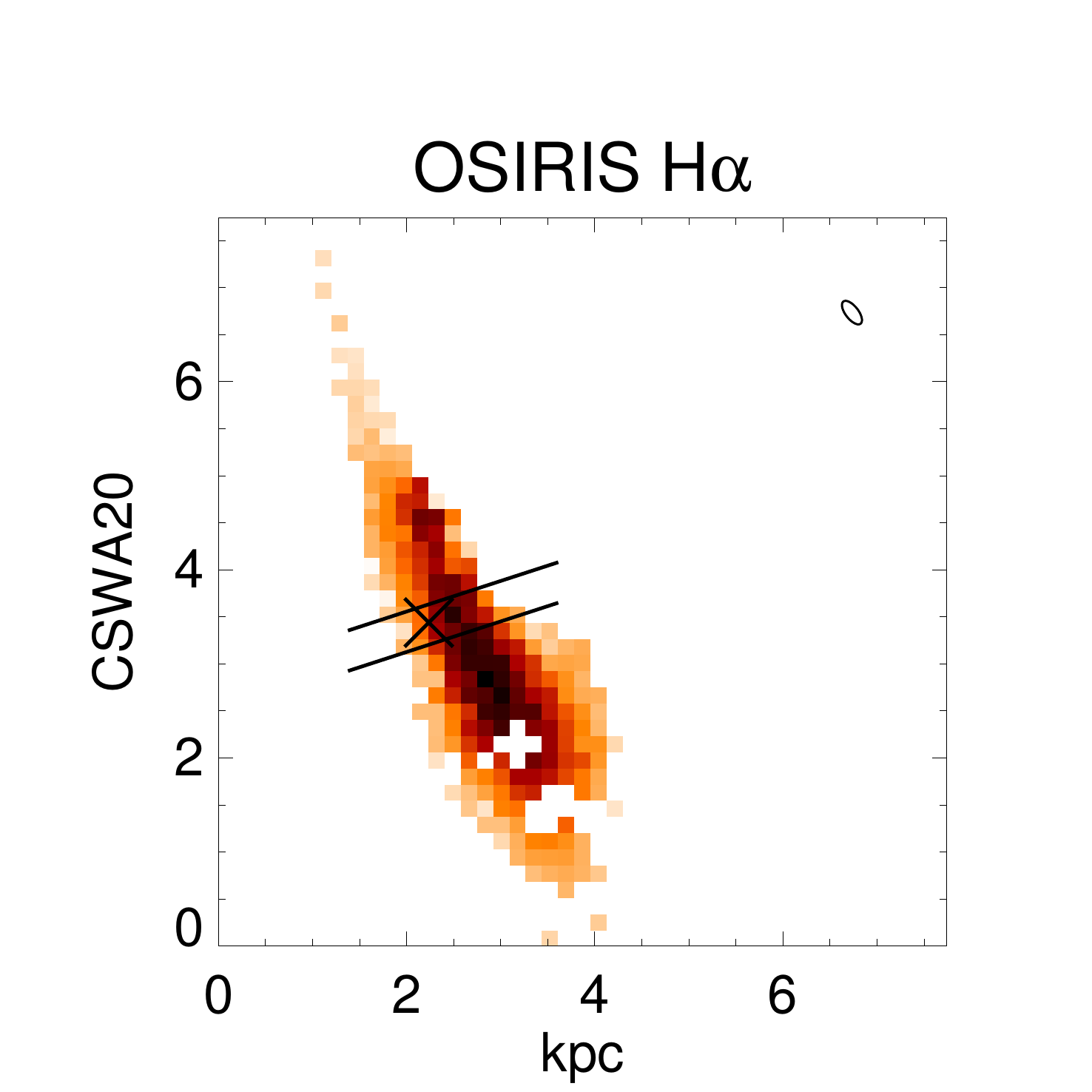}
	\includegraphics[width=4.5cm]{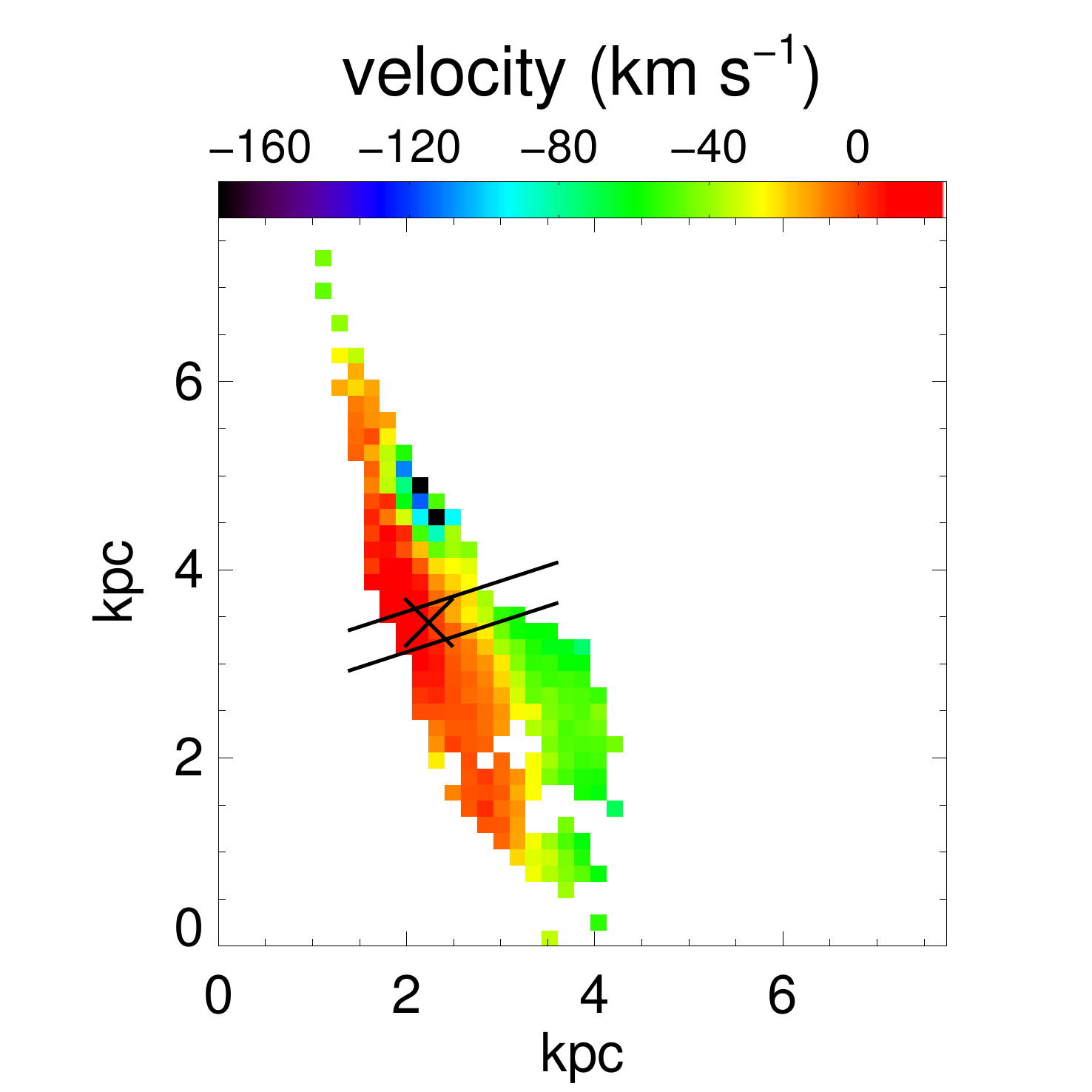}
	\includegraphics[width=5cm]{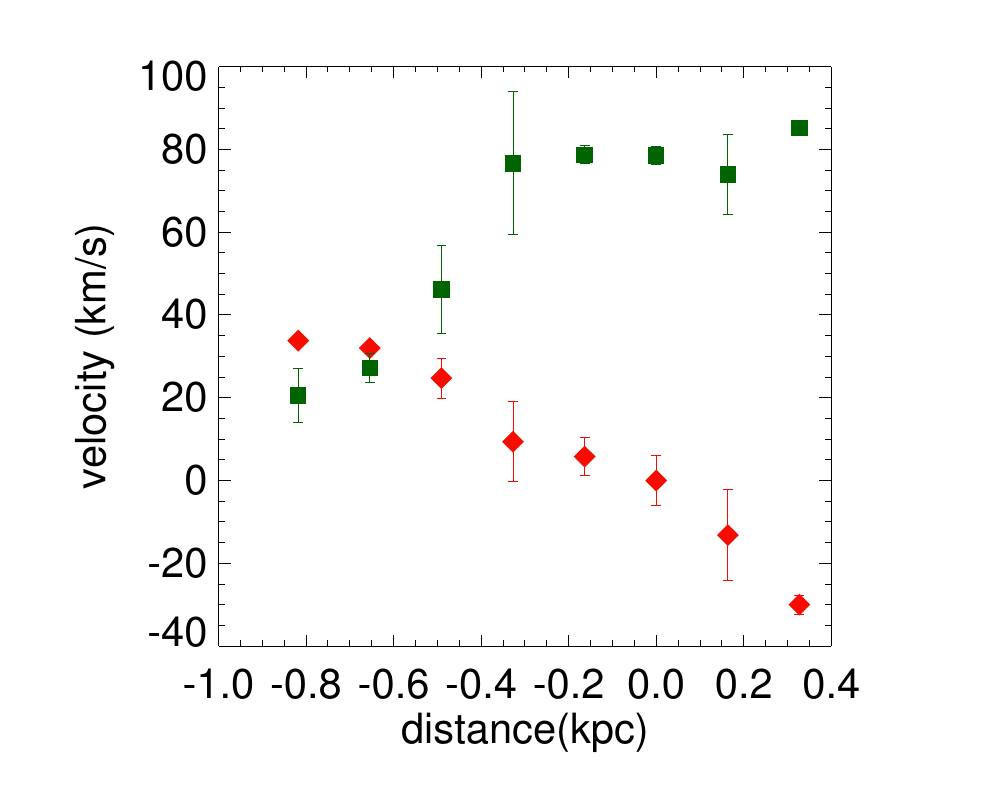}\\
	\includegraphics[width=4.5cm]{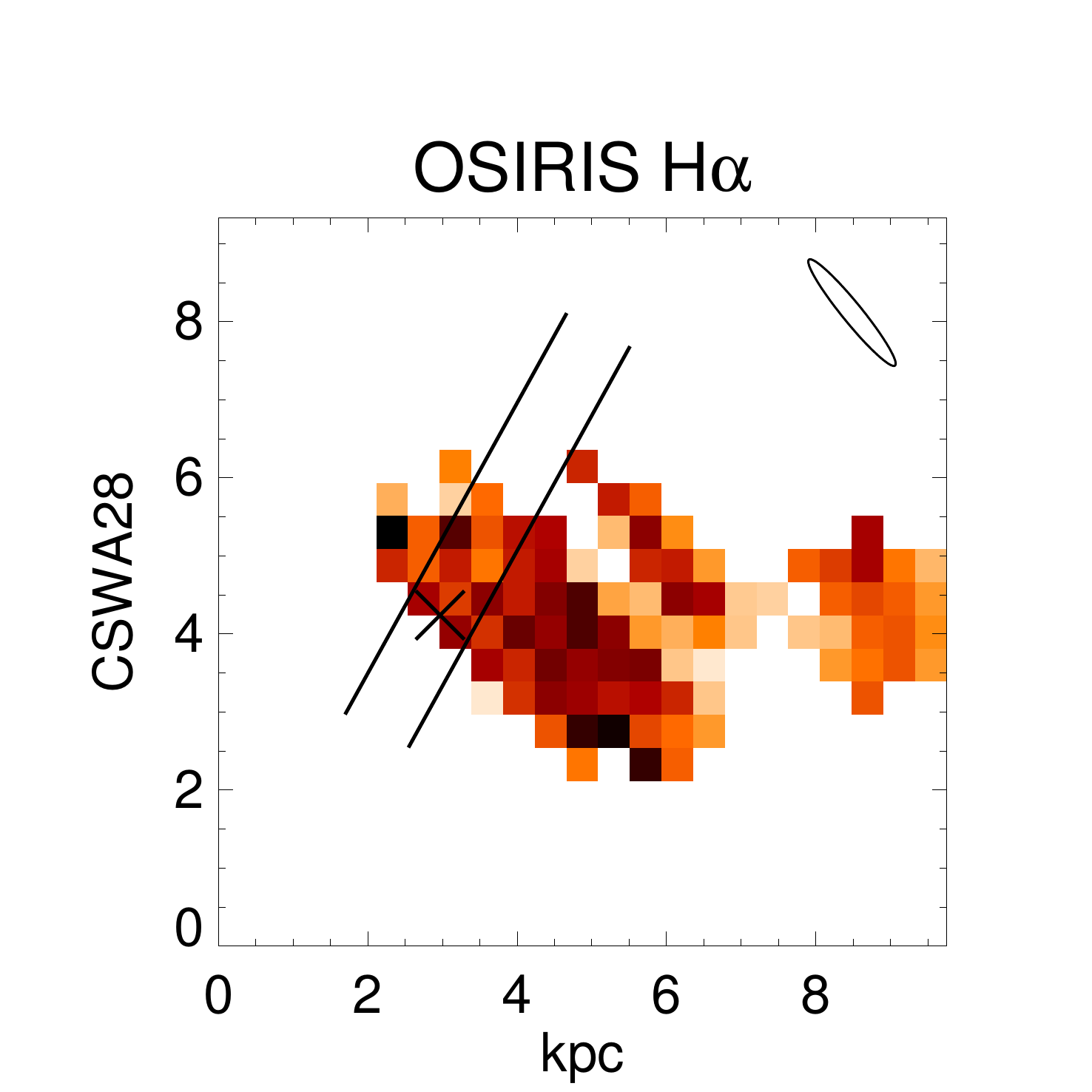}	
	\includegraphics[width=4.5cm]{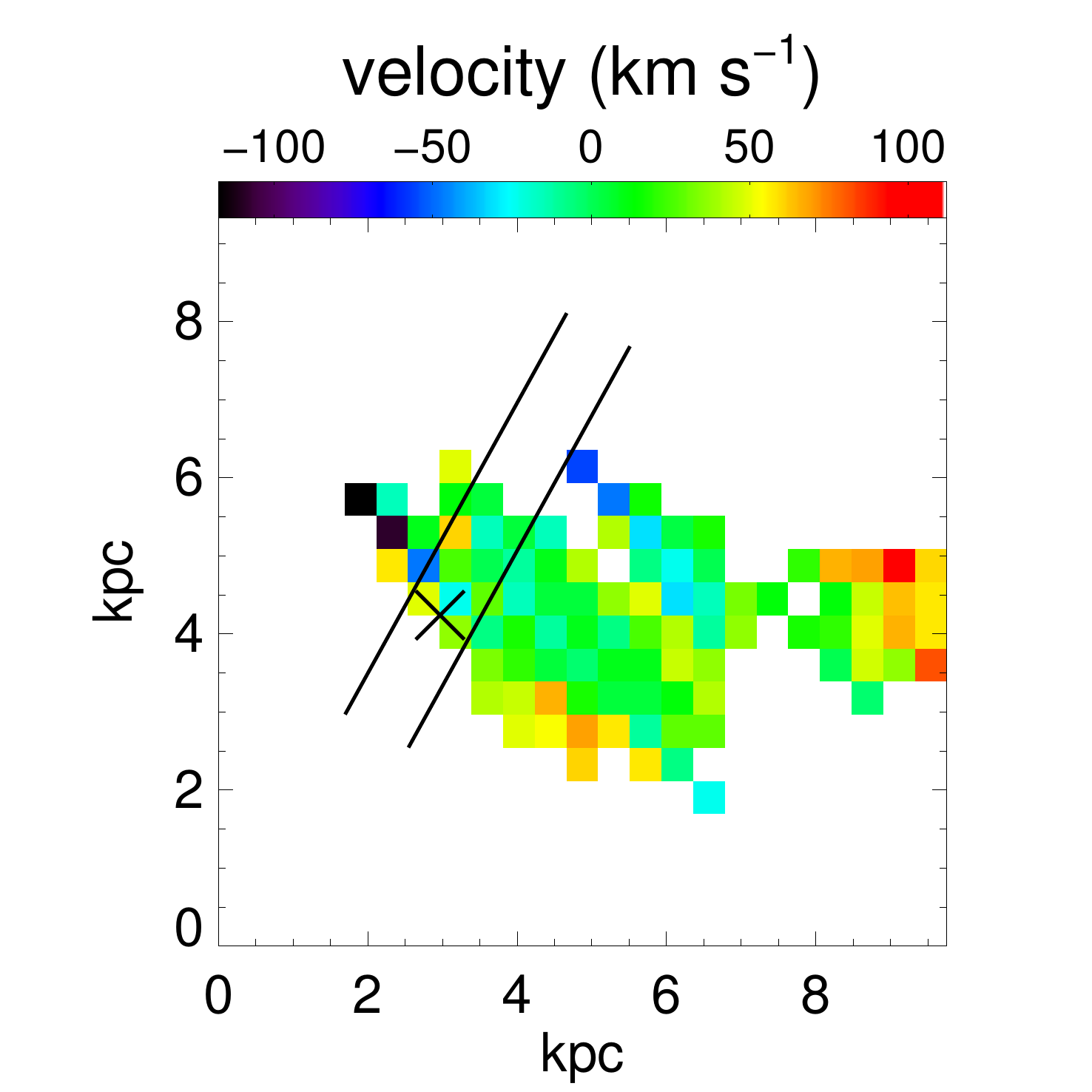}
	\includegraphics[width=5cm]{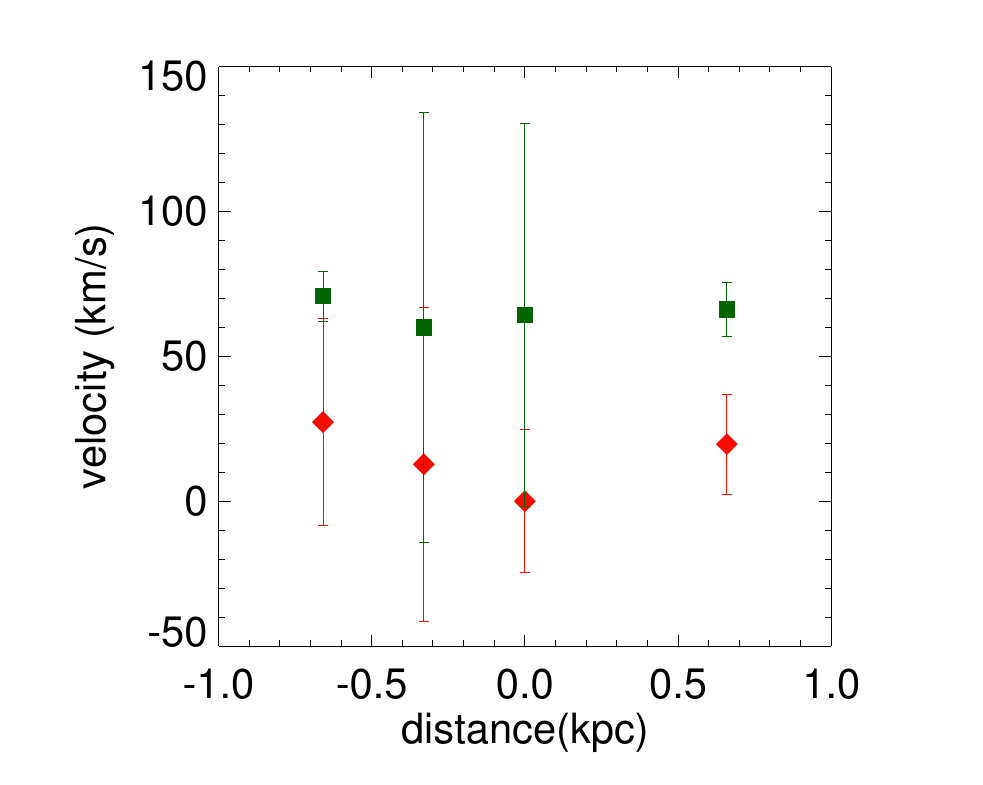}\\
			
\caption{From left to right: source plane H$\alpha$ emission map, two-dimensional velocity field, and one-dimensional velocity (red diamonds) and dispersion (green squares) for each of our targets. Ellipses represent the source-plane PSFs. Solid lines show the slits used to extract the one-dimensional velocity and dispersion with position angles determined from our simple disk model fits (see text in Section \ref{sec:kinematics} for details). Black crosses mark the adopted centers of each galaxy. }
\label{fig:velmaps}
\end{figure*}
\begin{figure*}Figure9
\ContinuedFloat
\centering
	\includegraphics[width=4.5cm]{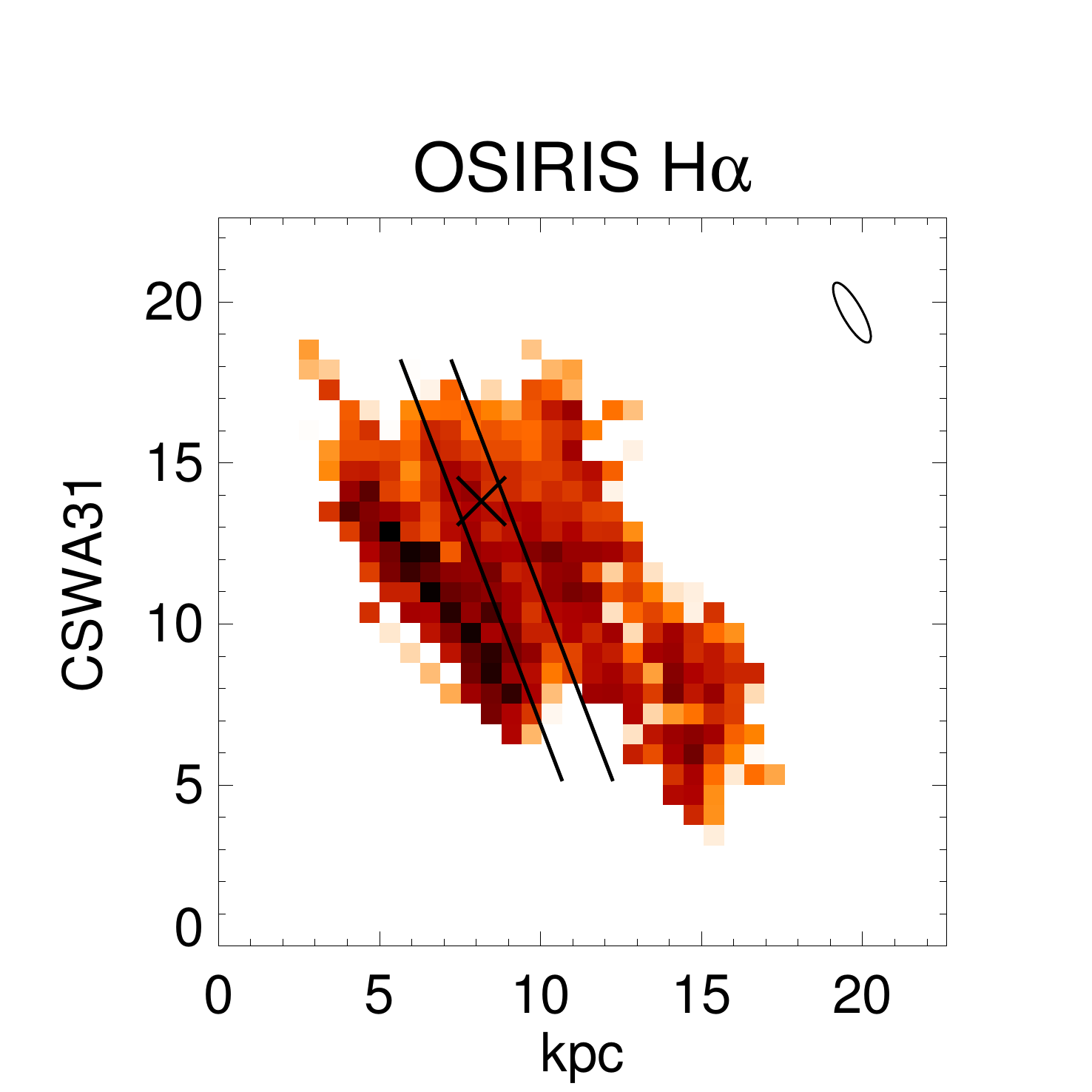}	
	\includegraphics[width=4.5cm]{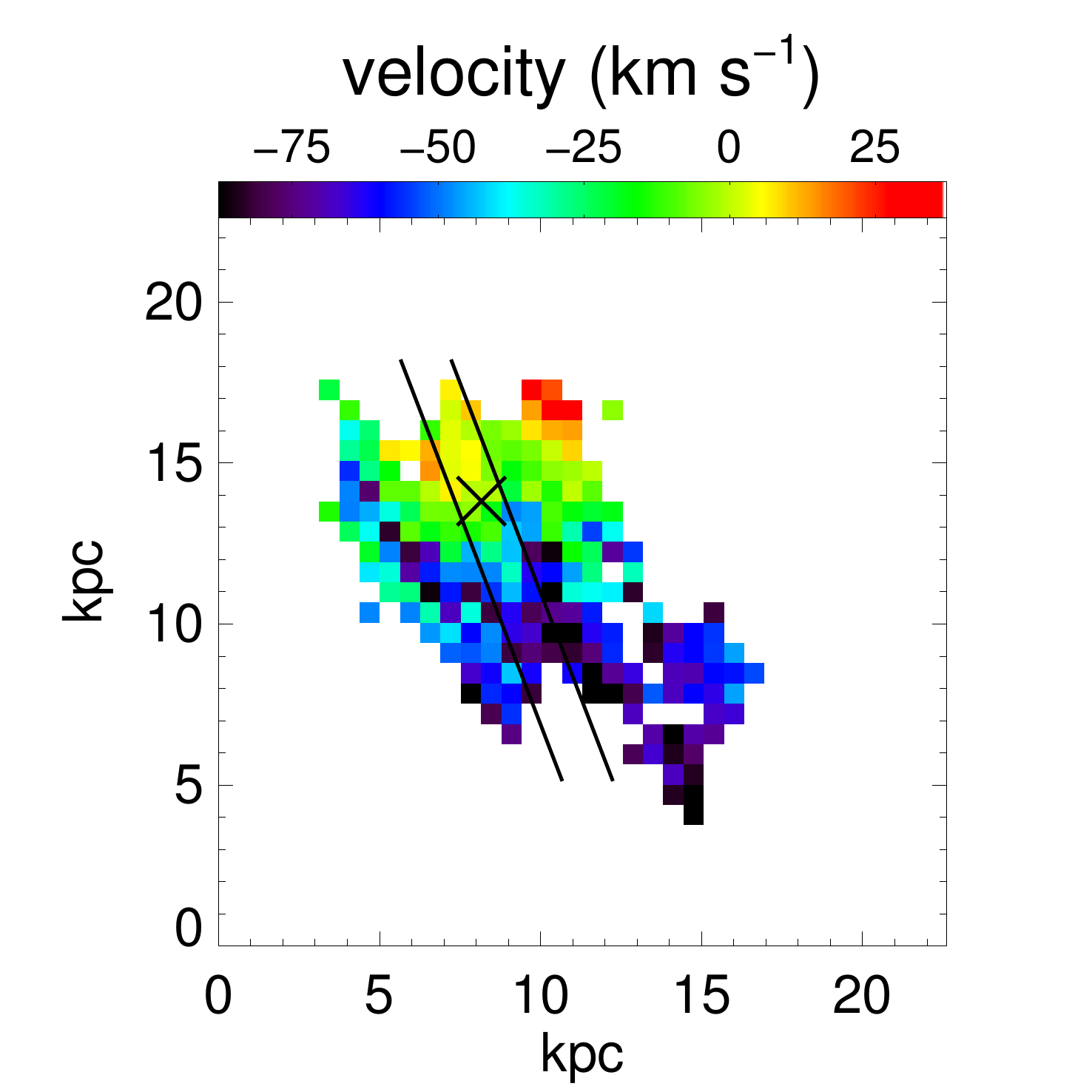}
	\includegraphics[width=5cm]{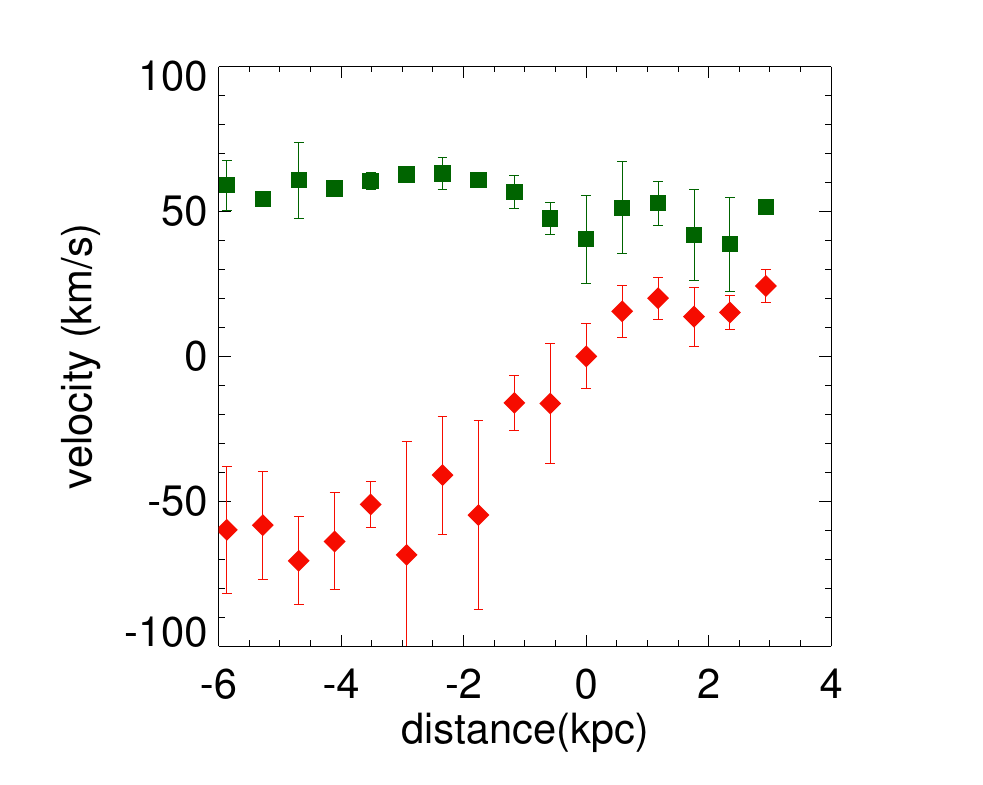}\\
	\includegraphics[width=4.5cm]{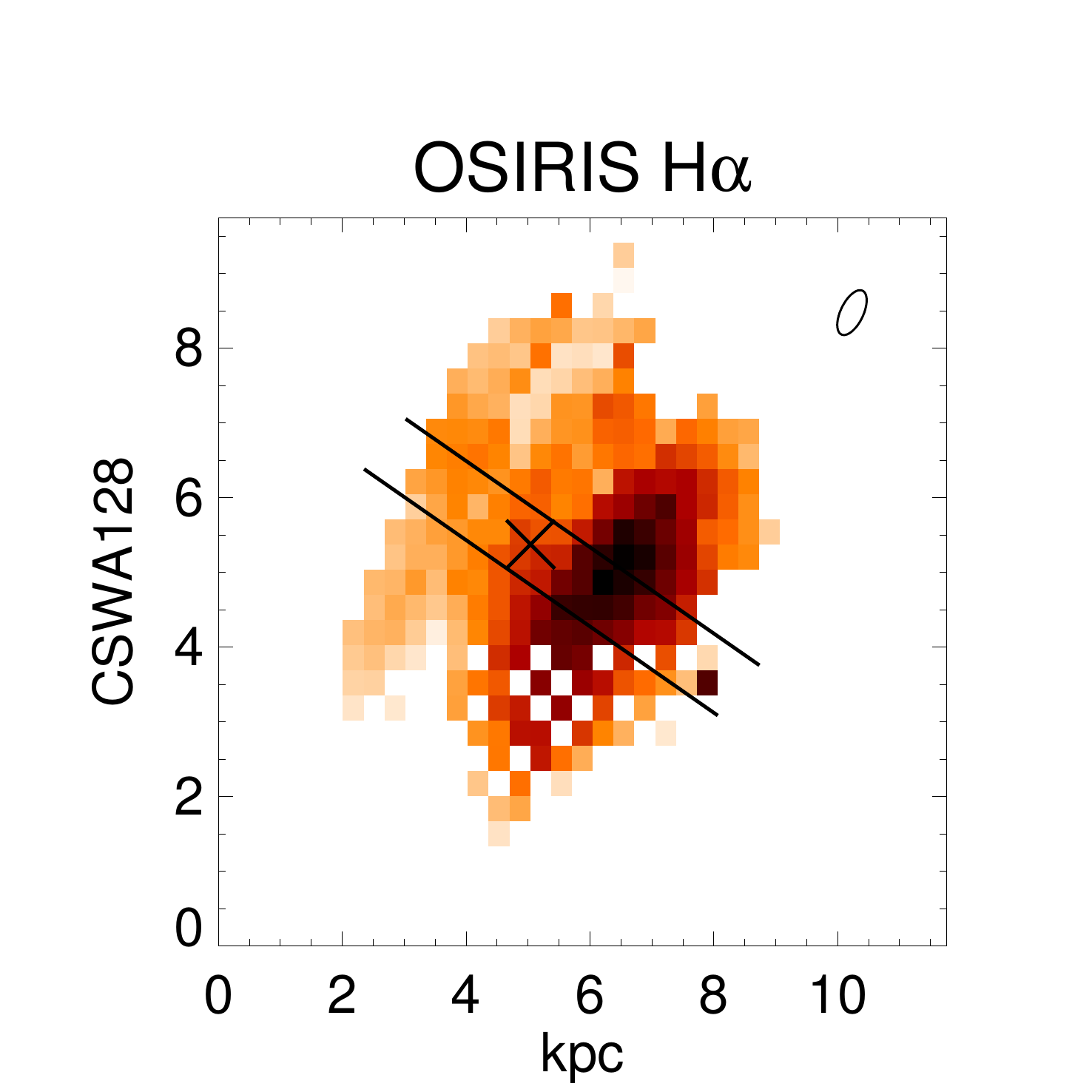}	
	\includegraphics[width=4.5cm]{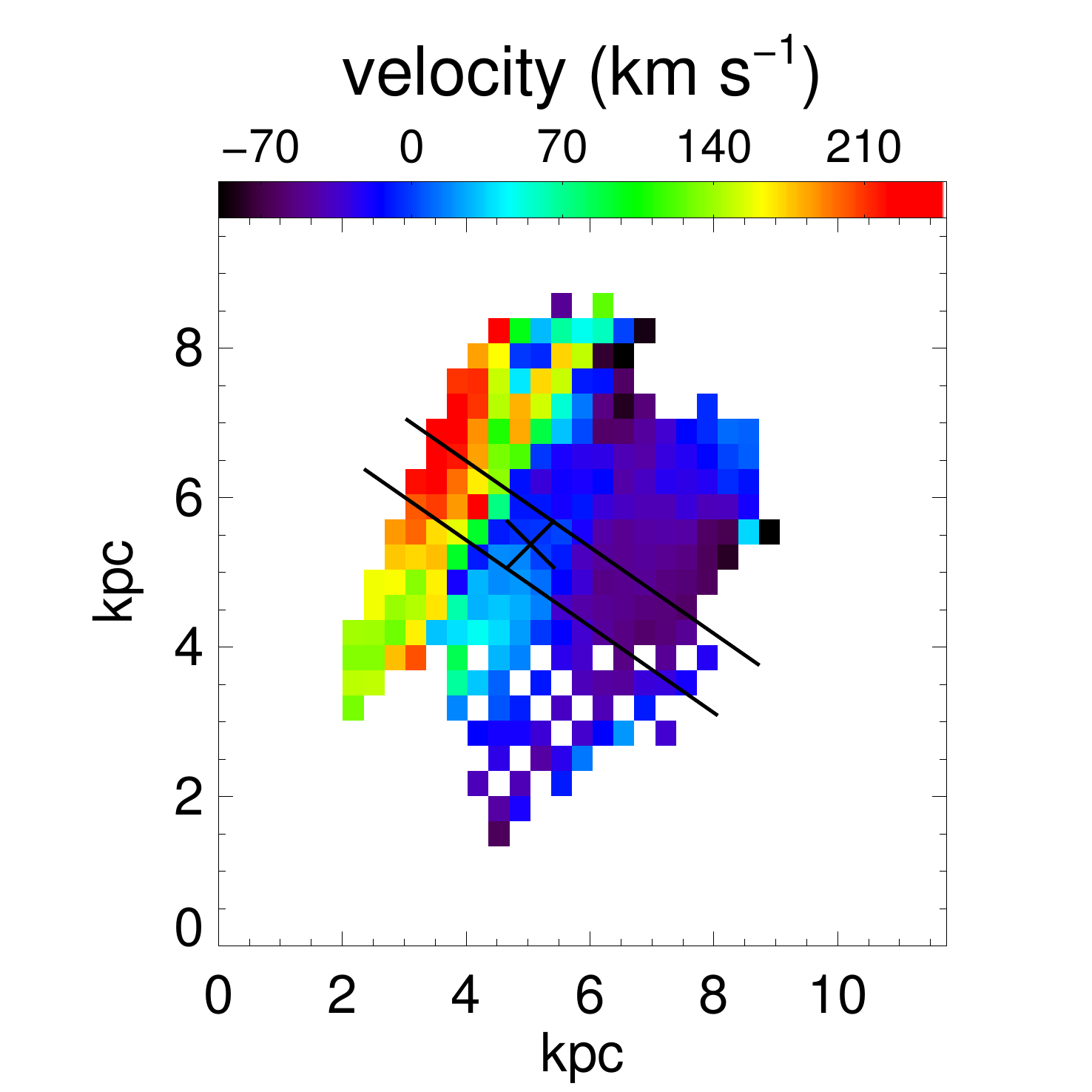}
	\includegraphics[width=5cm]{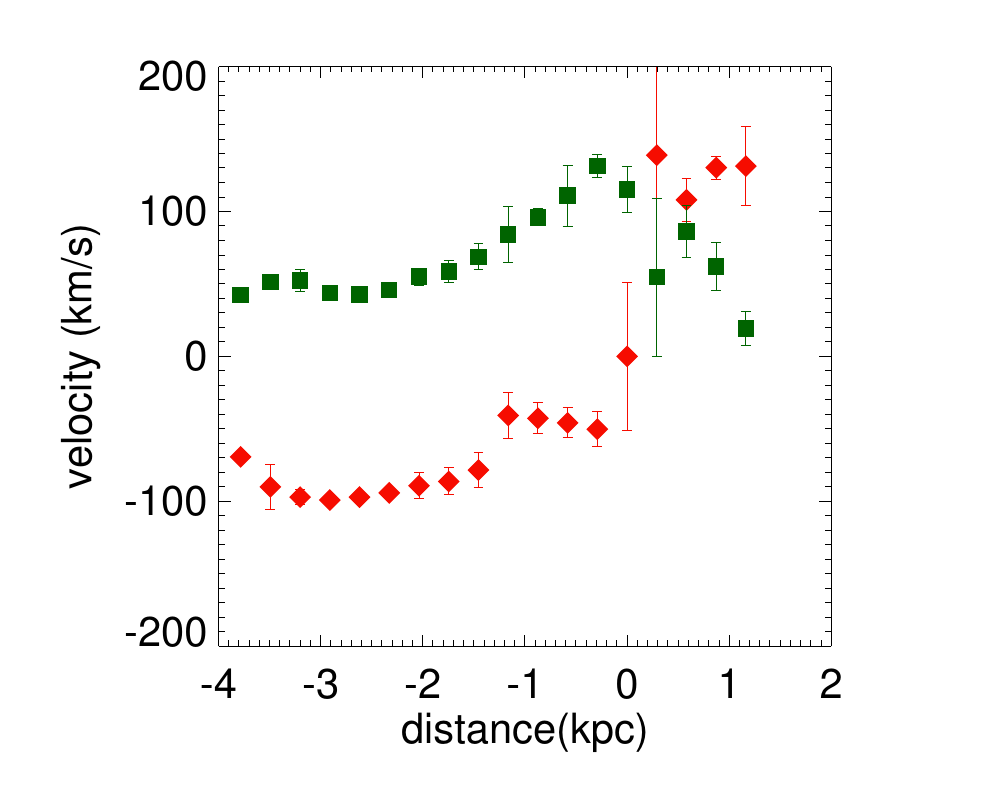}\\	
	\includegraphics[width=4.5cm]{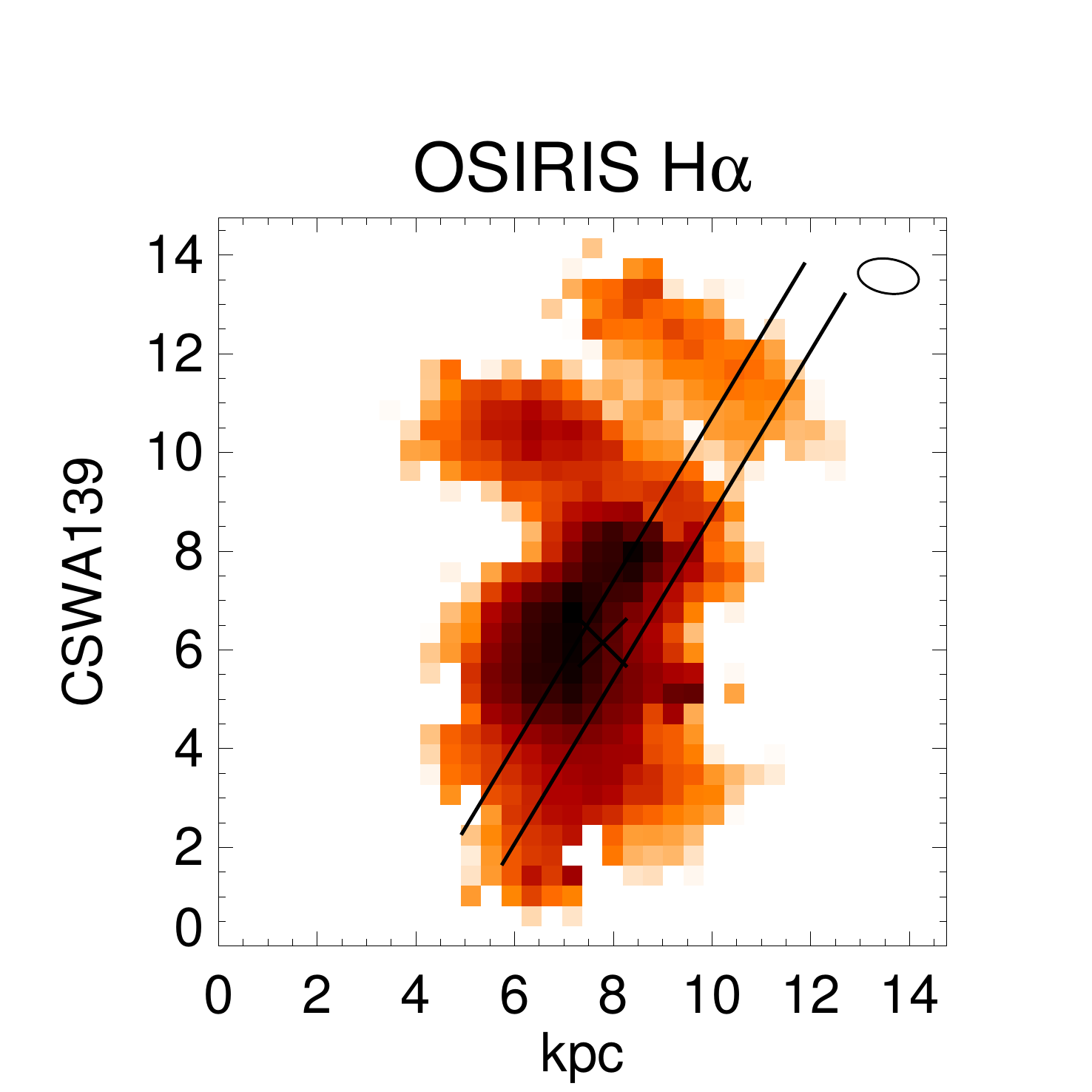}		
	\includegraphics[width=4.5cm]{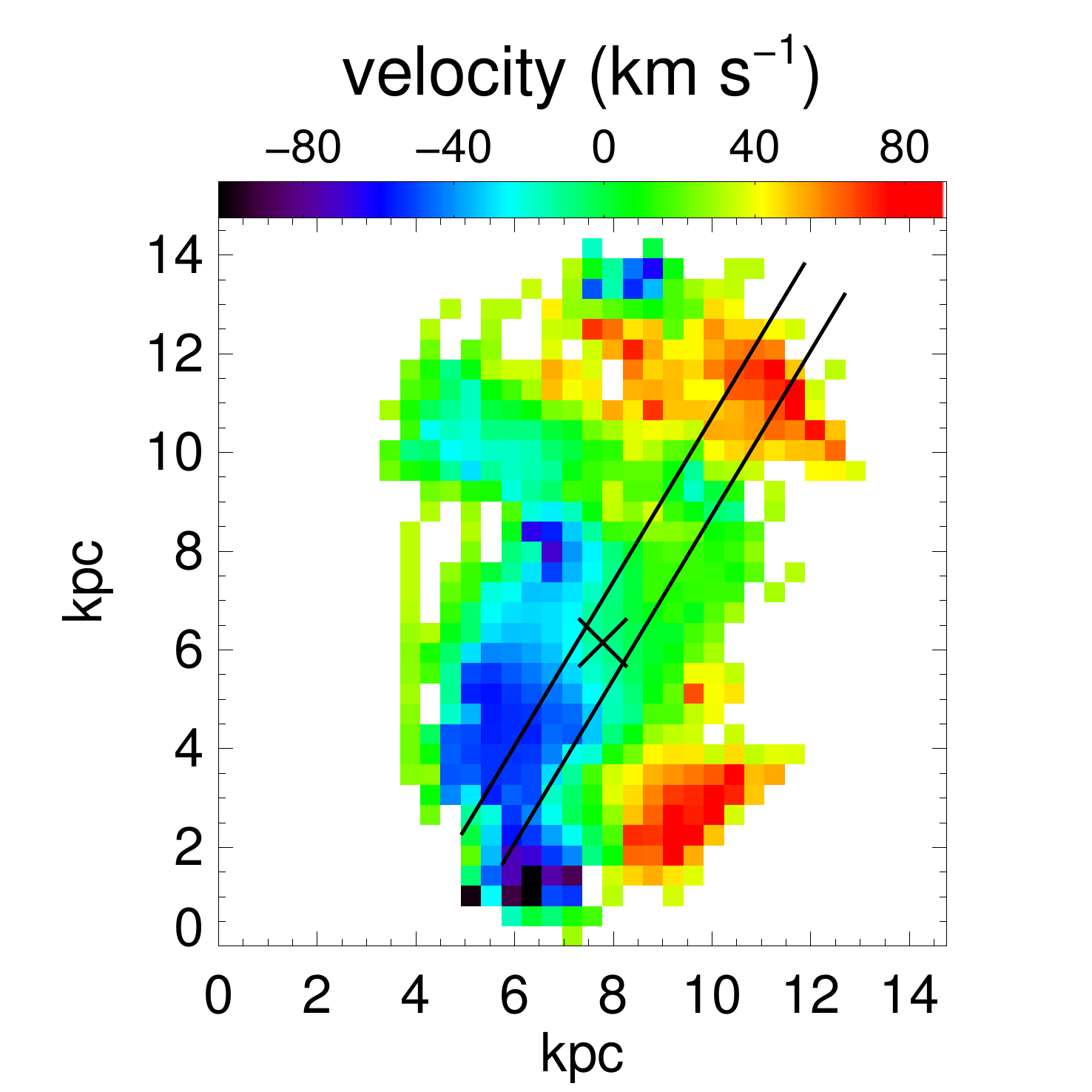}
	\includegraphics[width=5cm]{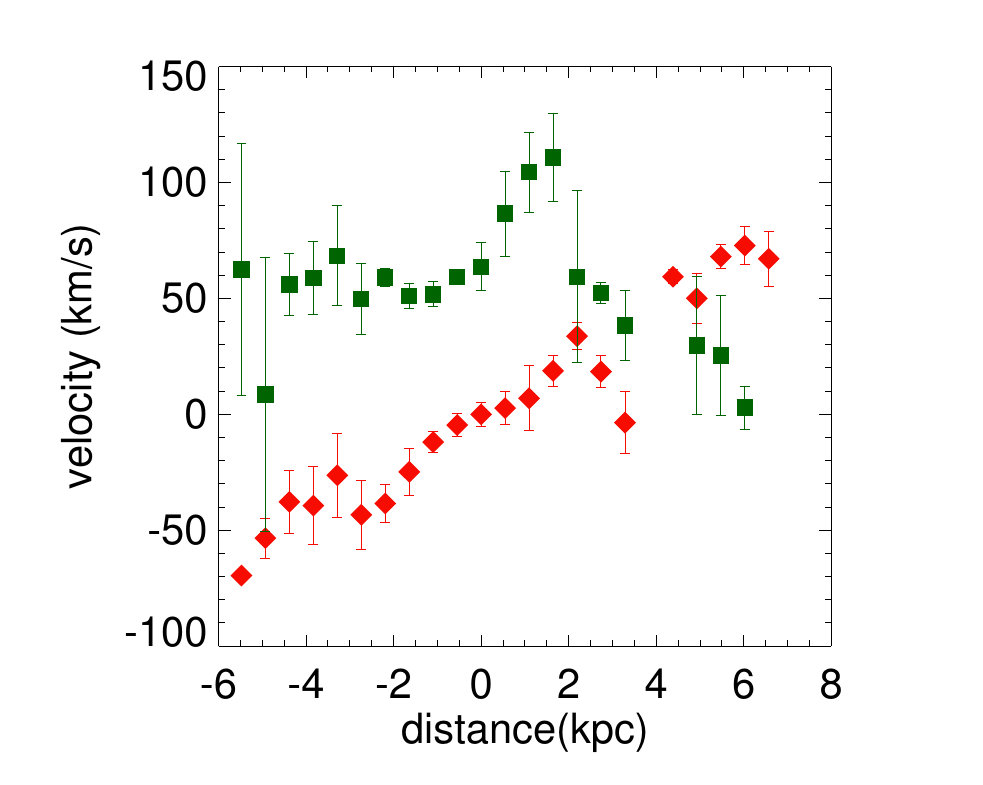}\\
	\includegraphics[width=4.5cm]{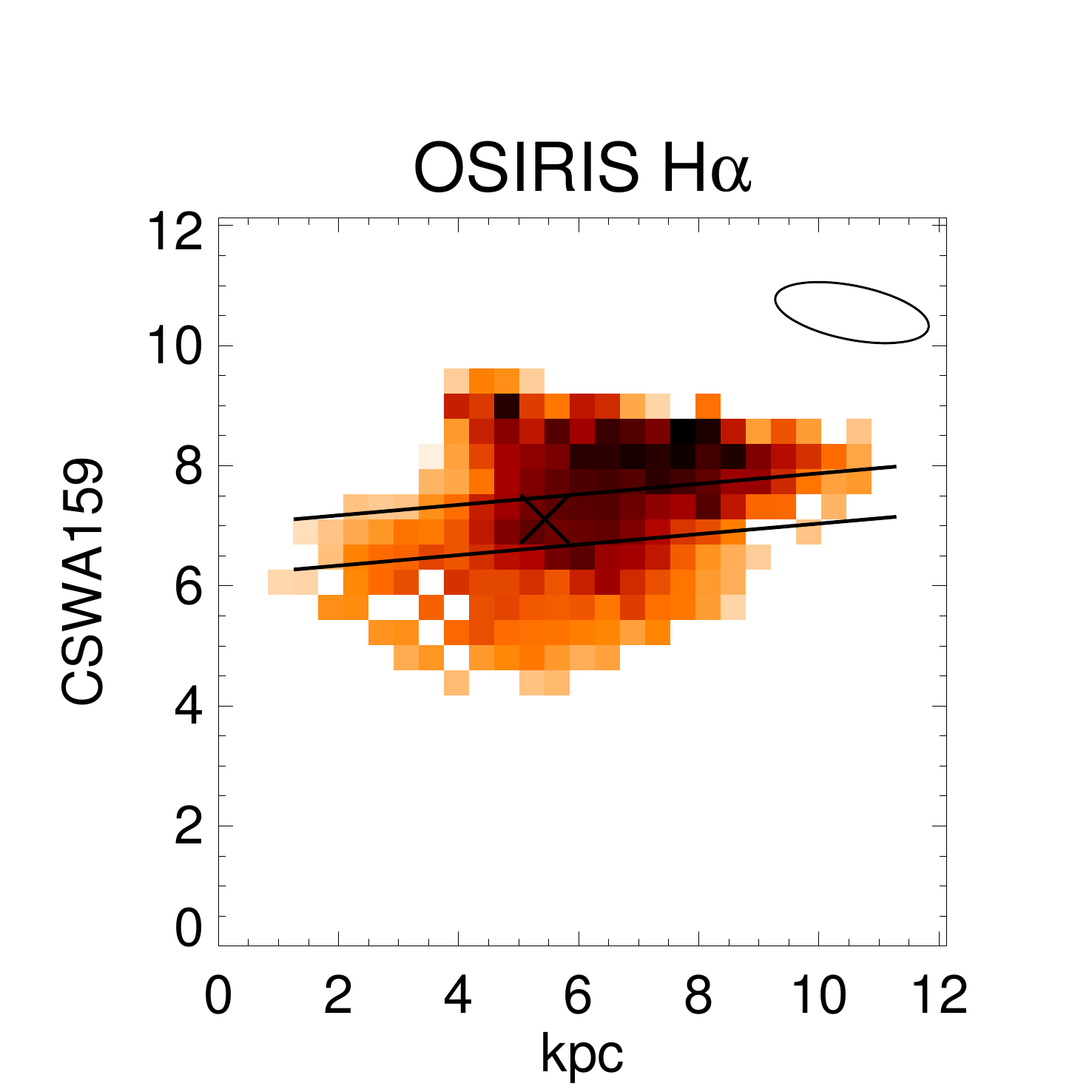}		
	\includegraphics[width=4.5cm]{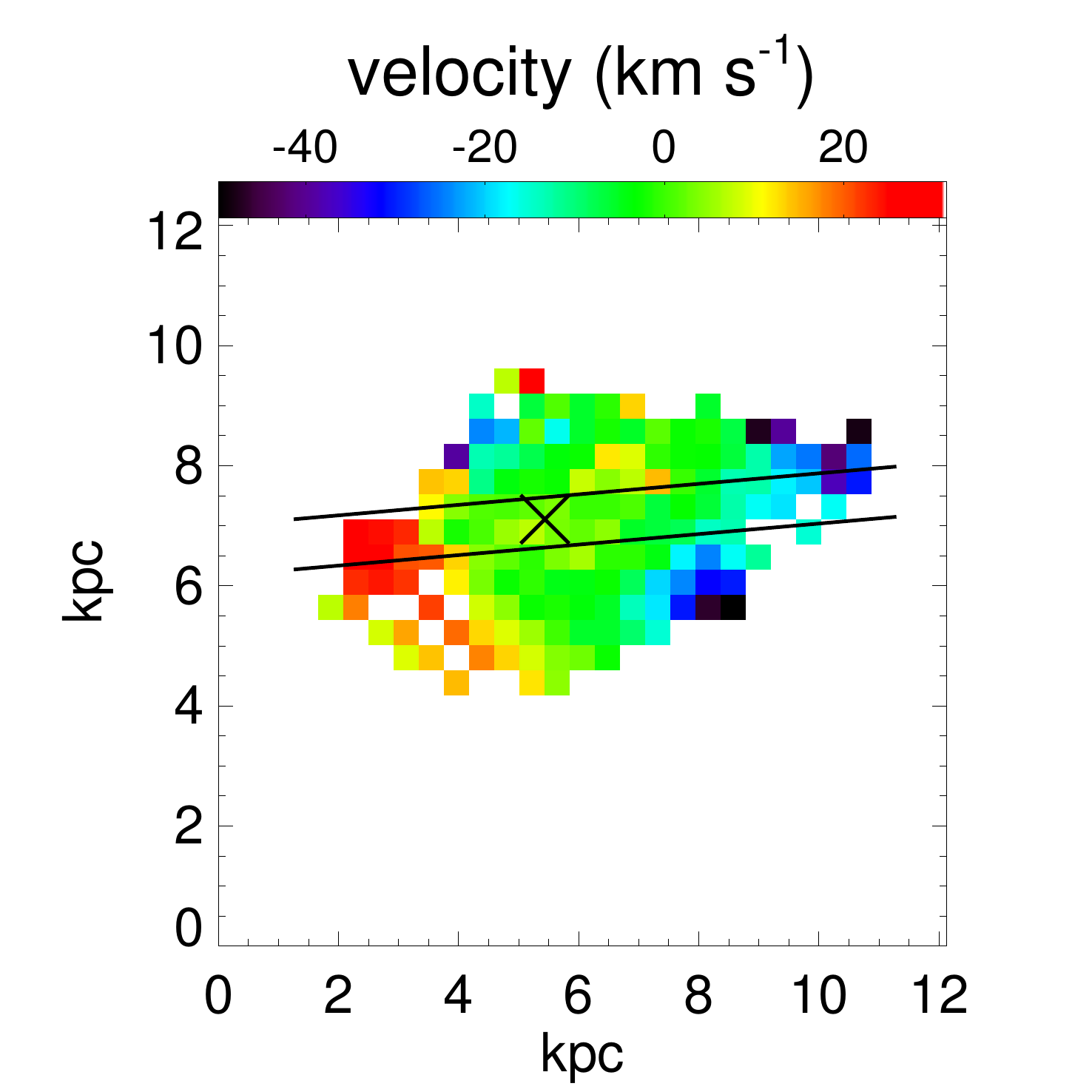}
	\includegraphics[width=5cm]{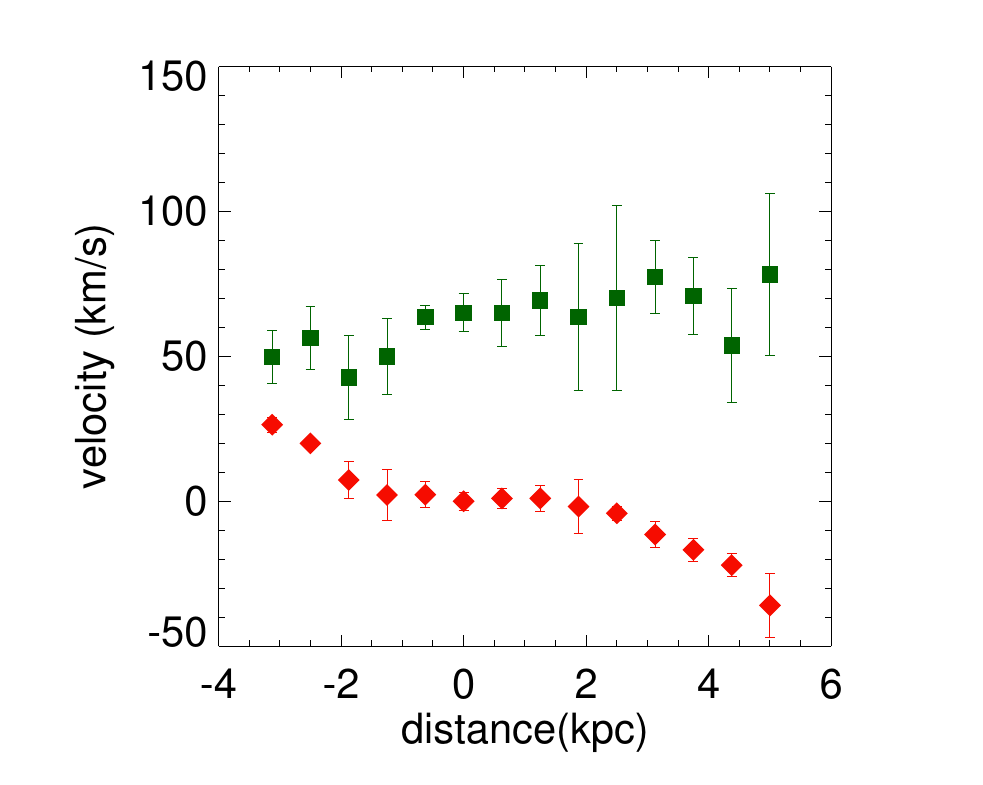}\\
	\includegraphics[width=4.5cm]{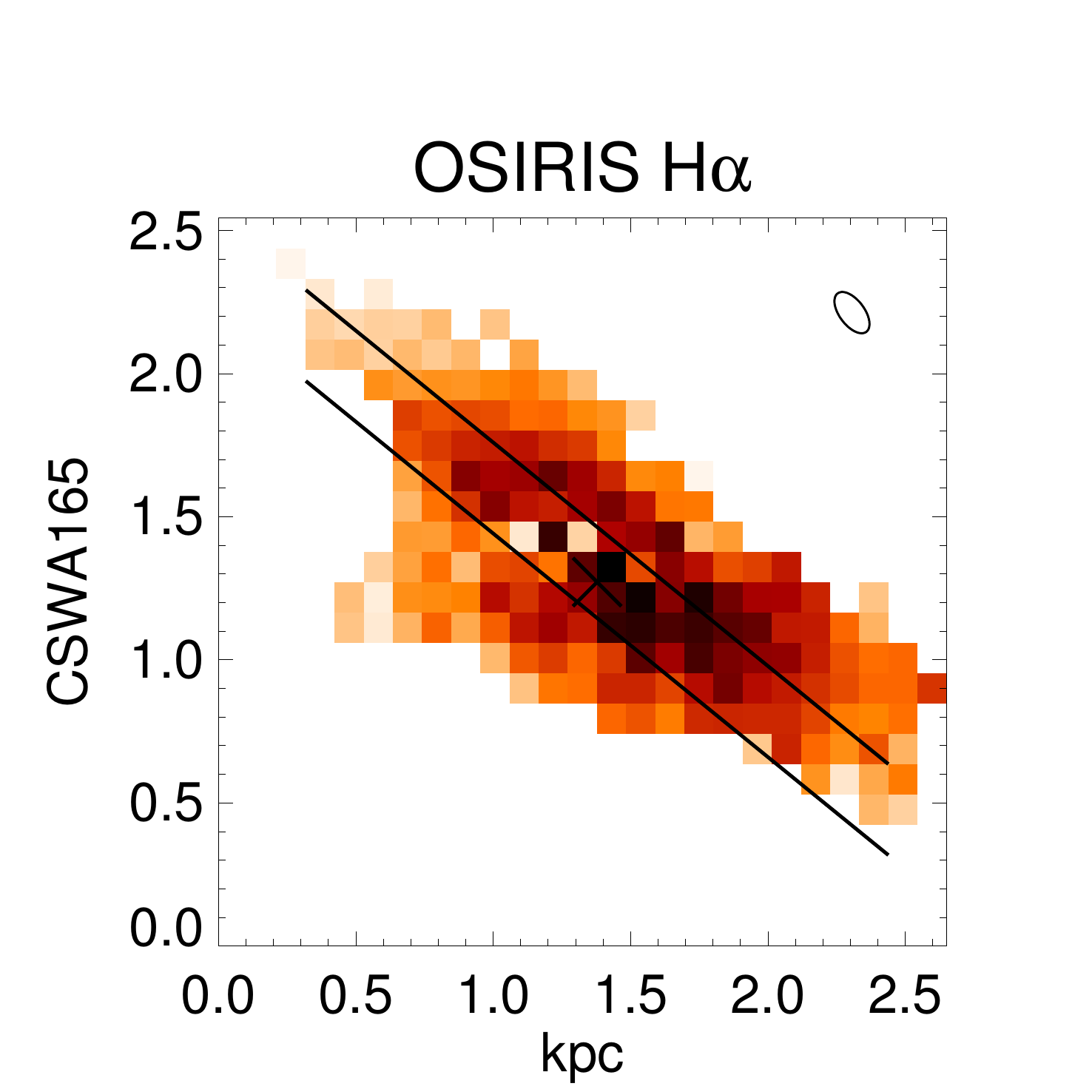}			
	\includegraphics[width=4.5cm]{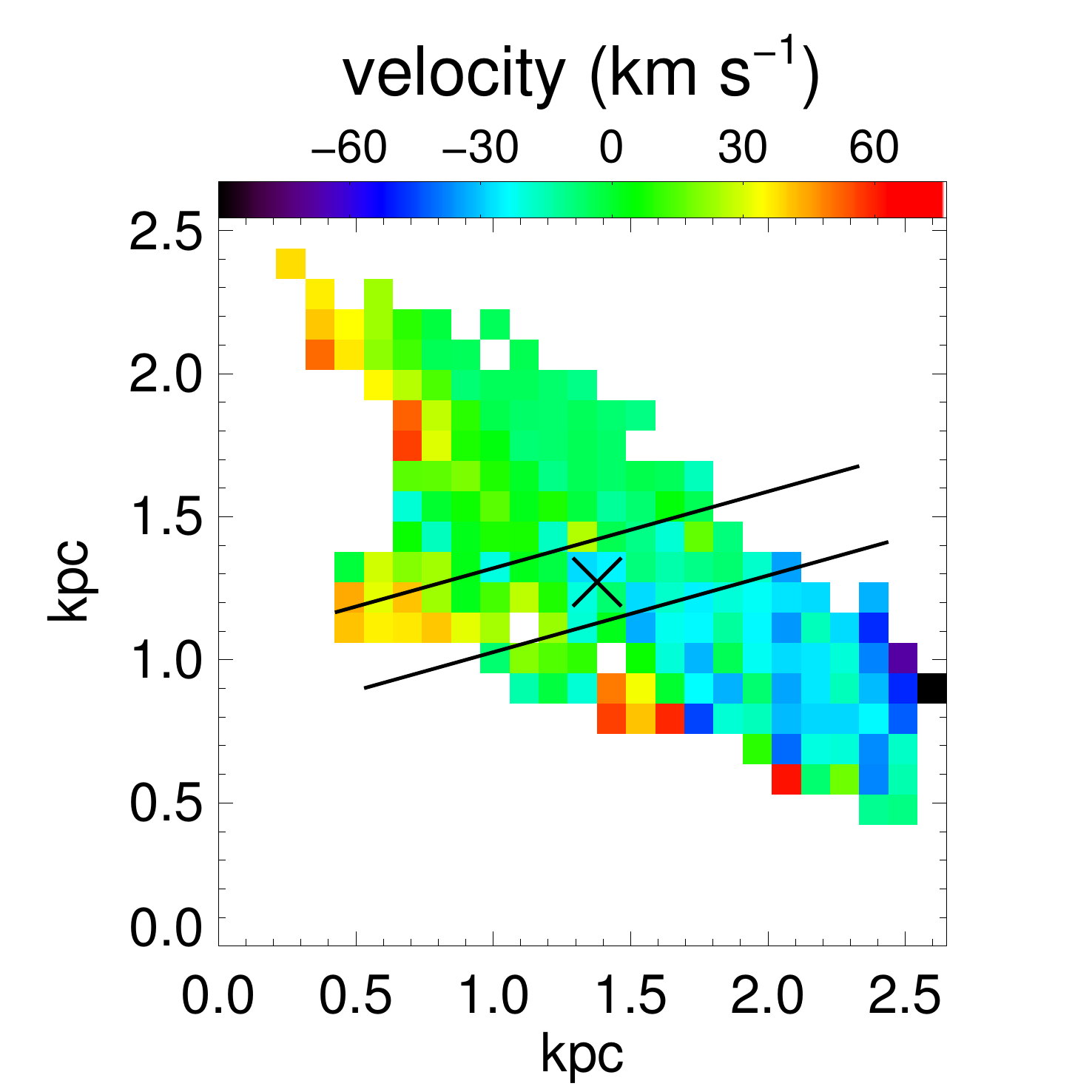}
	\includegraphics[width=5cm]{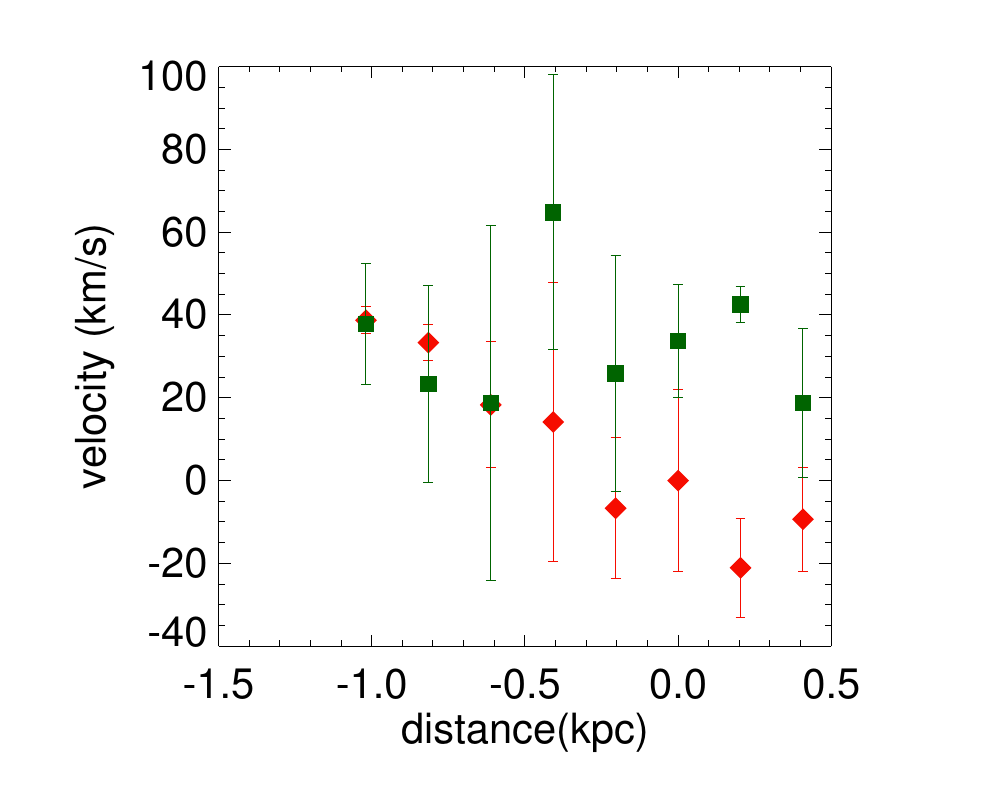}
		\caption{continued}
\end{figure*}
\begin{figure*}
\ContinuedFloat
\centering
	\includegraphics[width=4.5cm]{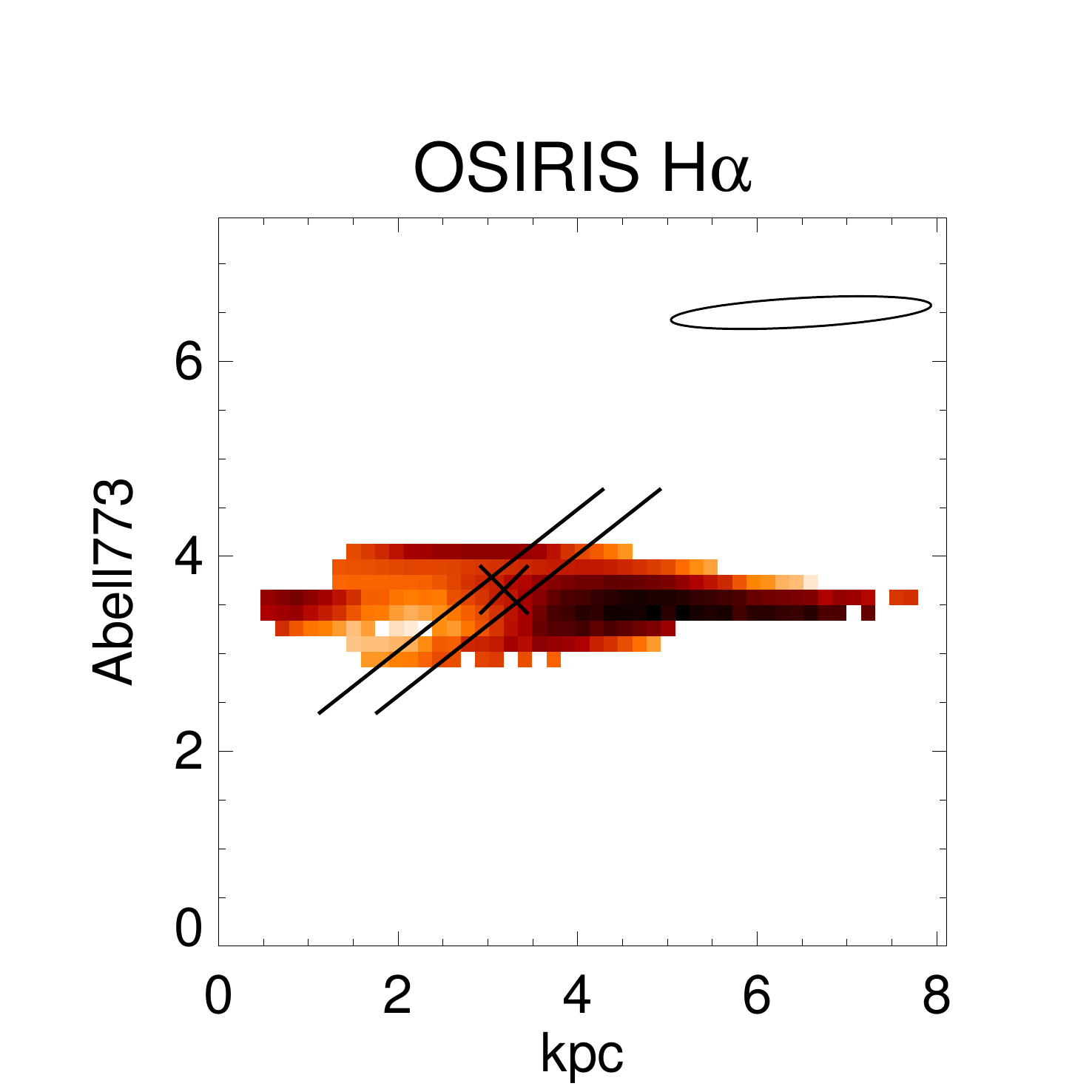}			
	\includegraphics[width=4.5cm]{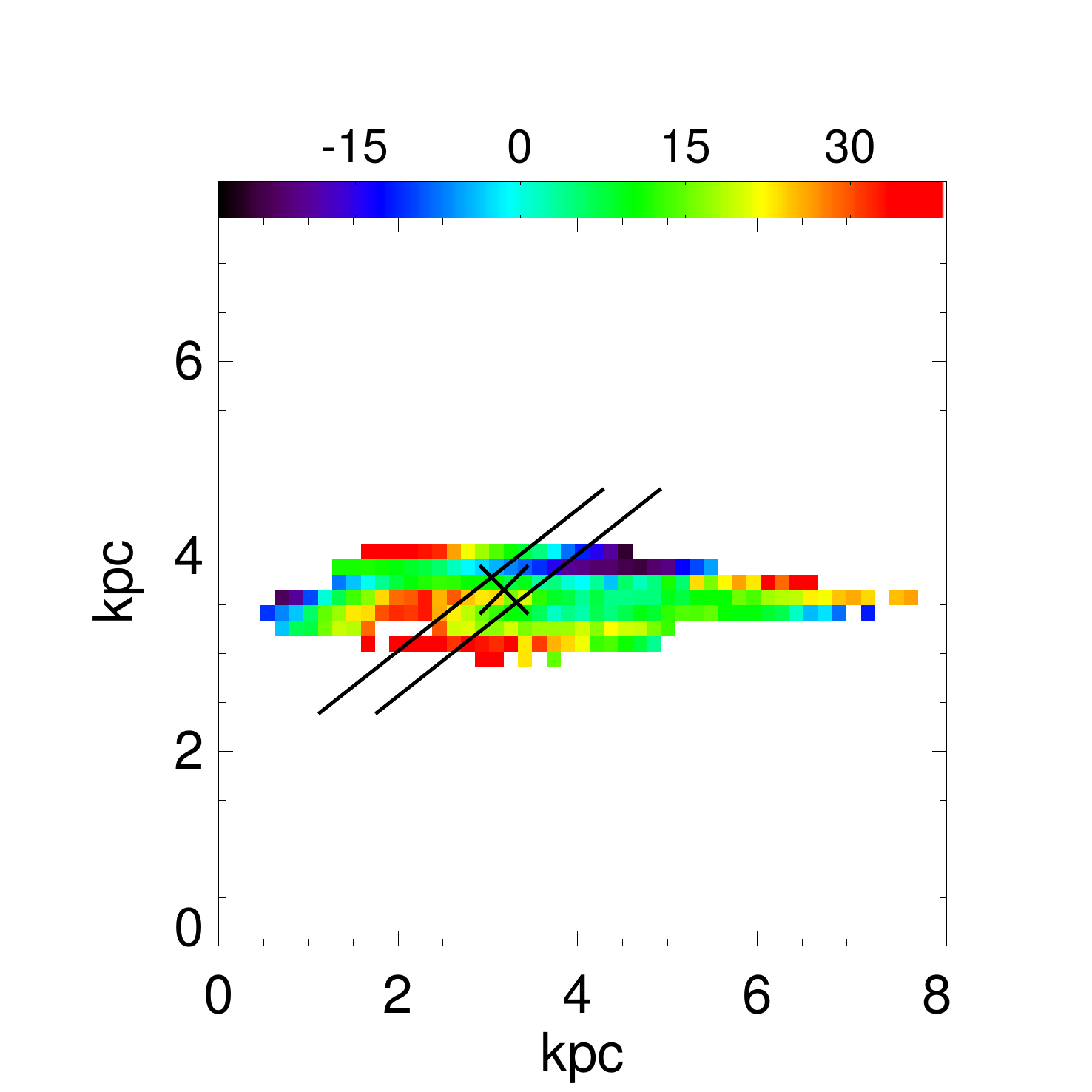}
	\includegraphics[width=5cm]{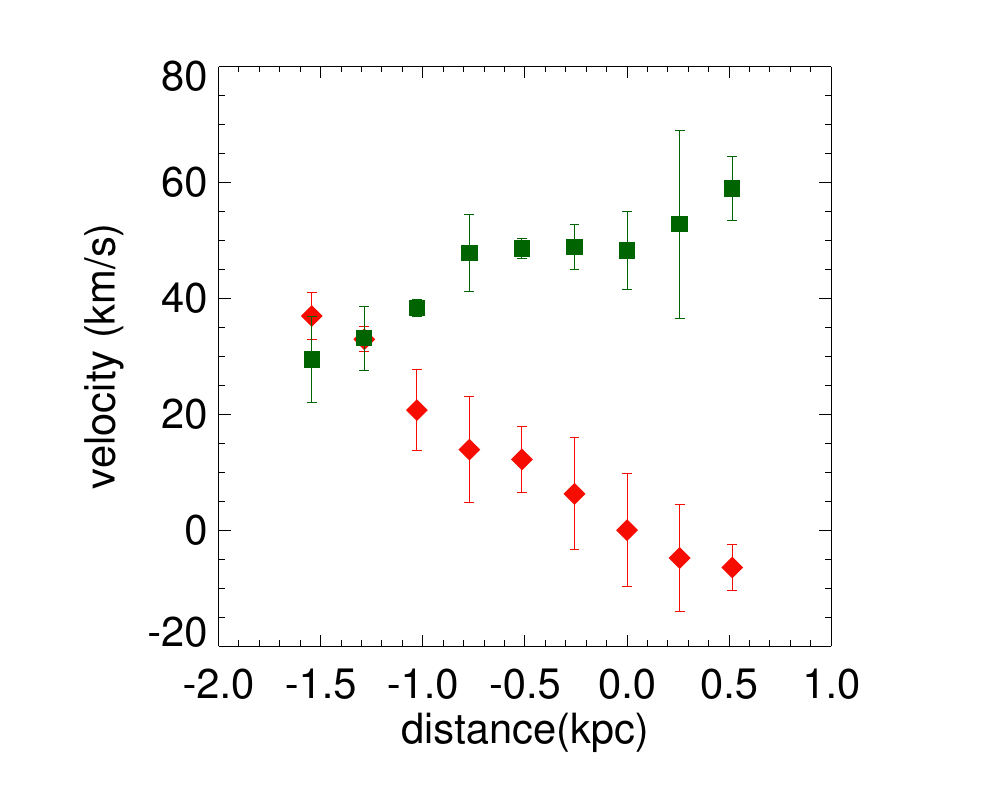}
	\caption{continued}
\end{figure*}

The improved spatial sampling made possible by studying gravitationally-lensed sources offers the key to resolving several uncertainties of interpretation of the kinematic and morphological properties of high redshift star-forming galaxies. These include discriminating between single rotating systems and close merging pairs as well as the validity of well-ordered disk models claimed to fit a high proportion of objects studied in the literature. For this study we also seek to examine possible correlations between the ratio of ordered and random motions ($v/\sigma$ - typically used as a measure
of the extent to which a system is or is not dispersion-dominated) with the presence of a metallicity gradient (see \S5) as this might indicate important diagnostics of how disks assemble and mature over a key period in cosmic history. For these and other applications, clearly determining the reliability of the derived values of the rotational velocity and the dispersion is crucially important.

Early LGSAO-based data on lensed sources already highlighted the importance of securing well-sampled data \citep{Jones10a,Jones10b}. Prior to these studies it was commonly believed that systemic rotation was confined to the galaxies with larger stellar masses \citep{Genzel06,Law09}. However, due to the small number of lensed sources available, limited statistics has remained a problem. \citet{Jones10b} presented spatially-resolved dynamics for 6 lensed systems claiming 4/6 showed well-ordered velocity fields - a fraction consistent with that determined for larger, more luminous sources. Recently, \citet{Livermore15} collated the available data on 17 lensed sources including 10 studied with natural guide star-assisted AO using SINFONI on the VLT and 6 with LGSAO and Keck's OSIRIS, mostly from \citet{Jones10b,Jones13}, similarly concluding $\simeq$60\% of these systems with stellar masses in the range $4\times10^8< M/M_{\odot}<6\times10^{10}$ are consistent with rotating disks. This fraction is lower than that deduced from the KMOS$^{3D}$ survey \citep{Wisnioski15} based on coarser sampled data of a larger ($\simeq$30) sample of $1.9<z<2.7$ galaxies where, over their entire sample, it is claimed 83\% of the sample are rotation-dominated with $v/\sigma>$1 and at least 70\% can be considered `disk-like' systems. 

Our newly-enlarged set of well-sampled resolved spectroscopic data provides a valuable test of the above claims and we now examine both the validity of a simple disk model as a fit to our source plane velocity fields as well as the extent to which our systems are rotation or dispersion-dominated. Our H$\alpha$ emission maps and their associated radial velocities fields are presented in Figure~\ref{fig:velmaps}. While our H$\alpha$-based radial velocity fields are in many cases indicative of velocity gradients, the associated H$\alpha$ surface brightness distributions have irregular, asymmetric, and non-disk like shapes. Such irregular emission line morphologies are common at these redshifts and star-formation rates and contrast with more regular distributions in broad-band filters sampling the stellar distributions. To verify this, for each galaxy, we attempted to extract the total continuum spectrum adjacent to the H$\alpha$ line. Although the resulting signal to noise of the continuum is limited beyond a scale of $\simeq$1 kpc, it provides a valuable indicator of the center of each galaxy.
 
First we assess the validity of a rotating disk as a representation of the observed H$\alpha$ radial velocity field following the method adopted in \citet{Jones10b}. Those authors used an arctangent function as the simplest model, viz:
\begin{equation}
V(R) = V_0+\frac{2}{\pi}V_c \mbox{ arctan }\frac{R}{R_t}
\label{eq:disc_model}
\end{equation}
where R is the radius from the disc center,
\begin{equation}
R^2= (x\cos\theta-y\sin\theta)^2+(\frac{x\sin\theta+y\cos\theta}{\sin i})^2 \nonumber
\end{equation}
The simple model has seven unknown parameters: the inclination $i$, position angle $\theta$, the position of the disk center $(\alpha_c,\delta_C)$, the scale radius $R_i$, an asymptotic velocity $V_c$, and an overall systemic velocity $V_0$. The variables $x$ and $y$ represent the distance of each pixel from the disc center. The resulting asymptotic velocity is corrected for the effect of inclination.

We use a Markov-chain Monte Carlo (MCMC) method to find the best fitting disk model parameters. We fit the observed H$\alpha$ radial velocity map with the model convolved with an elliptical point spread function based on the reconstructed tip-tilt star image. In the MCMC fitting, we set flat priors within the parameter ranges as follows.  The disk center is set to lie within the peak of the stellar continuum light distribution (which does not necessarily coincide with the H$\alpha$ emission peak). The scale radius is confined to lie within the range of the observed H$\alpha$ map. The asymptotic velocity is $\lesssim 3 \times \Delta V$ where $\Delta V$ is the maximum velocity difference along the slit that passes through the disk center for all position angles. We allow the full range for inclination and position angles i.e. $0^{\circ} \leq i\leq 90^{\circ}$ and $0^{\circ} \leq \theta \leq 360^{\circ}$. The uncertainty in each derived parameter is calculated from its posterior probability distribution.  The results from these simple disk model fits are summarized in Table~\ref{tab:kinematics}.

\subsection{Results}

Our data reveals many cases where there are significant kinematic deviations from this simple rotating disk model. Because our spatial resolution is $<$500pc, we find the velocity field for the H$\alpha$ emission line samples the local bulk motion of gas and is not smeared over a large area as would be the case in coarser-sampled data. Typically our resolution elements have velocity uncertainties of only a few kms sec$^{-1}$ which often results in large reduced  $\chi^2$  despite similar r.m.s velocity residuals of $\big< v_\textrm{model}-v_\textrm{data}\big>\sim30$ km s$^{-1}$ compared to studies sampling $\sim$ kpc scales \citep[e.g.][]{Swinbank12}. 

These significant velocity deviations suggest that several of our systems are undergoing various stages of a merging process.  One-dimensional velocity profiles in Figure~\ref{fig:velmaps} show many galaxies in our sample can be considered as a perturbed disk where the large-scale velocity map exhibits a clear gradient but the velocity dispersion peak is offset from our adopted center. In several cases (e.g. CSWA19, and CSWA139), two distinct H$\alpha$ sources of different velocities can be discerned, consistent with early stages of a merging pair. There is good evidence from either the velocity profiles or the velocity deviations from our disk model, that CSWA19, 31, 128, 139 and 165 represent merging systems.

\begin{figure*}
\centering
\includegraphics[width=4.cm]{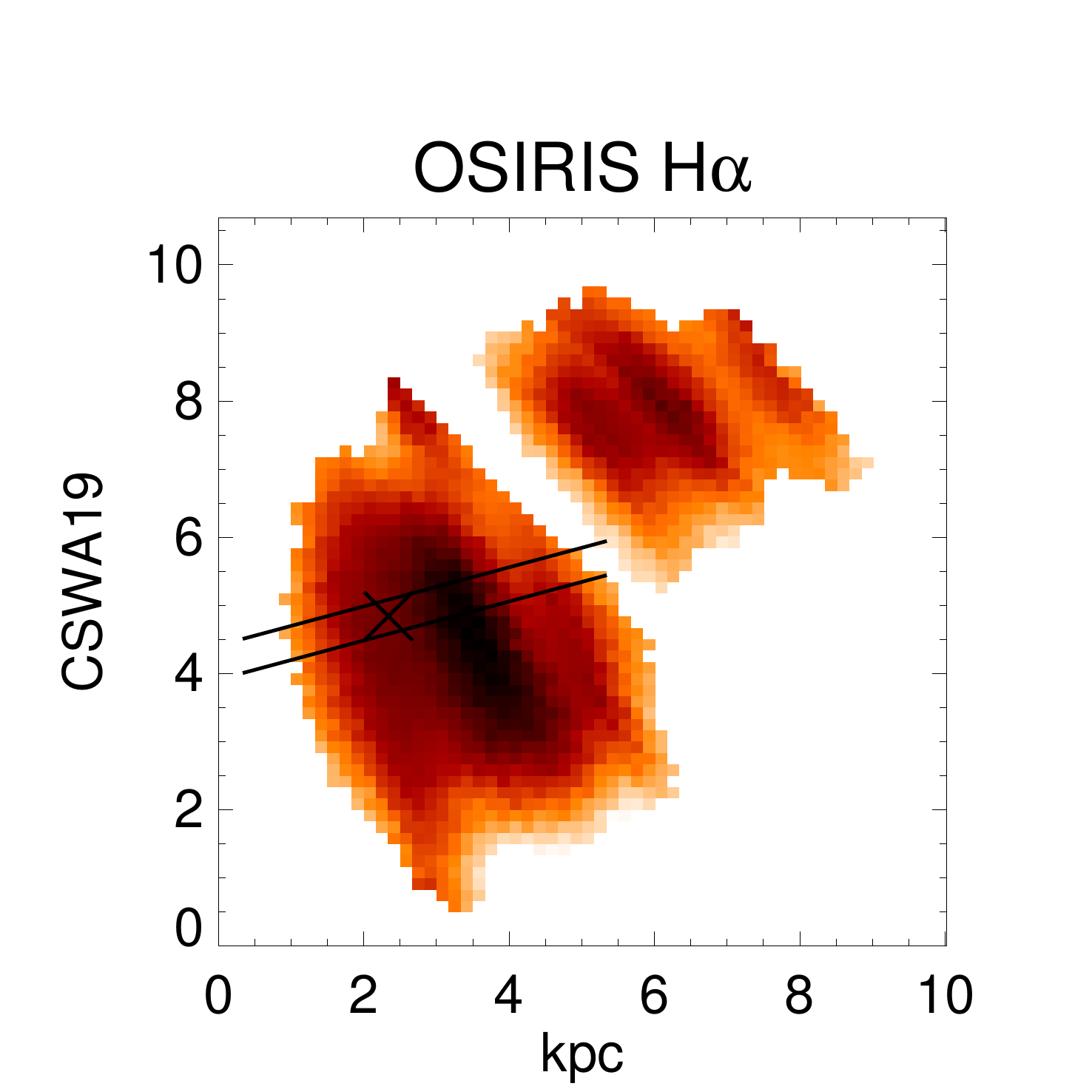}
\includegraphics[width=4.cm]{cswa19sourcekinematic_pretty_10sigma_vel}
\includegraphics[width=4.cm]{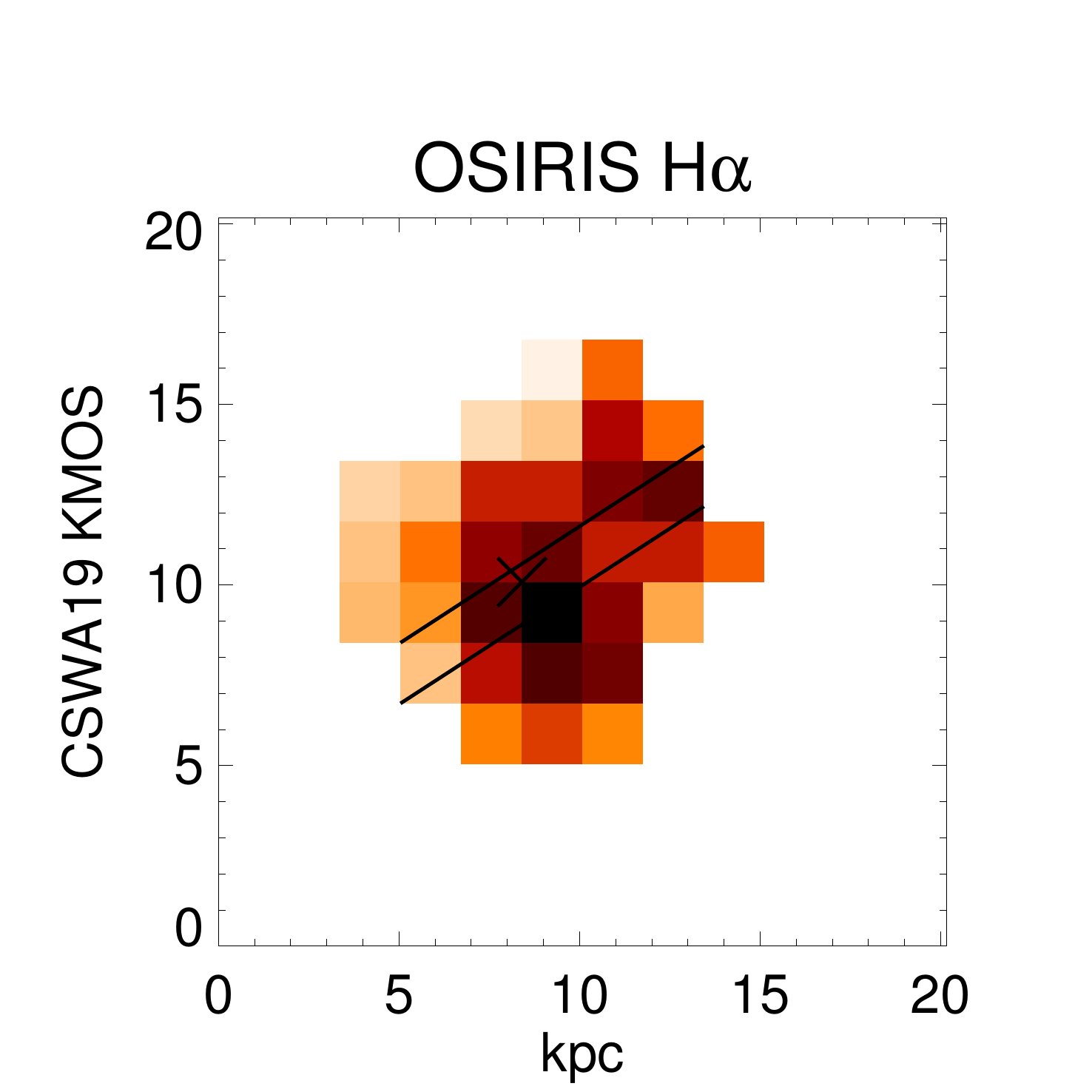}
\includegraphics[width=4.cm]{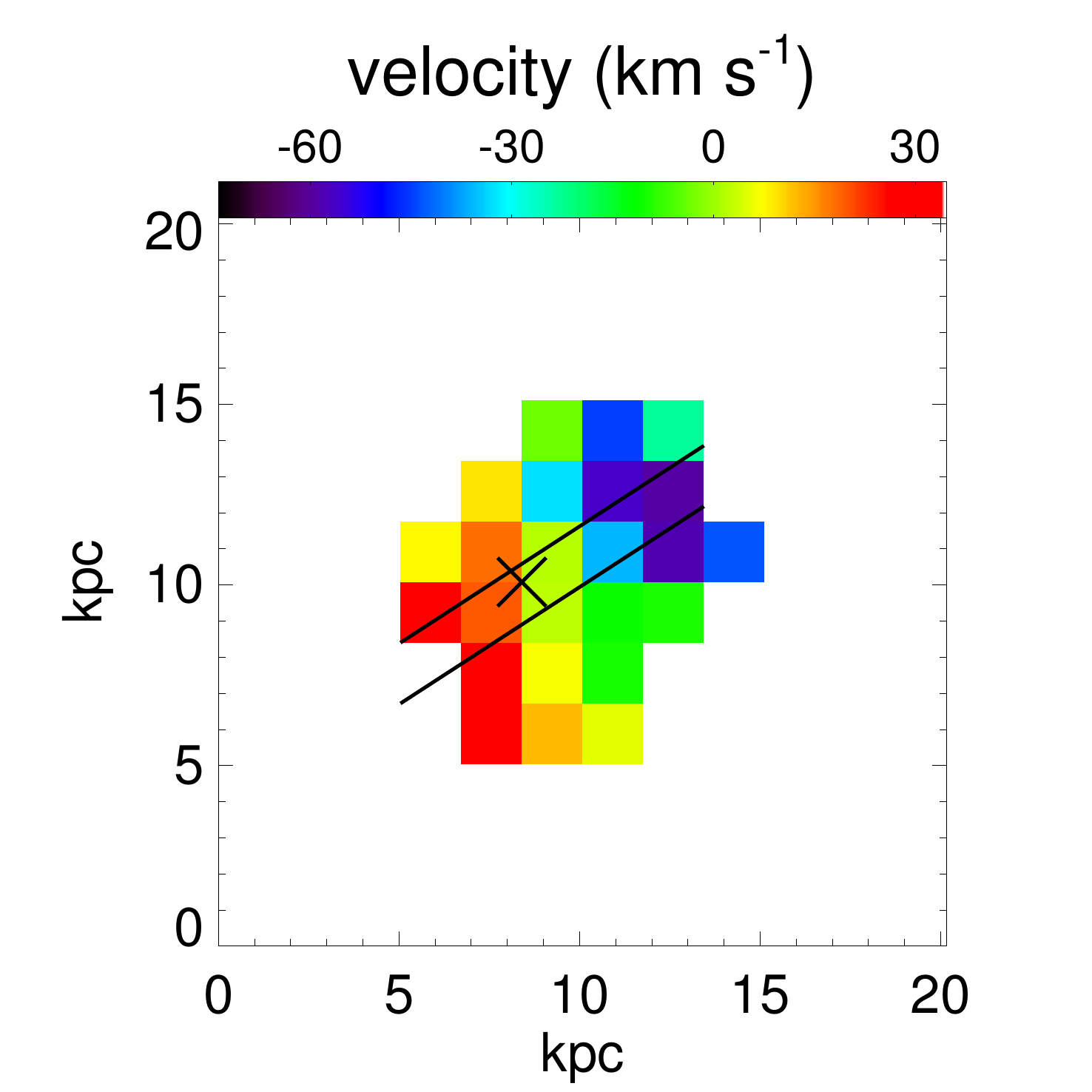} \\
\includegraphics[width=4.cm]{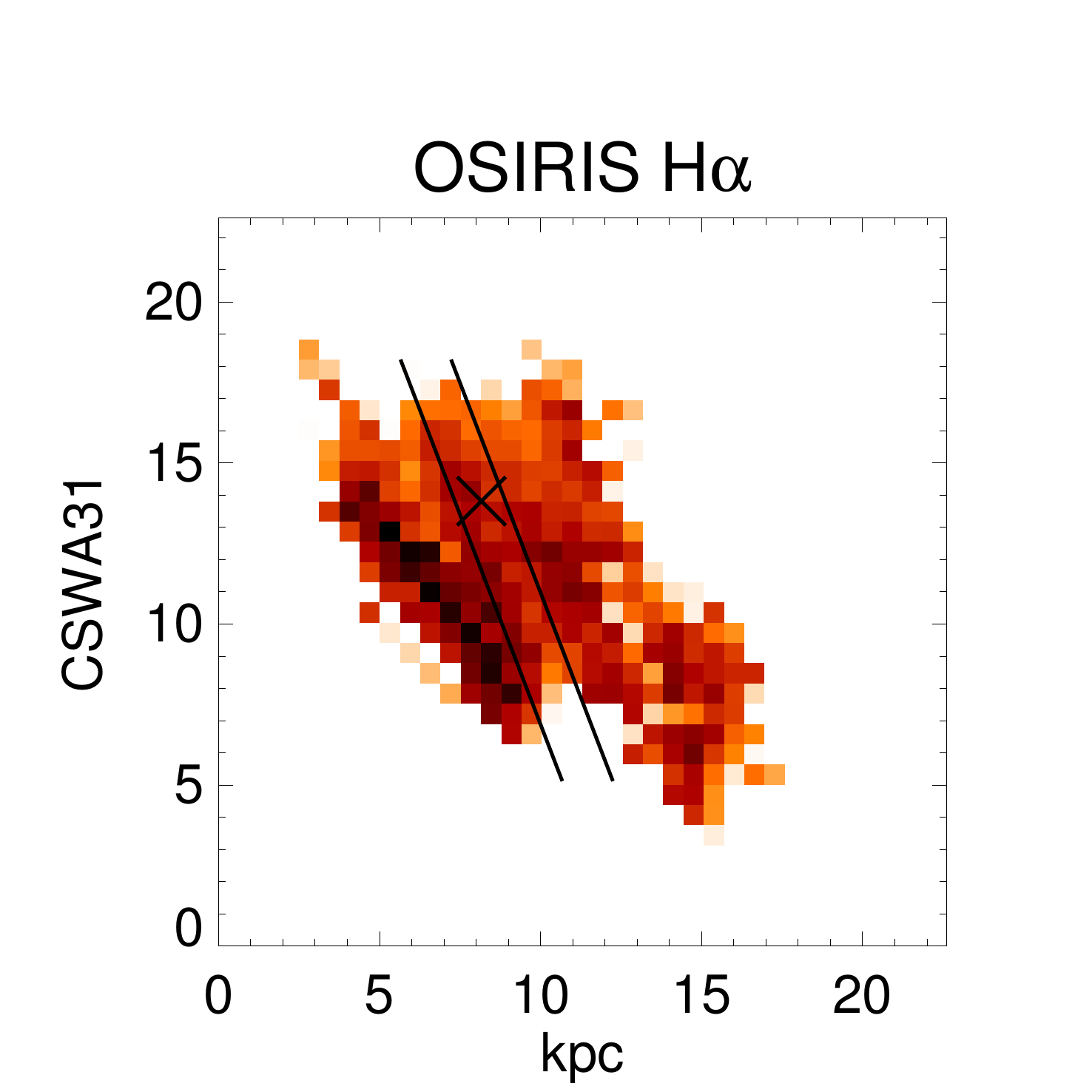}	
\includegraphics[width=4.cm]{cswa31_slitmap_vel}
\includegraphics[width=4cm]{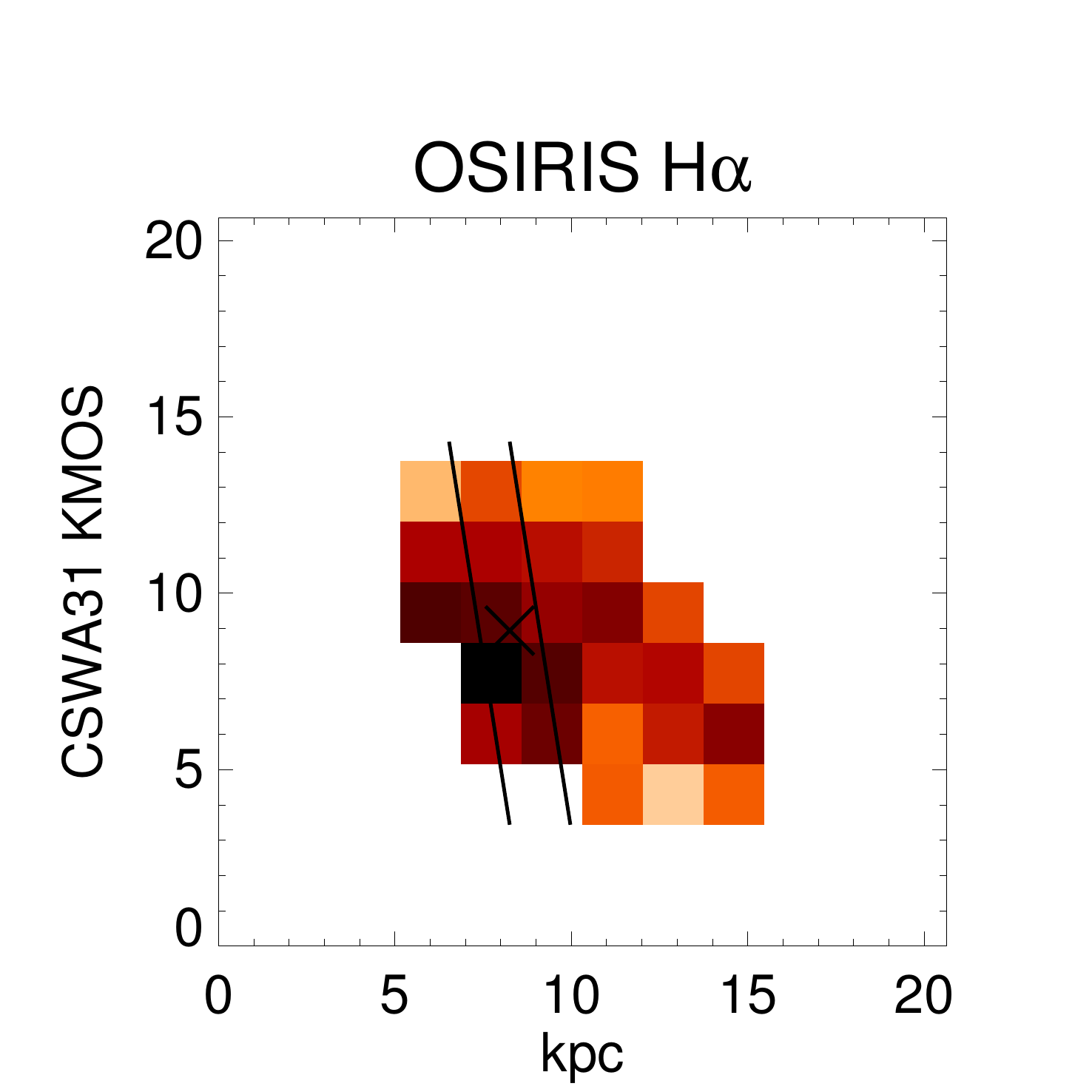}
\includegraphics[width=4cm]{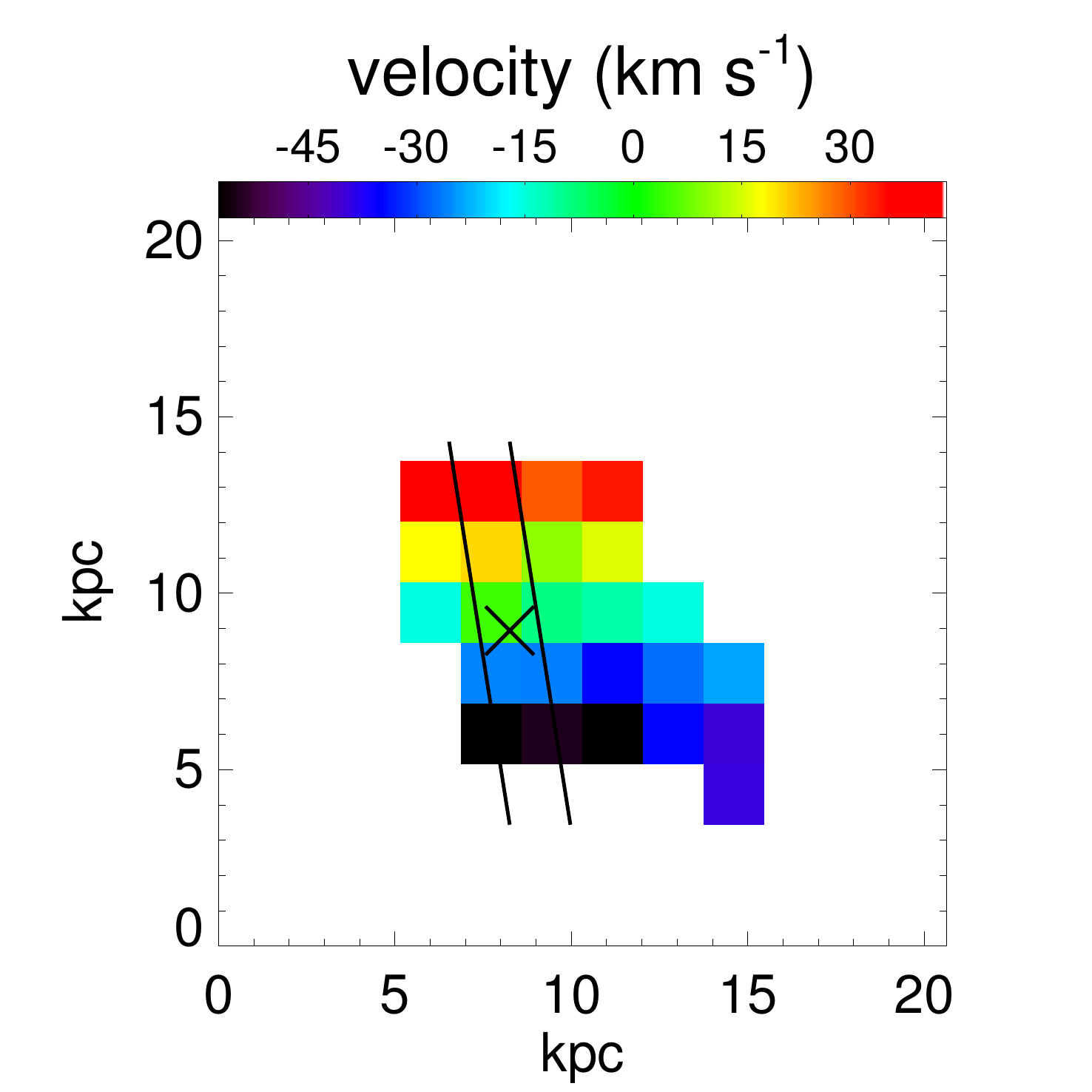}
\caption{The potential dangers of interpreting kinematic data with poor sampling. {\it Left:} Source plane H$\alpha$ intensity map and velocity field in this study for CSWA19 (top row) and CSWA31 (bottom row). {\it Right:} Respective simulated H$\alpha$ intensity map and velocity field indicating the deterioration in resolution equivalent to that for non-lensed sources studied with KMOS. In both cases, the complex morphology and velocity structure is lost in the poorer sampled data leading to the erroneous conclusion of kinematically well-ordered systems.}
\label{fig:kmostest}
\end{figure*}

At this point, it is interesting to conjecture whether such complex kinematic patterns would have been discernible with a lower spatial resolution. As an experiment we re-analyzed two systems, CSWA19 and CSWA31, both clearly poor candidates for pure rotation-dominated systems, as if they had been observed at lower angular resolution.  The source plane velocity fields were first smoothed with a typical PSF of $0.6''$, rebinned to a coarser resolution of $0.2''$, comparable to that achievable with the KMOS$^{\textrm{3D}}$ survey\citep{Wisnioski15}, and re-analyzed with our disk model. The results of this experiment are shown in Figure \ref{fig:kmostest}. In both cases the $\chi^2$ fit is significantly improved and there is little evidence of any departure from a pure disk model. This accentuates the importance of optimally-sampled data for a robust interpretation of the velocity field.

The second question posed above relates to the precision of the derived rotational velocity and, particularly, the ratio of systemic to random motions which is such an important diagnostic of the maturity of early disk systems. To examine this, we first adopt the inclination and position angle values obtained from the disk model fitting. Despite the velocity deviations from the disk model discussed above, the posterior probability distributions of inclination and position angle are generally Gaussian in form suggesting reasonably good estimates of these parameters.  Other methods for determining the inclination e.g. analyzing the elliptical shape of the H$\alpha$ or near-infrared light distributions\citep[e.g.][]{Newman13,Wisnioski15} are less applicable to our less luminous systems with irregular forms.  As for the position angle, our disk model fits are generally consistent (within $2\sigma$) with values  obtained from a `peak-to-peak' velocity method where a pseudo-slit placed at the center is rotated to find to yield the largest velocity shear. With the exception in CSWA165 where the discrepancy between the two methods is large and the disk model fit is particularly poor, we adopt the position angle from the peak-to-peak method ($\theta'$ in Table~\ref{tab:kinematics}).

\begin{figure}
\centering
\includegraphics[width=8.5cm]{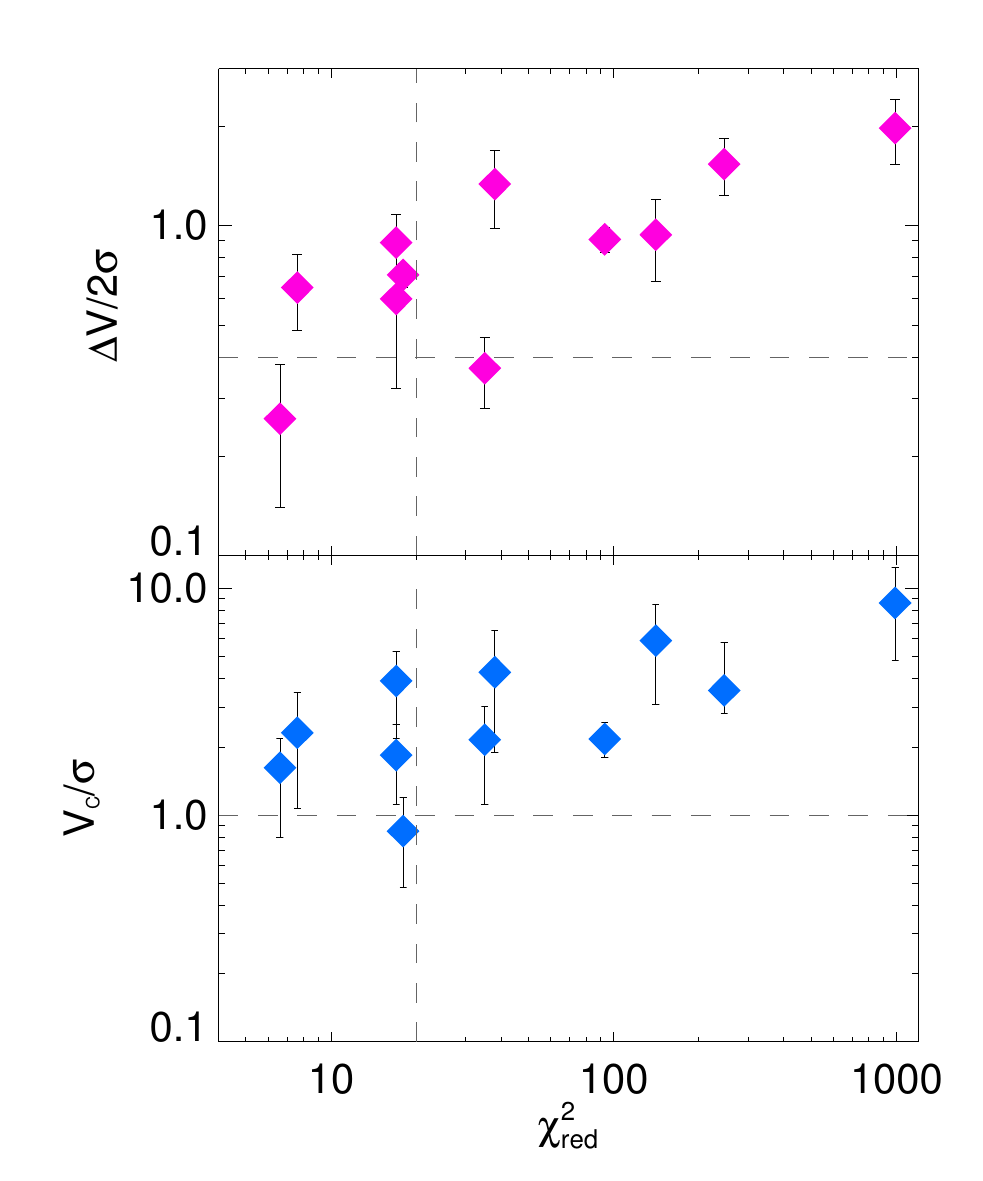}
\caption{A comparison of two methods for determining the degree of rotational support in our sample plotted against the goodness-of-fit in the disk model. $v_c/\sigma$ from our disk fits and $\Delta V/2\sigma$ from the peak-to-peak method is considered in the context of the reduced $\chi^2$. Horizontal dashed  lines represent the usually adopted thresholds, $\Delta V/2\sigma \ge 0.4$ and $v_c/\sigma>1$ for rotationally-supported systems and the vertical dashed line represents $\chi^2=20$, below which we consider the disk model to be appropriate. Without adequate sampling, a larger fraction of our sample would be incorrectly considered to be rotationally-supported.}
\label{fig:vc_delv}
\end{figure}

Our sample can then be split in two subsets. For five galaxies (CSWA11, 15, 28, A773, 159) where the disk model represents an acceptable fit ($\chi_{red}^2<20$), the ratio $v_c/\sigma$ is probably a valid indicator of whether the system is dynamically supported. Adopting the standard often-used threshold $v_c/\sigma \geq 1$ and calculating the intrinsic velocity dispersion $\sigma$ from the weighted velocity dispersions of each spaxel along the major axis, all but CSWA15 could be considered to be rotationally-supported systems (see Table \ref{tab:kinematics}). However, the $v_c/\sigma$ ratio is clearly inappropriate for the remaining galaxies where the disk model is a much poorer fit. An alternative statistic sometimes used is the observed velocity ratio $\Delta V/2\sigma$ ratio where $\Delta V$ is the maximum velocity shear obtained from the peak-to-peak method. Applying the threshold criterion of  $\Delta V/2\sigma \ge 0.4$ for rotationally-supported systems adopted by \citet{Forster09} to the entire sample, 9 out of 11 of our sources would then be classified as rotation-dominated systems. Figure \ref{fig:vc_delv} shows the correspondence between these two measures ($v_c/\sigma$, $\Delta V/2\sigma$) and the goodness of the disk model  ($\chi_{red}^2$) for our sample. Broadly speaking, both above criteria are reasonably consistent in selecting rotationally-supported systems. However, without the highly-sampled data that allows an adequate test of the goodness of fit, a much larger number would be incorrectly placed in this category.
 
We perform the same kinematic fitting method to the 4 galaxies published in \citet{Jones13} and find consistent categorization using the $\chi_{red}^2$ criterion above. 
\citet{Jones10a} already performed kinetic model fitting for J0744 and J1206 and we find very similar results. J1148, which \citet{Jones13} considered as a rotating system using a 1-D velocity profile, has a $\chi^2_{red}\approx 6.5$ with a very low velocity residual root-mean-square of 7 km/s suggesting a well-ordered disk. Finally, J1038 which is visibly a merger, has a $\chi^2_{red}\approx380$ and is categorized as a perturbed disk. 
  
It is illustrative, therefore, to compare our conclusions to the first results of the KMOS$^{\textrm{3D}}$ Survey which reports kinematics result of an IFU survey of $\sim 200$ galaxies at $0.7< z <  2.7$ \citep{Wisnioski15}. Our galaxy sample has an average $\Delta V/\mbox{sin}i$ of 148 km s$^{-1}$ and average velocity dispersion of 60 km s$^{-1}$, both of which are broadly consistent with the range of $\Delta V/\mbox{sin}i \sim100-200$  and $\sigma \sim 30-80$ km s$^-1$ in the KMOS$^{\textrm{3D}}$ data.  However, according to Figure \ref{fig:vc_delv} only 4/11 ($\sim36\%$) of our sample would be classified as rotationally-supported galaxies via the combination of both criteria in Figure \ref{fig:vc_delv}, whereas in the KMOS$^{\textrm{3D}}$$74\%$ of $z\sim2$ galaxies were classified as being rotationally-supported. The higher percentages in the KMOS$^{\textrm{3D}}$ analysis may arise at least in part from a combination of the lower spatial resolution where late merging systems are mistaken as a regular rotating system (Figure \ref{fig:kmostest}) as well as the difficulty of fitting a disk model in such circumstances. 

\begin{deluxetable*}{lccccccccccc}
\tablewidth{\linewidth}
\tablecaption{Kinematic Properties of the Samples}
\tablehead{
\colhead{}&\multicolumn{6}{c}{Simple-disk model results}&\multicolumn{4}{c}{Peak-to-peak method results}\\
\colhead{ID} & \colhead{$\chi^2_\textrm{red}$} &\colhead{$\big<V_{\textrm{model}}-V_{\textrm{data}}\big>_{\textrm{rms}}$}& \colhead{$i (^{\circ})$}& \colhead{$\theta(^{\circ})$}& \colhead{V$_c(\mbox{km s}^{-1})$} &  \colhead{$V_{\textrm{c}}/\sigma$} & \colhead{ $\Delta V(\mbox{km s}^{-1})$ }& \colhead{$\sigma(\mbox{km s}^{-1})$} & \colhead{$\theta'(^{\circ})$}& \colhead{$\Delta V/2\sigma$} }
\tablecolumns{11}
\startdata
cswa11		&17		&40.2	&$64^{69}_{56}$	&$16\pm12$	&$173^{216}_{125}$& $1.8^{2.5}_{1.1}$	&$167\pm53$ 	&$94\pm26$  	& 48	&$0.89\pm0.19$\\[1.2mm]
cswa15		&18		&12.9	&$25^{29}_{19}$	&$194\pm10$	&$29^{40}_{17}$	& $0.8^{1.2}_{0.5}$	&$48\pm3$ 	&$34\pm5$ 	& 205& $0.71\pm0.06$\\[1.2mm]
cswa19$^a$	&9.6		&13.6	&$49^{56}_{43}$	&$155\pm4$	&$126^{159}_{83}$	&$1.6^{2.1}_{1.0}$	&$89\pm2$	& $78\pm14$	&153& $0.57\pm0.05$\\[1.2mm]
cswa19$^b$	&246	&42.3	&$75^{77}_{65}$	&$156\pm4$	&$377^{442}_{248}$&$3.6^{5.8}_{2.8}$	&$120\pm7$	& $78\pm14$	&150& $1.54\pm0.30$\\[1.2mm]
cswa20		&35		&21.8	&$40^{47}_{32}$	&$162\pm2$	&$133^{146}_{99}$ &$2.2^{3.0}_{1.1}$	&$46\pm14$	& $62\pm25$	&168& $0.37\pm0.09$\\[1.2mm]
cswa28		&6.6		&25.3	&$23^{47}_{18}$	&$129\pm10$	&$91^{116}_{49}$	&$1.6^{2.2}_{0.8}$	&$29\pm25$	& $56\pm12$	&127& $0.26\pm0.12$\\[1.2mm]
cswa31		&93		&20.8	&$22^{27}_{19}$	&$249\pm3$	&$122^{135}_{110}$&$2.2^{2.6}_{1.8}$	&$102\pm12$   & $56\pm8$	&244& $0.91\pm0.08$\\[1.2mm]
cswa128		&990	&69.9	&$57\pm3	$		&$210\pm2$	&$570^{578}_{554}$&$8.6^{12.4}_{4.8}$	&$262\pm8$	& $66\pm29$	&208& $1.98\pm0.44$\\[1.2mm]
cswa139		&38		&25.6	&$24\pm4$		&$297\pm15$	&$265^{290}_{211}$&$4.3^{6.5}_{1.9}$	&$166\pm27$	& $62\pm32$	&294& $1.34\pm0.36$\\[1.2mm]
cswa159		&7.6		&12.7	&$13^{19}_{10}$	&$175\pm9$	&$141^{191}_{85} $	& $2.3^{3.5}_{1.1}$	&$79\pm29$	& $61\pm22$	&164& $0.65\pm0.17$\\[1.2mm]
cswa165		&141	&17.3	&$57\pm5$		&$164\pm5$	&$359^{384}_{307}$&$5.9^{8.5}_{3.1}$	&$83\pm28$	& $44\pm19$	&218& $0.94\pm0.26$\\[1.2mm]
a773			&17		&53.7	&$22^{37}_{17}$	&$144\pm12$	&$176^{225}_{108}$&$3.9^{5.3}_{2.2}$	&$54\pm22$	& $45\pm10$	&158  & $0.60\pm0.28$\\[1.2mm]
\enddata
\tablecomments{ $^a$ Results from fitting to the main galaxy only. $^b$ Results from fitting to both mergers together as one system. \label{tab:kinematics}}
\end{deluxetable*}

\section{Metal Gradients}
\label{sec:gradients}

\begin{table*}
	\centering
	\caption{Metallicity Gradients \label{tab:gradients}}
	\begin{tabular}{lcccccccc}
	\toprule
	{ID} &{[N II]/H$\alpha$} &{Central 12+log(O/H)} &{N2 PP04} &{N2 S+14}&{N2 M+08}&{O3N2PP04} &{O3N2 S+14}&{O3N2 M+08} \\
& &{(from N2 PP04)}&{dex kpc$^{-1}$}&{dex kpc$^{-1}$}&{dex kpc$^{-1}$}&{dex kpc$^{-1}$}&{dex kpc$^{-1}$}&{dex kpc$^{-1}$}\\
\hline
\multicolumn{9}{c}{Objects classified as rotationally-supported systems ($\chi^2<20$)} \\
\hline
cswa11	&$0.14\pm0.01$	&$8.54\pm0.06$	& $-0.07\pm0.02$      & $-0.05\pm0.01$	 & $-0.10\pm0.03$ & \nodata &\nodata &\nodata\\
		&				&$8.60\pm0.06$	& $-0.11\pm0.02$      & $-0.07\pm0.01$	 & $-0.16\pm0.02$ &\nodata&\nodata &\nodata\\
\tableline
cswa15	&$0.03\pm0.01$	&$8.16\pm0.01$ 	& $-0.04\pm0.01$	& $-0.02\pm0.01$	& $-0.05\pm0.02$	& $-0.02\pm0.01$	& $-0.02\pm0.01$   &$-0.01\pm0.01$ \\
 	         	&				&$8.17\pm0.01$ 	& $-0.03\pm0.01$	& $-0.02\pm0.01$	& $-0.05\pm0.01$	& $-0.02\pm0.01$	& $-0.01\pm0.01$   &$-0.00\pm0.01$ \\
\tableline
cswa28	& $0.09\pm0.01$       &$8.10\pm0.01$	& $0.11\pm0.03$	&  $0.07\pm0.24$	& $0.17\pm0.59$ &\nodata&\nodata &\nodata\\
		&				&$8.33\pm0.12$	& $0.04\pm0.03$	&  $0.02\pm0.01$	& $0.07\pm0.04$&\nodata&\nodata &\nodata\\
\tableline
cswa159	&$0.08\pm0.01$	&$8.50\pm0.01$	& $-0.01\pm0.01$     &$-0.01\pm0.01$	& $-0.03\pm0.02$	& $0.00\pm0.01$	&$0.00\pm0.01$     &$-0.01\pm0.01$\\
		&				&$8.49\pm0.03$ 	& $-0.01\pm0.01$     &$0.00\pm0.01$	& $-0.02\pm0.01$	& $0.00\pm0.01$	&$0.00\pm0.01$     &$-0.01\pm0.01$\\
\tableline
a773		&$0.17\pm0.07$	&$8.42\pm0.10$	& $-0.13\pm0.18$	& $-0.08\pm0.06$     &$-0.19\pm0.28$      &$-0.25\pm0.41$      & $-0.22\pm0.35$    &$0.01\pm0.30$\\		
                   & 				&$8.43\pm0.05$	& $-0.05\pm0.02$	& $-0.03\pm0.01$     &$-0.07\pm0.03$      &$-0.03\pm0.03$      & $-0.02\pm0.02$    &$-0.08\pm0.02$\\
\tableline
\multicolumn{9}{c}{Objects with $\chi^2>20$} \\
\tableline
cswa19	&$0.09\pm0.01$	&$8.30\pm0.01$ 	& $-0.01\pm0.01$	& $0.01\pm0.01$	& $-0.01\pm0.07$	& $0.12\pm0.04$	& $0.10\pm0.01$   &$0.07\pm0.05$\\
		&				&$8.26\pm0.01$ 	& $0.02\pm0.01$	& $0.01\pm0.01$	& $0.02\pm0.01$	& $0.01\pm0.01$	& $0.01\pm$ 0.01  &$0.03\pm0.01$\\
\tableline
cswa20	&$0.05\pm0.01$	&$7.89\pm0.02$ 	&$-0.15\pm0.05$	& $-0.10\pm0.07$	& $-0.29\pm0.45$	&$-0.41\pm0.19$	& $-0.36\pm0.17$  &$-1.00\pm0.24$\\
		&				&$7.97\pm0.04$ 	& $0.05\pm0.02 $	& $0.03\pm 0.01$	& $0.09\pm0.03$	&$ 0.03\pm0.02$	& $0.02\pm0.02$   &$0.25\pm0.05$\\
\tableline
cswa31	&$0.33\pm0.06$	&$8.50\pm0.03$	& $0.02\pm0.01$	&  $0.01\pm0.01$	& $0.03\pm0.01$ &\nodata&\nodata &\nodata\\
		&				&$8.54\pm0.01$	& $0.00\pm0.01$	&  $0.00\pm0.01$	& $0.00\pm0.01$ &\nodata&\nodata &\nodata\\
\tableline
cswa128	&$0.10\pm0.02$	&$8.57\pm0.03$ 	& $-0.10\pm0.02$	& $-0.06\pm0.01$	& $-0.12\pm0.02$	& $-0.09\pm0.02$	& $-0.08\pm0.02$   &$-0.15\pm0.02$\\
		&				&$8.51\pm0.01$ 	& $-0.04\pm0.01$	& $-0.02\pm0.01$	& $-0.05\pm0.01$	& $-0.04\pm0.01$	& $-0.03\pm0.01$   &$-0.06\pm0.01$\\
\tableline
cswa139	&$0.17\pm0.08$	&$8.25\pm0.01$ 	& $-0.01\pm0.01$	& $-0.01\pm0.01$	& $-0.02\pm0.02$	& $0.03\pm0.01$	& $0.03\pm0.01$    &$0.02\pm0.01$\\
		&				&$8.12\pm0.03$ 	& $0.04\pm0.01$	& $0.02\pm0.01$	& $0.05\pm0.01$	& $0.02\pm0.01$	& $0.01\pm0.01$    &$0.04\pm0.01$\\		
\tableline
cswa165	&$0.26\pm0.06$	&$8.53\pm0.01$	& $0.03\pm0.01$	& $0.02\pm0.02$     &$0.04\pm0.05$	& $0.13\pm0.01$	& $0.11\pm0.01$    &$0.16\pm0.05$\\
		&				&$8.53\pm0.01$	& $0.01\pm0.02$	& $0.01\pm0.01$      &$0.01\pm0.04$	& $0.01\pm0.02$	& $0.00\pm0.01$    &$0.13\pm0.03$\\
\tableline
	\end{tabular}
	\tablecomments{For each object, the top line contains values derived along each galaxy's `major axis'. The bottom line contains values derived from radial binning.}
\end{table*}

Given the capability of the current integral field instruments, the only practical method to measure metallicity in high-redshift galaxies is to use strong-line metallicity calibrators. The direct measurement of metallicity ($T_e$ method) is not feasible for individual galaxies at this redshift since it relies on the measurement of emission lines that are too faint to detect. In this work, we use the [N II] $\lambda6584$/H$\alpha$ ratio (N2) and the  ([O III] $\lambda$5008/H$\beta$)/([N II] $\lambda$6584/H$\alpha$) ratio (O3N2) with the calibrations from \citet{pp04} - PP04, \citet{Steidel14} - S14, and \citet{Maiolino08} - M08:
\begin{equation}
12+\log\textrm{(O/H)} = 8.90+0.57\times \textrm{N2} \tag{N2, PP04}
\end{equation}
\begin{equation}
12+\log\textrm{(O/H)} = 8.62+0.36\times \textrm{N2} \tag{N2, S14}
\end{equation}
\begin{equation}
12+\log\textrm{(O/H)} = 8.73-0.32\times \textrm{O3N2} \tag{O3N2, PP04}
\end{equation}
\begin{equation}
12+\log\textrm{(O/H)} = 8.66-0.28\times \textrm{O3N2} \tag{O3N2, S14}
\end{equation}
where N2$\equiv \log(\textrm{[N II]} \lambda6584/\textrm{H}\alpha)$ and O3N2$ \equiv \log\textrm{(([O III] }\lambda5008/\textrm{H}\beta)/\textrm{([N II]}\lambda6584\textrm{/H}\alpha))$. 

The PP04 relations were primarily calibrated from the $T_e$ method based on nearby extragalactic HII regions. The M08 relations were based on local galaxies that cover a wider range of metallicity compared to the PP04 relations. The S14 relations were based on a subset of the same nearby extragalactic HII regions sample in PP04 relations that are limited to the range of N2 and O3N2 observed in $z\sim2.3$ sample which have shown to lower the systematic offset between the two indicators when applied to high redshift sample. To counter the possible bias of the inferred metallicity due to the higher N/O ratios in high redshift galaxies than the ratios in local galaxies, we will compare our measured gradients from the N2 calibrators to the gradients from the O3N2 indicators which have been shown to be less dependent on the N/O ratio than the N2 indicator  \citep{Steidel14}. 

For each calibrator, we calculate the metallicity gradient using two methods: from the pixels along the major axis and from all pixels binned radially. The galaxy center, position angle, inclination, and pseudo slit along major axis are the same as those in Section \ref{sec:kinematics}. An advantage of the metallicity gradient measured from the pixels along the major axis is that it is independent of the inclination and thus less model-dependent. However, it is based on fewer measurements than the metallicity gradient measured from all pixels and sometimes not feasible when most of the HII regions do not lie along the major axis (e.g. CSWA20). The best-fit gradients and central metallicities are shown in Table \ref{tab:gradients} and plots are shown in Figure \ref{fig:NIIHa}.

In detail, we calculate an average N2 ratio (or O3N2 ratio) for each radial bin along the major axis (or along the annuli) for PP04 and S14 calibrators.  We assume that metallicity $12+\log\textrm{(O/H)}$ is a linear function of radius. Hence, the line ratio is an exponential function of radius, metallicity gradient, and central metallicity. We fit the N2 ratios (or O3N2 ratios) with the weighted least-square regression to find the metallicity gradient and central metallicity for each galaxy. 

For the M08 calibrator, we calculate the N2 ratio (and [O III]/H$\beta$ ratio, if possible) for each radial bin and find the metallicity for that radial bin with a brute force maximum likelihood estimation. We then obtain the metallicity gradient and the central metallicity with a linear fit. 

We check the consistency between the metallicity gradients derived from pixels along major axis and radial annuli obtained in section \ref{sec:kinematics}. We found that the results from the two methods are consistent with each other within one standard deviation especially when the rotation is relatively well-described by a simple disk rotation. The deviation is larger when the major axis is short and the measurement along the major axis suffers from a poor sampling e.g. CSWA20 and Abell773. 

Metallicity gradients derived from the O3N2 calibrators are consistent with the gradients derived from the N2 calibrators when using radial binning method and comparing between the same set of calibrators e.g. PP04 N2 with PP04 O3N2. Although O3N2 calibrators give more accurate metallicity measurement than N2 calibrators, the consistency in the gradients derived from the two calibrators suggests that the metallicity gradients measured with N2 calibrators should not be heavily affected by a possible variation in N/O ratios in high redshift galaxies. Hence, for the three galaxies that we do not have information on [O III]/H$\beta$, the derived metallicity gradients should be relatively reliable. When compared between the S14 and PP04 measurements, the metallicity gradients derived with S14 calibrators are typically flatter than those derived with PP04 calibrators. 

In further analysis, we use the metallicity gradients derived from PP04 in radial binning because we can avoid the bias from poor sampling in short major axis cases. The PP04 calibrator is chosen so that we can easily compared with other observed gradients from previous observations. 

\section{AGN Contamination and Bias from Low Signal to Noise Data}
\label{sec:agn}
We now examine whether our sample is affected by line emission from AGN. Since our metallicity calibrators are only applicable to bounded HII regions ionized by hot stars, any emission from nuclear activity would give misleading results for central metallicities and gradients. Furthermore the presence or absence of AGN is interesting in terms of understanding the relationship between growth of galaxies and their central supermassive black holes. Optical spectra of AGN are characterized by high ratios of collisional to Balmer lines, and by broad line widths arising from outflows and/or broad line region kinematics. These signatures have been confirmed in the nuclear regions of many galaxies at $z\simeq1-2$ with spatially resolved spectroscopy \citep{Wright10, Newman14,Forster14, Genzel14}. Notably, \citet{Genzel14} show that the fraction of galaxies showing signatures of AGN is strongly correlated with stellar mass.

\begin{figure}
\begin{center}
\includegraphics[width=6cm]{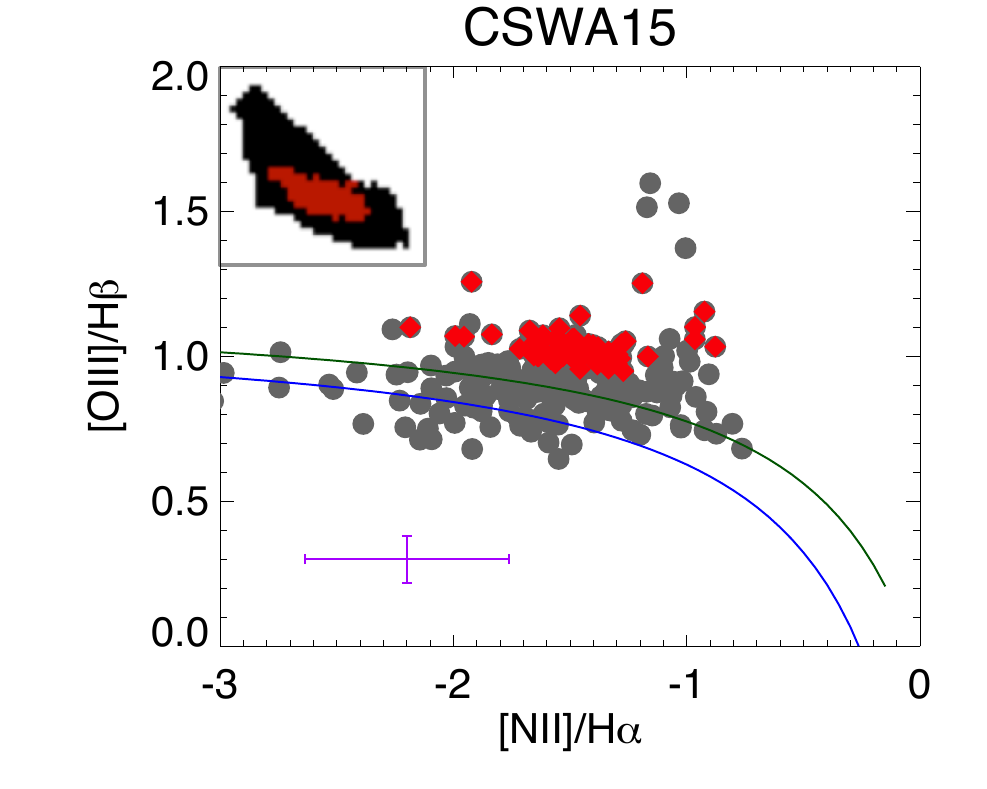}
\includegraphics[width=6cm]{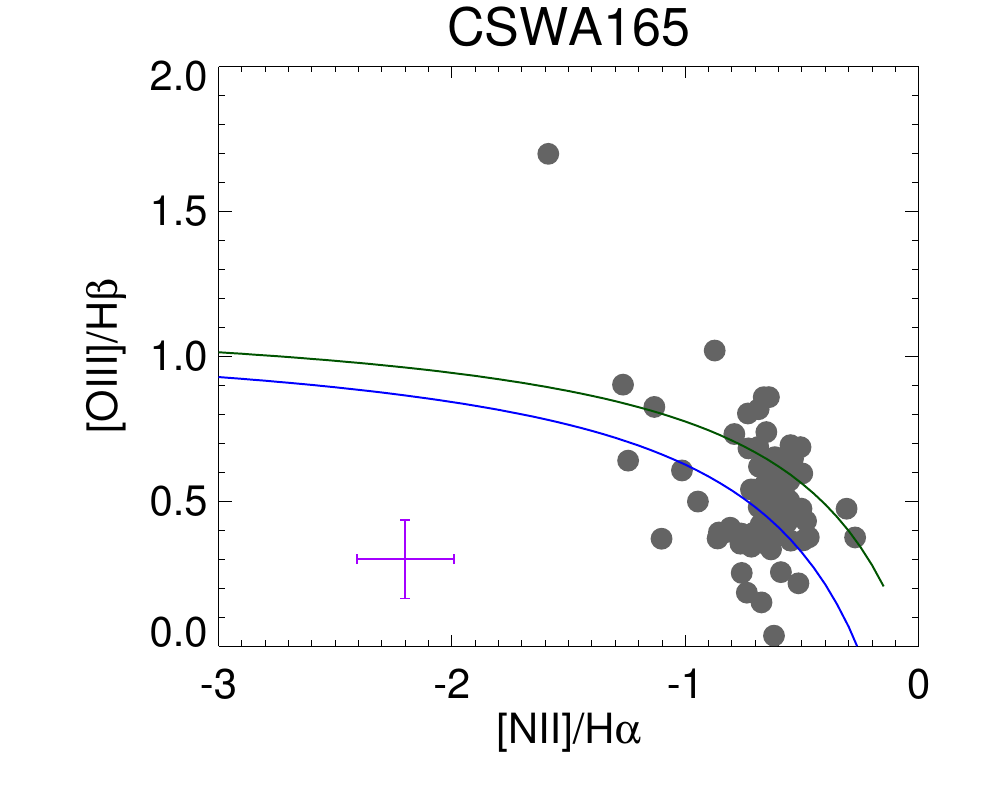}
\end{center}
\caption{BPT diagrams for CSWA15 (upper) and CSWA165 (lower). Gray points are from individual pixels in each galaxy. Red points are from pixels whose values lie significantly above the maximum starburst line (green curve, \citet{Kewley01}). The corresponding region to the red points is shown in red in the map at the corner. Blue curve is the best fitted line for the $z\sim2.3$ samples in \citet{Steidel14}. Median values in uncertainties are shown in the purple error bars. A few pixels have large error bars e.g. the data point at the top of the lower plot.}
\label{fig:bpt}
\end{figure}

For each of the eight new galaxies in this paper with [O III]/H$\beta$ information, we construct a ``BPT":[N II]/H$\alpha$ and [O III]/H$\beta$ diagram from the spatially resolved spectra. Most show a distribution of values entirely consistent with star-forming regions as seen in other $z\sim2$ galaxies \citep{Steidel14,Shapley15}. We thus only show two examples where the interpretation is less clear (Figure \ref{fig:bpt}). CSWA15 shows a central region in whose strong line ratios significantly lie at an average of $0.16\pm0.06$ dex above the maximum starburst classification line \citep{Kewley01} as shown in  the upper panel of Figure \ref{fig:bpt}. However, the central [N II]/H$\alpha = 0.03\pm0.01$ and [S II]$\lambda\lambda6716,6731$/H$\alpha=0.13\pm0.03$ are both low and atypical of AGN. These ratios are at the extreme low end probed by current $z\simeq2$ samples \citep[e.g.][]{Steidel14} and broadly consistent with an extension of the star-forming locus seen in these surveys.  Alternatively the line ratios are consistent with theoretical expectations for an AGN with gas metallicity of $\simeq0.5$ solar \citep{Groves06}, but the surrounding (and clearly star-forming) regions of CSWA15 indicate significantly lower metallicity. If the central line ratios are affected by AGN then the gas-phase metallicity gradient would be significantly steeper than we infer assuming stellar ionization.

As an additional check for nuclear activity, we fit H$\alpha$ from the central regions of CSWA15 with a double Gaussian profile. The best-fit broad component has a line width $\sigma=153$ km s$^{-1}$ and comprises 29\% of the total flux (c.f. $\sigma=59$ km s$^{-1}$ for the dominant narrow component). This is similar to stacked spectra of outer disk regions in $z\simeq2$ galaxies ($\sigma\sim200$  km s$^{-1}$ and $\sim$40\% of flux in the broad component; \citealt{Newman13}, \citealt{Genzel14}) and is not indicative of AGN-driven outflows ($\sigma\gtrsim500$ km s$^{-1}$) or broad-line regions ($\sigma\gtrsim1000$ km s$^{-1}$). The spectrum of CSWA15 is therefore consistent with expectations for emission from star formation and associated outflows, and does not show strong evidence of an AGN.

The incidence of AGN in our sample is consistent with expectations given their stellar masses. We find no clear signatures of AGN and only one possible case discussed above out of the 12 galaxies with available BPT diagnostics, including those from \citet{Jones13}. An example of typical BPT diagrams that are compliant with star-forming regions in our sample, CSWA165, is shown in the lower panel of Figure \ref{fig:bpt}. The remaining galaxies in our sample have low central [N II]/H$\alpha$ and [S II]/H$\alpha$ ratios indicating no evidence for AGN. \citet{Genzel14} find similar results using the same methods: zero secure and only two potential AGN in a sample of 17 $z\simeq1$--2 galaxies with stellar mass $\log{M_*/M_\odot}=9.4$--10.3, comparable to that of our sample. They likewise find an AGN fraction $<10$\% at $\log{M_*/M_{\odot}}<10.3$ based on independent X-ray, infrared, and radio indicators. Therefore the absence of AGN in our sample is not unexpected given the stellar mass range probed.

\begin{figure}
\begin{center}
\includegraphics[width=7cm]{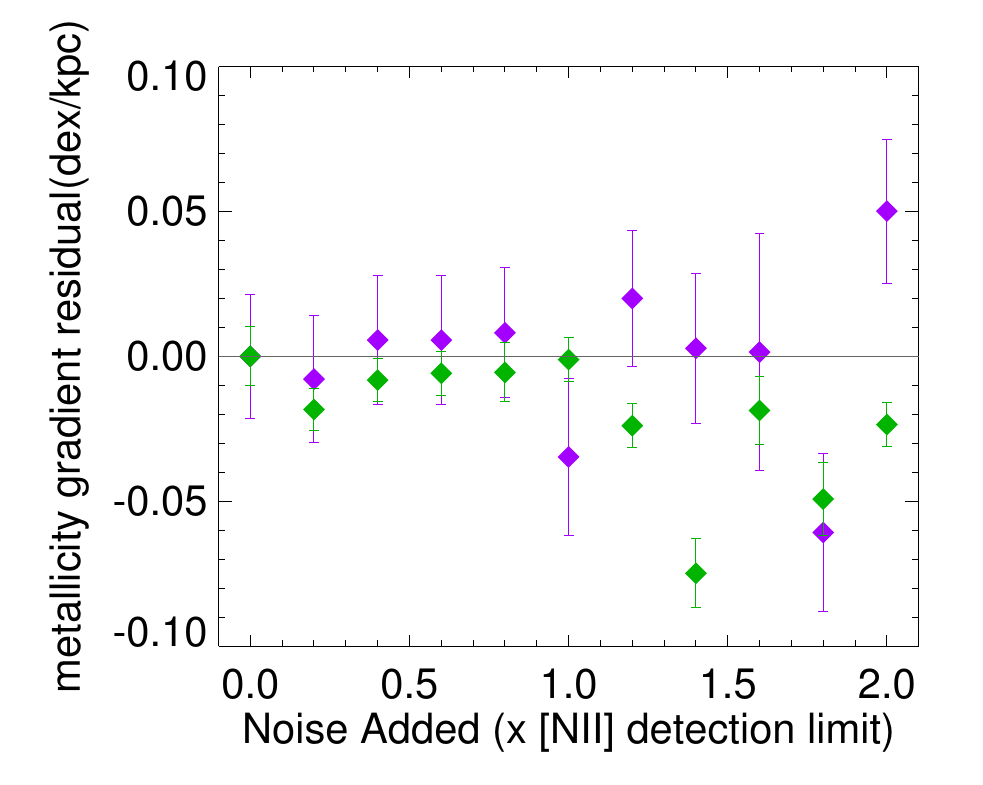}
\end{center}
\caption{Verifying the validity of our derived metal gradients for low signal/noise [N II] data.  Metallicity gradient residuals are shown for CSWA128 as a function of the amount of synthetic noise added (see text in Section \ref{sec:agn} for technical details). Gradients are derived with PP04 method both along the galaxy 'major axis' (purple) and using radial binning(green).}
\label{fig:slope_noise}
\end{figure}

We then consider the possible bias in measured metallicity gradients due to possibly low signal-to-noise ratio of [N II] emission lines. We select CSWA128 to be a fiducial sample in this analysis due to its moderately steep gradient  so that we can examine possible degradation of this gradient as we increase the noise. We adopt the $2\sigma$ detection limit of the [N II] emission line for CSWA128;the limits for other galaxies are similar. We then synthetically add Gaussian noise to the data cube up to twice the detection limit and re-measure the metallicity gradient with the same procedure. We compare the measured gradients when the noise is added to the data cube to the gradients measured from the original data cube in Figure \ref{fig:slope_noise}. We find that the gradients measured from the noise-added data cubes are consistent with the original measurement up to doubling the noise or a quarter of the exposure time for CSWA128. Accordingly we can be confident that the measured gradients in other galaxies are not likely to be affected given our [N II] detection limits. 

\section{Discussion}
\label{sec:discussion}

\begin{figure}
 	\includegraphics[width=8.5cm]{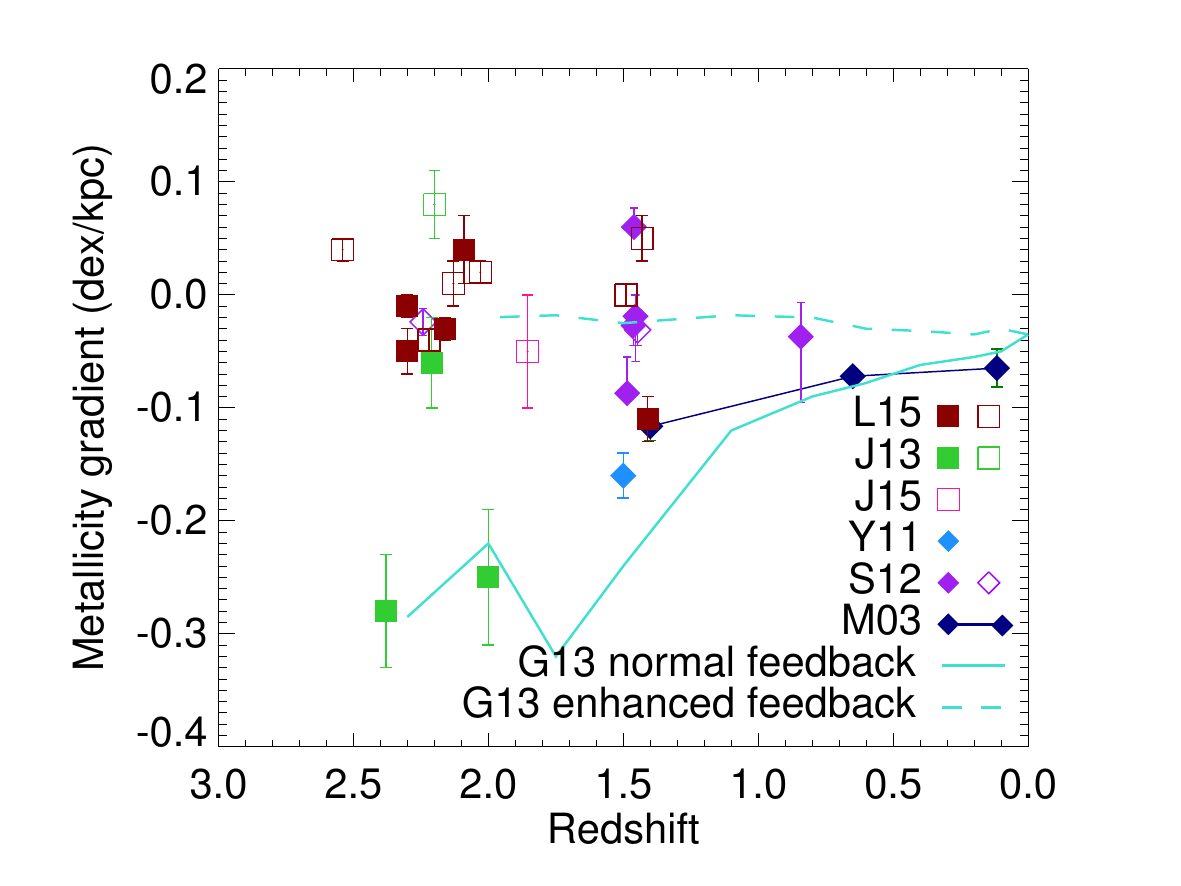}
	\caption{The evolution of metallicity gradient with redshift. Red squares represent lensed galaxies from the present study (L15). Green squares are the four lensed galaxies in \citet[J13]{Jones13}. Other data points are measured from a lensed starburst dwarf \citep[J15]{Jones15}, a lensed galaxy \citep[Y11]{Yuan11}, non-lensed galaxies observed with adaptive optics \citep[S12]{Swinbank12}, and Milky WayÕs planetary nebulae \citep[M03]{Maciel03}. Open and filled symbols represent kinematically well-ordered and disturbed systems respectively. Turquoise lines show predictions from the hydrodynamical simulations of \citet[G13]{Gibson13} emphasizing the strong sensitivity to the incorporated feedback.}
	\label{fig:z_gradient}	
\end{figure}

Previous observations \citep{Yuan11,Jones13,Jones15} led to a simple picture whereby isolated galaxies at high redshifts tend to have steep metallicity gradients which gradually flatten with cosmic time, while interacting galaxies have no discernible gradients at all. Conceptually this could be understood if metal gradients only became properly established when galaxies are kinematically well-ordered, with gradients flattening as galaxies grow in size. However, the enlarged data set presented here shows a large diversity of gradients which is incompatible with this picture. The redshift-dependent behavior of metallicity gradients in our sample is summarized in Figure~\ref{fig:z_gradient}. As in previous work we find (negatively) steeper gradients in rotationally supported systems compared to the merging and dynamically immature systems. However, the gradients in rotationally supported galaxies are much flatter ($>-0.1$ dex/kpc) than those previously observed in \citet{Jones13} ($\sim-0.3$ dex/kpc), indicating less evolution in the gradient slope at $z<2$. We now seek to explain the diversity in observed gradient slopes especially among isolated rotating galaxies.

\begin{figure*}
	\includegraphics[width=8.5cm]{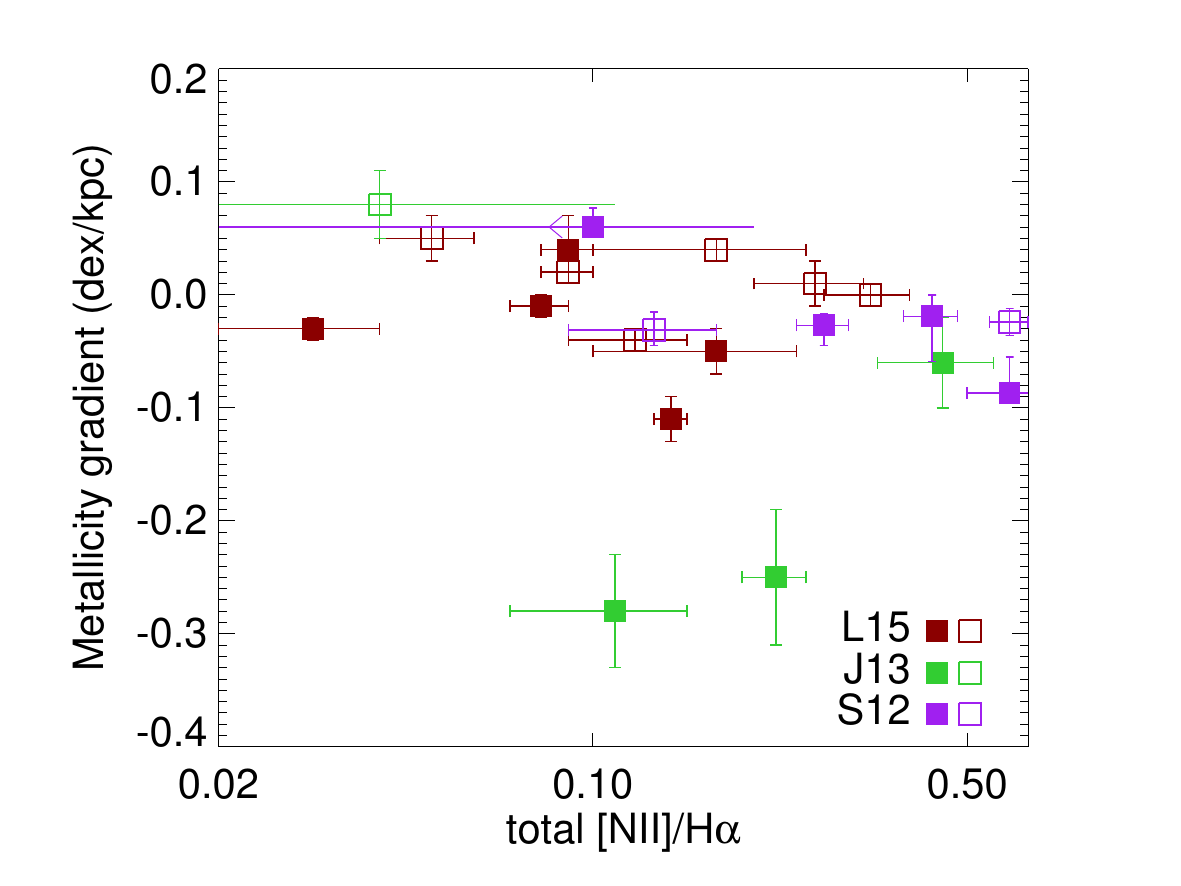}
	\includegraphics[width=8.5cm]{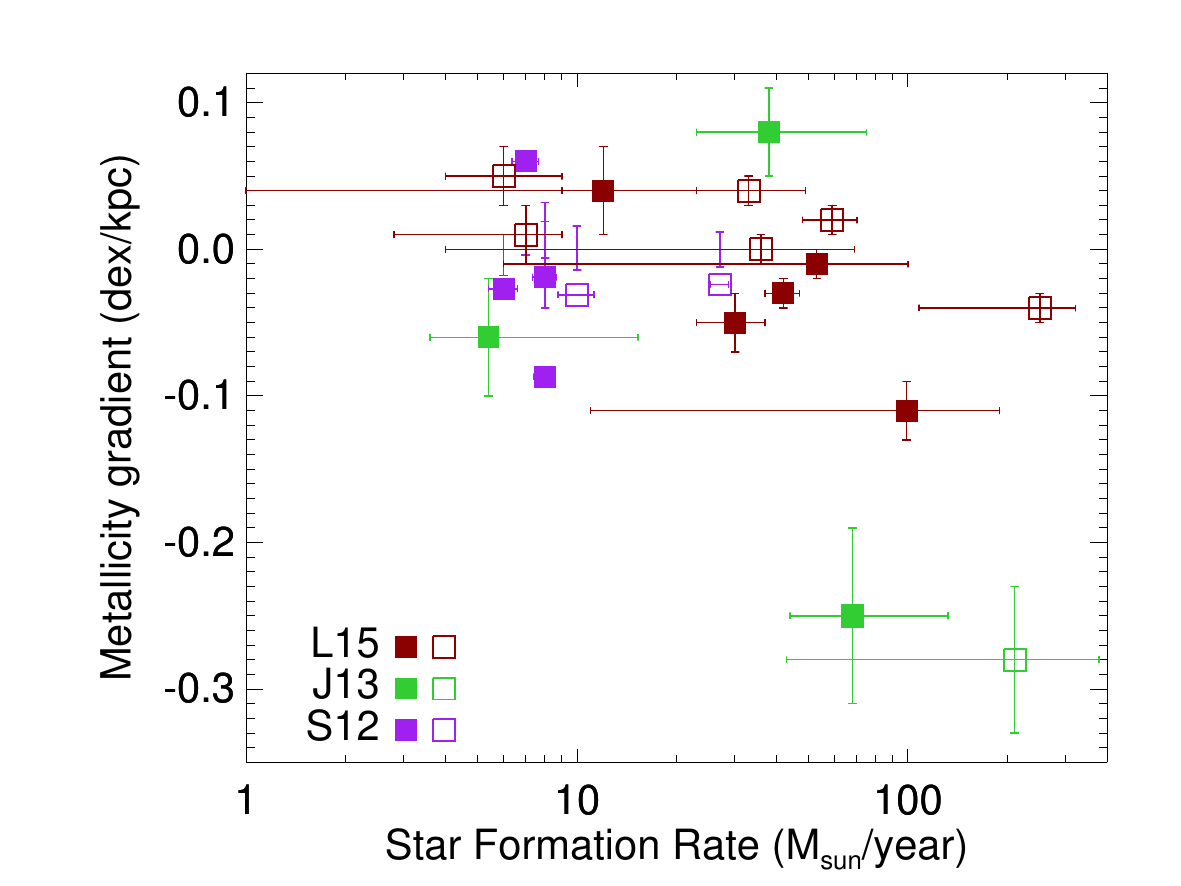}
	\caption{Correlation between the observed metallicity gradient for the present sample (L15), \citet[J13]{Jones13} and \citet[S12]{Swinbank12} and the integrated [N II]/H$\alpha$ ratio (left panel) and the star formation rate (right panel).  Filled symbols represent isolated galaxies; open symbols represent interacting and/or kinematically-immature galaxies.}
	\label{fig:totalmetal_gradient}
\end{figure*}

One possibility for the difference seen between this work and that of \citet{Jones13} could be procedural. In the present sample, we avoided a possible selection bias towards unusually active or metal-rich sources. In \citet{Jones13}, integrated [N II] and H$\alpha$ fluxes were first measured for a larger sample with a slit spectrograph and only those four sources with relatively bright emission lines (particularly [N II]) were selected for subsequent study with OSIRIS. Following upgrades to the OSIRIS grating and Keck I adaptive optics system, this pre-screening step was avoided in the current sample (with the exception of CSWA31). To test for this bias, the left panel of Figure~\ref{fig:totalmetal_gradient} correlates the metallicity gradient with the integrated [N II]/H$\alpha$ line ratio for both isolated or rotationally-supported and kinematically disturbed systems. All galaxies with [N II]/H$\alpha < 0.1$ have flat gradients ($< 0.1$ dex kpc$^{-1}$). Moreover, for galaxies with [N II]/H$\alpha > 0.1$ there is a divergence between isolated or rotationally-supported and merging or dynamically-immature systems with the latter showing zero gradients irrespective of the metallicity. The right panel of Figure~\ref{fig:totalmetal_gradient} examines the dependence on the integrated star formation rate to check whether the \citet{Jones13} sample is biased to more intense star-forming systems; however, no strong trend is revealed. 

A more likely possibility for the diversity of gradients at $z\simeq2$--2.5 is variation in feedback strength. Numerical simulations suggest that metallicity gradients are highly sensitive to feedback in the form of outflows and ``galactic fountains" \citep{Pilkington12,Gibson13,Angles-Alcazar14}, in the sense that stronger feedback (i.e., higher mass loading factors) leads to more gas mixing and therefore flatter gradients. This is illustrated in Figure \ref{fig:z_gradient} which shows evolution of the metallicity gradient in a simulated galaxy using two different feedback prescriptions. Simple analytical chemical evolution models show that stronger feedback also results in lower gas-phase metallicity \citep[e.g.][]{Jones13}. The correlation of gradient slope with integrated metallicity, as shown in the left panel of Figure~\ref{fig:totalmetal_gradient}, could therefore also be due to a variation in feedback strength. In fact, the enhanced feedback scheme in Figure \ref{fig:z_gradient} was preferred over the normal feedback scheme considering the stellar mass-halo mass relationship matching \citep{Stinson13}.

\begin{figure}
	\includegraphics[width=8.5cm]{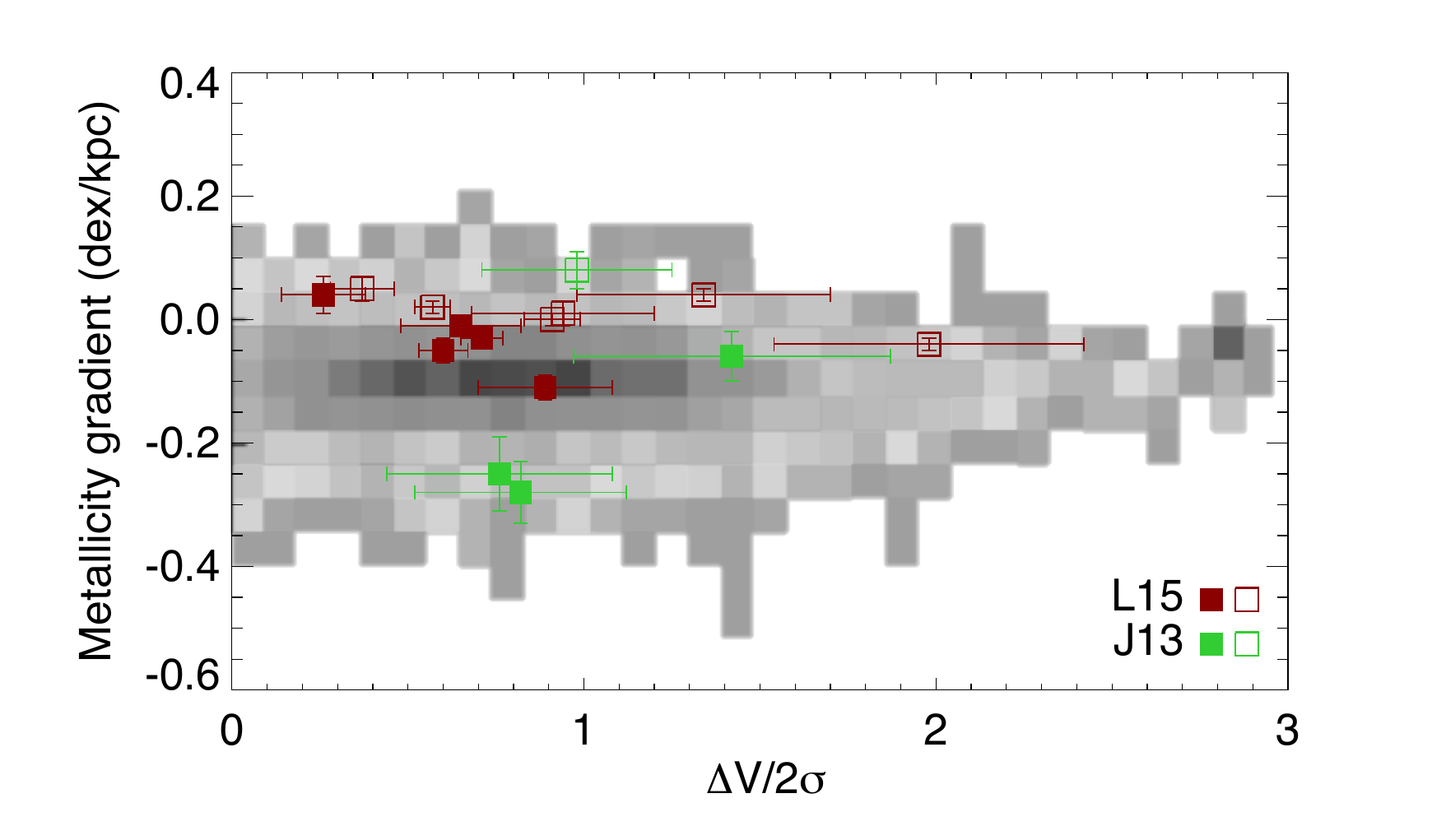}
	\caption{The absence of a correlation between the metallicity gradient and the degree of rotational support. Data points represent the present sample and the grey scale represent the output of the {\it Illustris} simulation for a sample of $z\simeq$2 galaxies \citep{Vogelsberger14, Nelson15}.}
	\label{fig:vel_gradient}
\end{figure}

While Figure~\ref{fig:totalmetal_gradient} provides evidence that gradient slopes are affected to different degrees by feedback, the trend has a large scatter suggesting that other factors are likely important. One obvious possibility is the degree of rotational support compared to random motions which would mix the gas and flatten any gradients. This effect is clearly seen in the sense that merging and interacting systems have flatter gradients compared to isolated galaxies. Among isolated galaxies, we would naively expect those with higher ratios of rotational to random motion (i.e., higher $v/\sigma$) to have stronger gradients in metallicity and perhaps other properties. However, Figure~\ref{fig:vel_gradient} shows very little dependence on $v/\sigma$. We compare our observational results with $z=2$ galaxies drawn from the {\it Illustris} simulation \citep{Vogelsberger14, Nelson15}\footnote{Illustris data is available through \url{http://www.illustris-project.org} }, a large-volume cosmological hydrodynamical simulation with baryonic feedback. Metallicity gradients in the simulation data were constructed based on a linear fit to all of the star forming gas within a radius of 10 kpc from each galaxyÕs center. The simulated galaxies have gas-phase kinematics and negative metallicity gradients which are similar to our observations and likewise show no clear trend between gradient slope and rotational support. In essence, galaxies with gas disks supported by rotation have radial metallicity gradients which are similar to systems dominated by random motions, in contrast to expectations that bulk velocity dispersion should flatten any gradients. This suggests that gradients are established on timescales of the order of a dynamical time, sufficiently short that prominent random motions are not effective at mixing the gas.

To summarize, we find a large diversity of metallicity gradients among rotating galaxies at $z\simeq2$. Most rotating galaxies in our sample have negative gradients while interacting systems have flatter gradients, as is found in local galaxy samples. Surprisingly we find that the degree of ordered rotation versus random bulk motion has no discernible effect on the gradient slope even among isolated galaxies, indicating that gradients are formed and destroyed on timescales comparable to the galaxy dynamical time ($\sim 10^8-10^9$ yr). The steepest observed gradients with slopes $\sim-0.3$ dex/kpc are less common than found in our previous work suggesting that this is a relatively rare or short-lived phase at high redshift. It would therefore be interesting to verify the time-dependence of steep gradients in the numerical simulations. Finally, we find a weak correlation between galaxy-integrated metallicity and gradient slope which we interpret as an effect of feedback: the same feedback which reduces the gas-phase metallicity evidently also flattens the gradients, as found in cosmological zoom-in simulations. The diversity of metallicity gradients at $z\simeq2$ is therefore likely caused in part by differences in the recent feedback history. As we argued in \cite{Jones13}, the evolution of metallicity gradients is evidently a sensitive probe of feedback in the form of gaseous outflows and galactic fountains.

\section{Summary}

We present spatially resolved kinematics and gas-phase metallicity gradient measurements for a total of 15 gravitationally lensed star-forming galaxies at $z\sim2$ based on the analysis of strong emission lines of H$\alpha$, [N II], [O III], and H$\beta$. 11 new sources were observed with the Keck laser-assisted adaptive optics systems and the upgraded OSIRIS integral field unit spectrograph to which we add  earlier data on four sources presented in \citet{Jones13}. With the aid of gravitational lensing, the typical source plane spatial resolution for each source is $<500$~pc which is considerably better than for other systems studied at similar redshifts.

We found the following key results:
\begin{itemize}

\item High spatial resolution is crucial in diagnosing the kinematic properties and dynamical maturity of $z\simeq2$ galaxies. We compare our observed velocity map of interacting pairs of merging galaxies with those simulated at the lower resolution typical in non-lensed surveys and demonstrate how easily such sources can be mistaken to represent rotationally-supported disks in poorer sampled data. Even for galaxies with no clear morphological sign of interaction, we often find significant large deviations in the velocity field compared to rotating disk models which would not be apparent in seeing-limited data. As a result we observe a significantly lower fraction (36\%) of rotationally-supported systems in our sample than has been claimed ($\simeq$74\%) from larger kinematic surveys undertaken with lower spatial resolution.

\item We find a much higher fraction of $z\simeq2$ galaxies have weak or non-existent metallicity gradients than in previous studies of smaller samples observed in the same redshift range. It seems unlikely that such a change arises as the result of a bias in the earlier sample which pre-selected sources with stronger [N II] lines and hence increased metallicity. We observe only a weak correlation between the presence of a gradient and the metallicity and none with the degree of rotational support, the latter being consistent with the recent predictions of the {\it Illustris} hydrodynamical simulations. We argue that variations in gas and metal mixing due to feedback most likely play the dominant role in modifying metal gradients and thus can explain the sizable scatter we see in our enlarged sample. The sensitivity of the observed metal gradient to the various modes of feedback indicates it will remain a promising tool for understanding galaxy assembly.  
\end{itemize}

 \begin{figure*}
\centering
	\includegraphics[width=4.3cm]{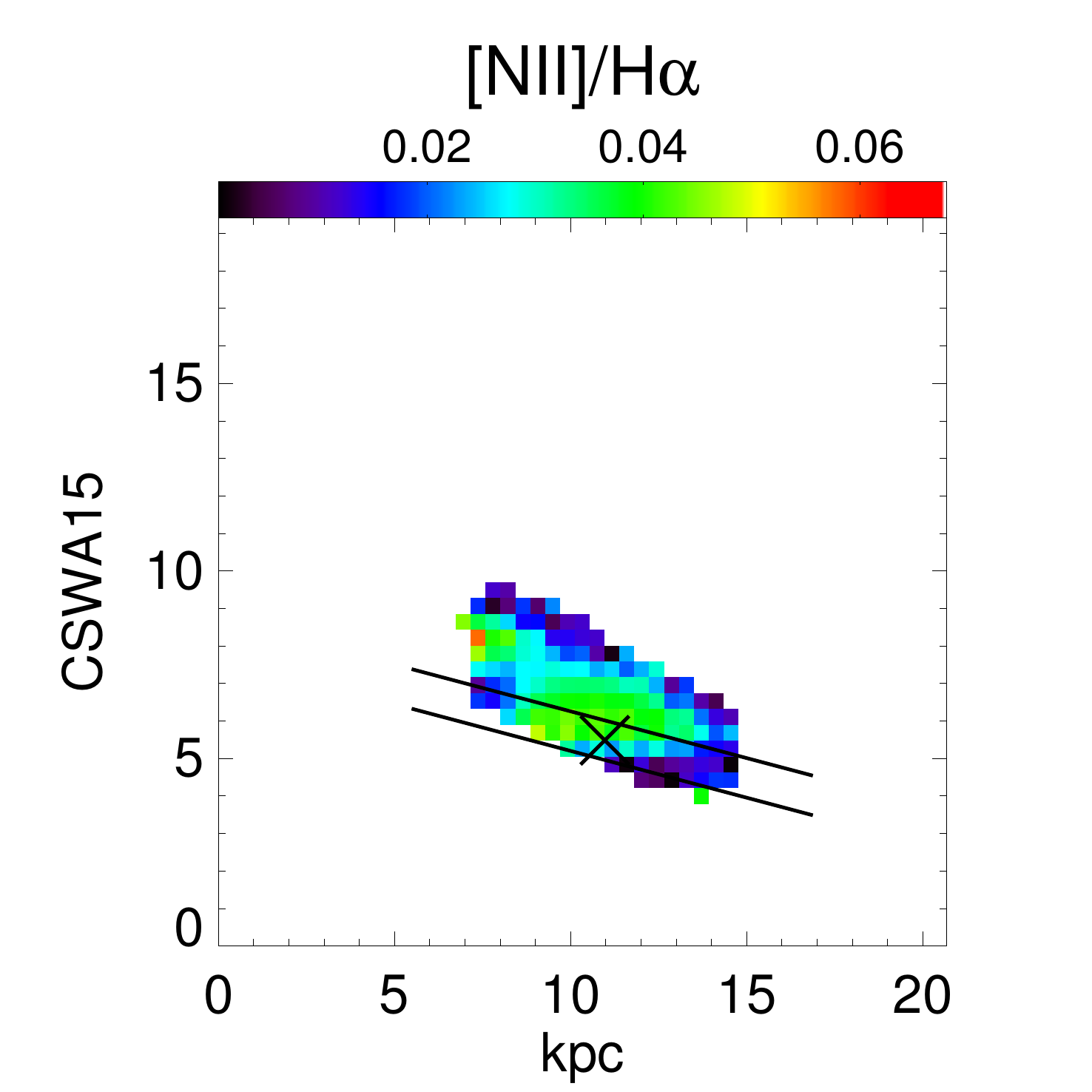}
	\includegraphics[width=5cm]{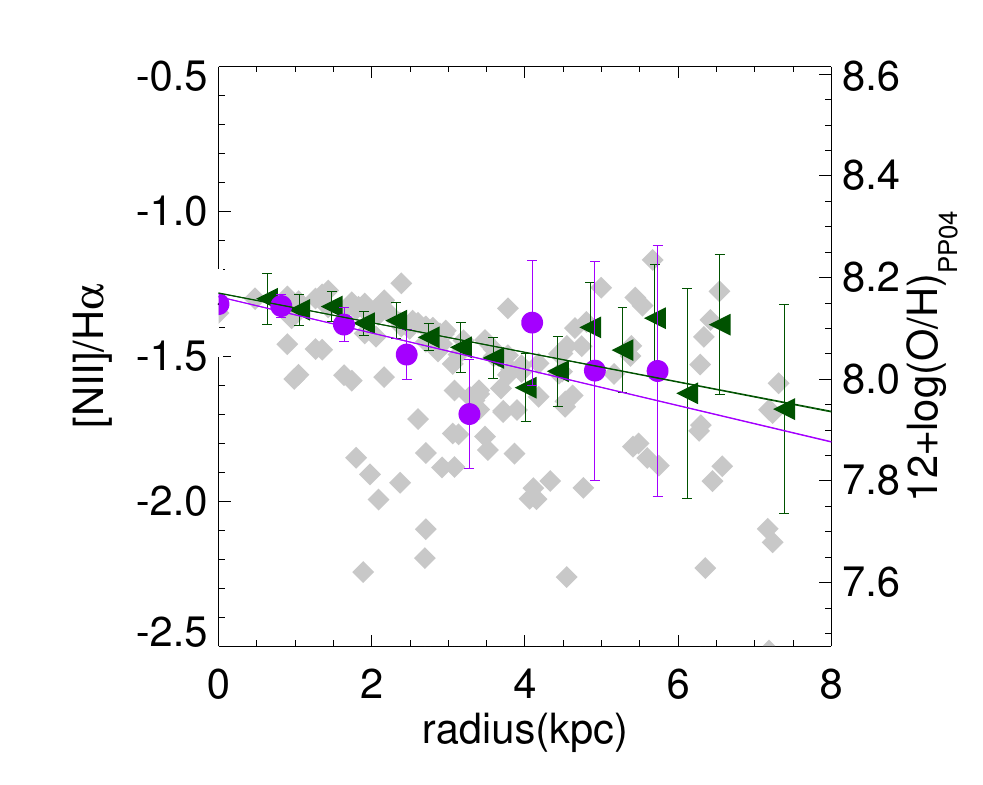}
	\includegraphics[width=5cm]{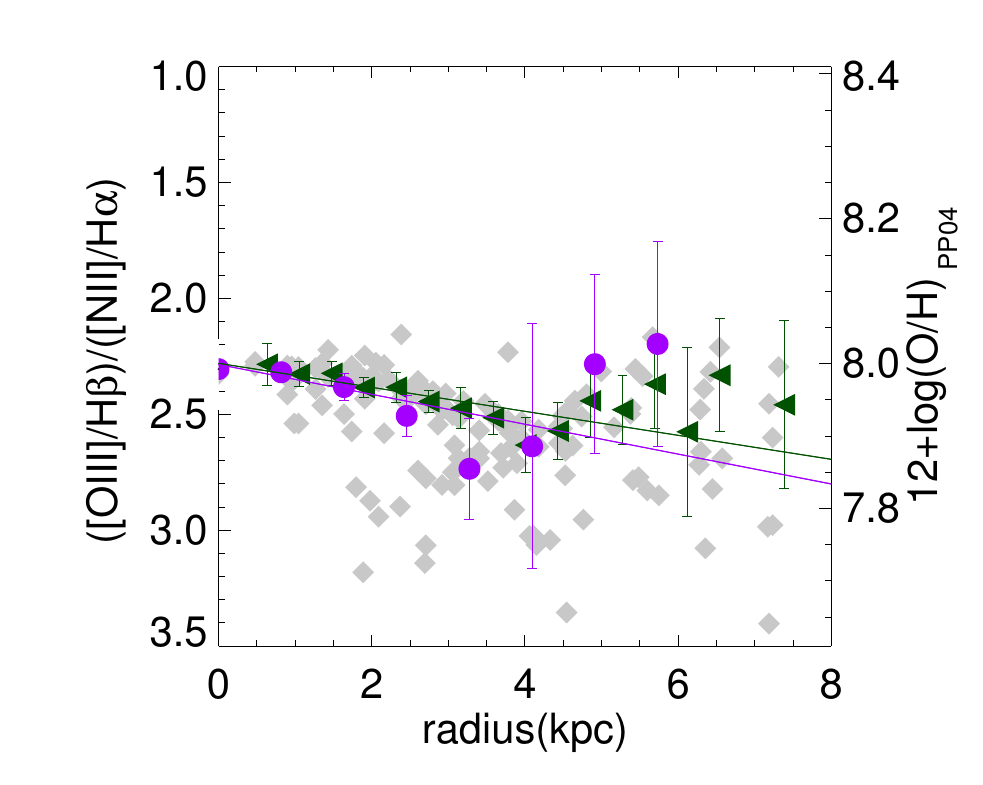}\\	
	\includegraphics[width=4.3cm]{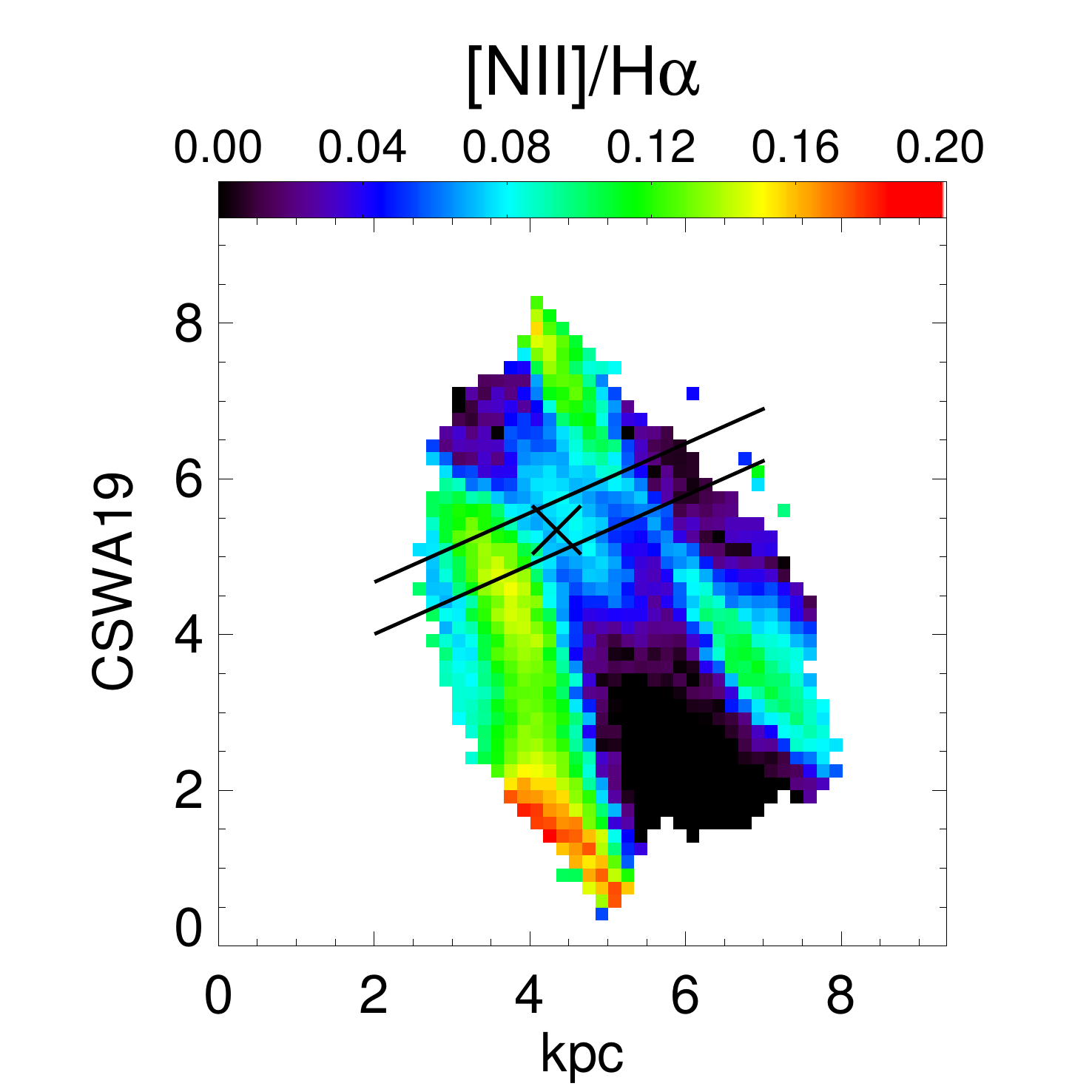}
	\includegraphics[width=5cm]{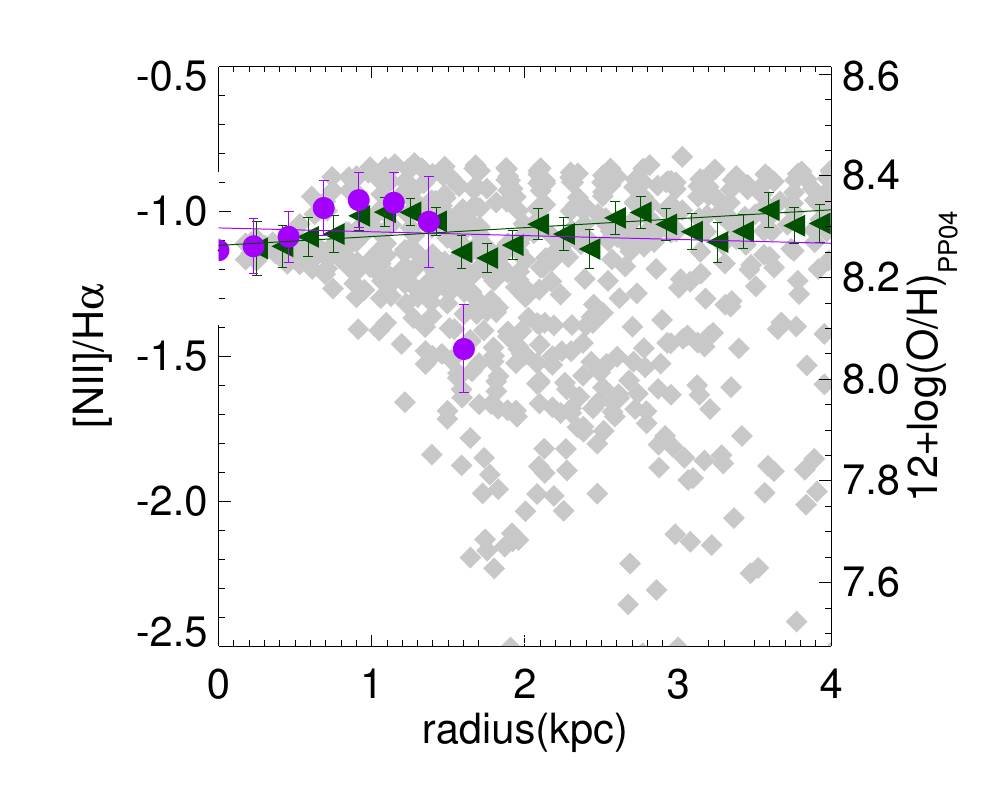}
	\includegraphics[width=5cm]{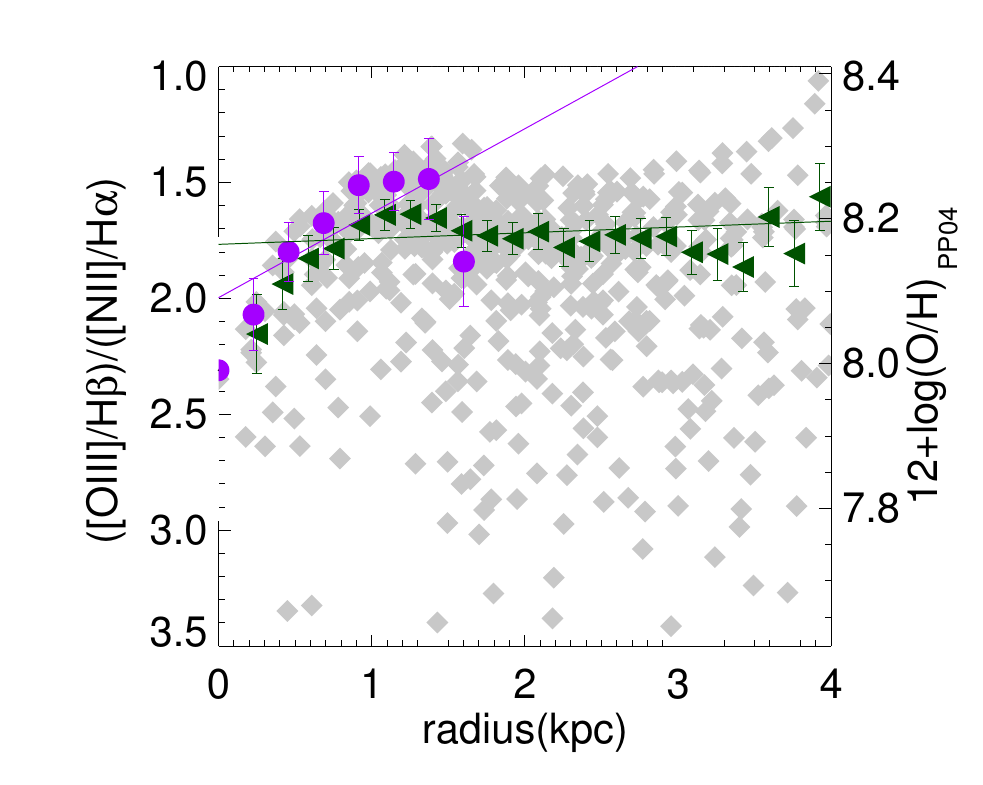}\\
	\includegraphics[width=4.3cm]{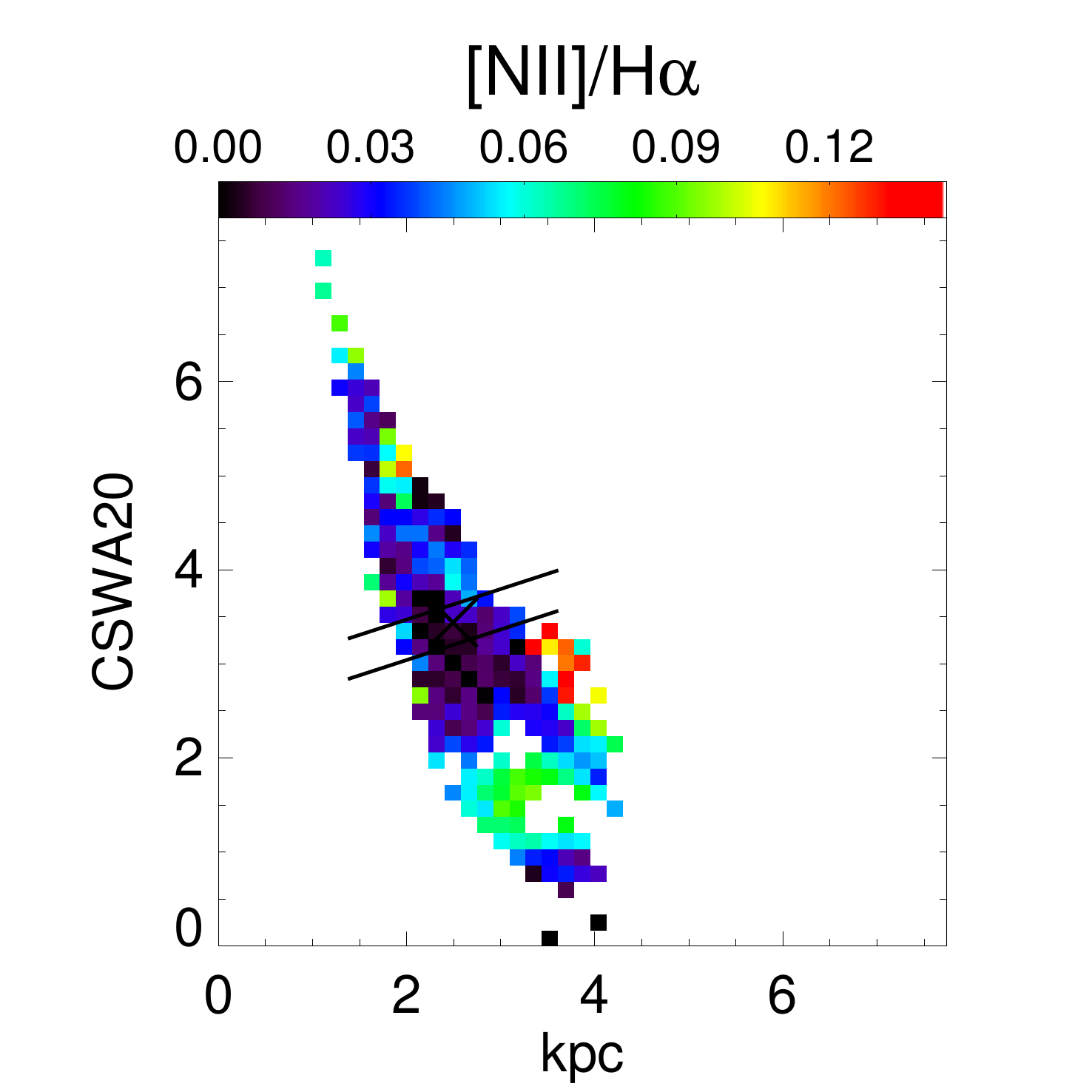}
	\includegraphics[width=5cm]{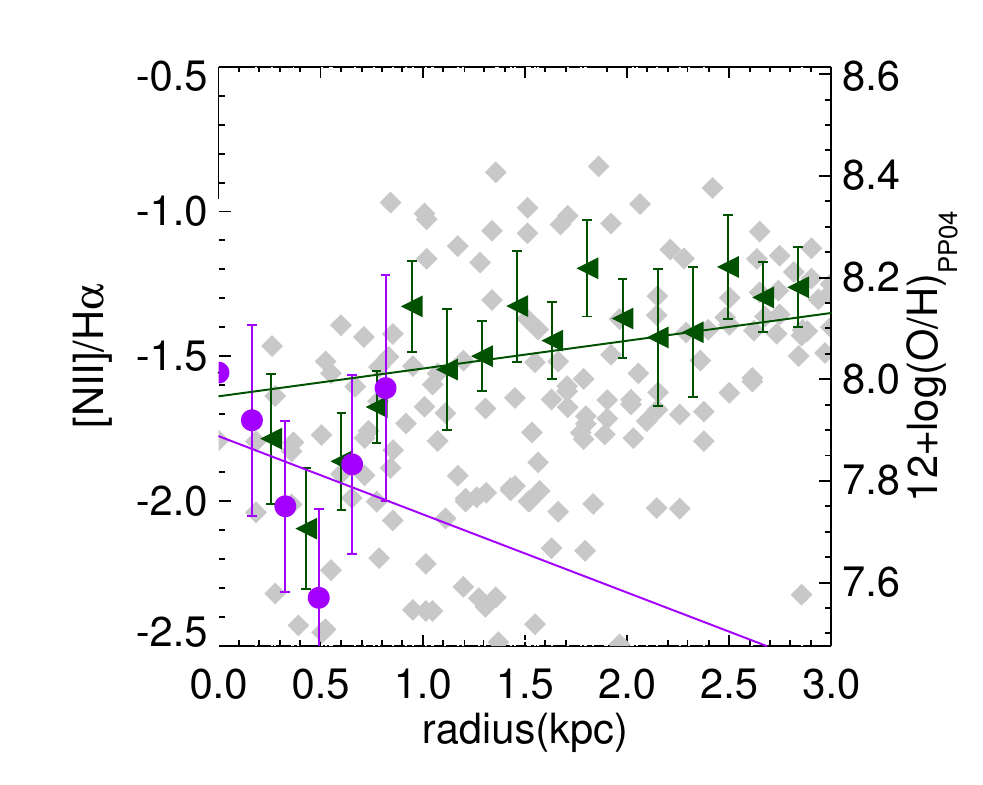}
	\includegraphics[width=5cm]{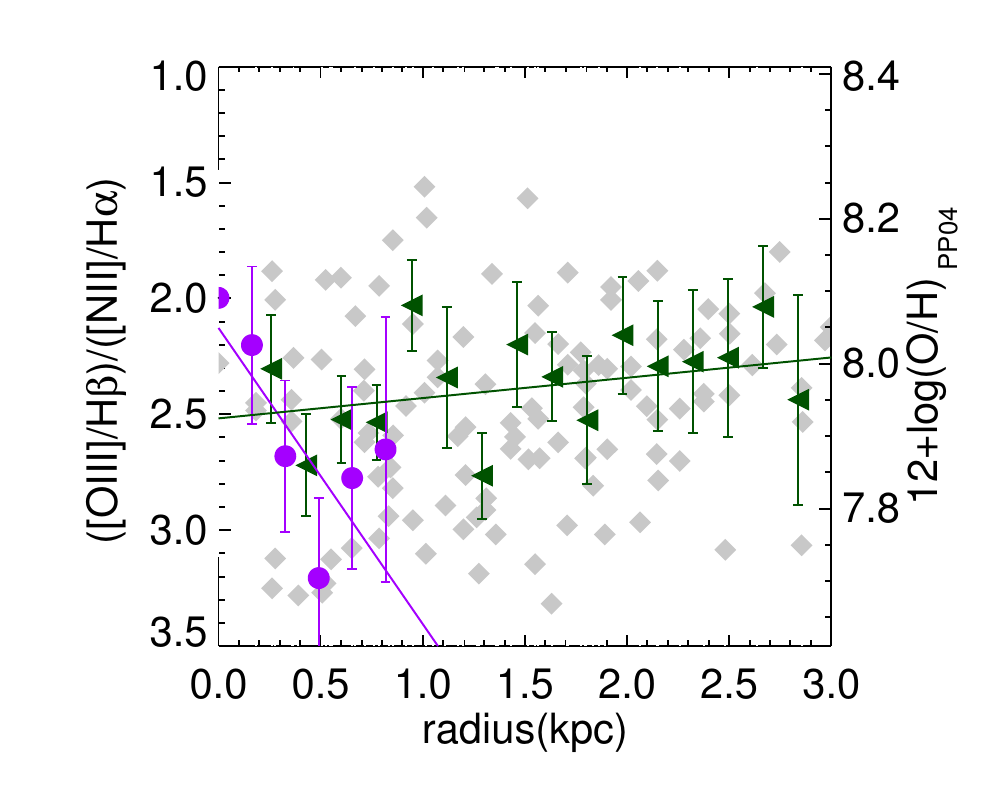}\\
	\includegraphics[width=4.3cm]{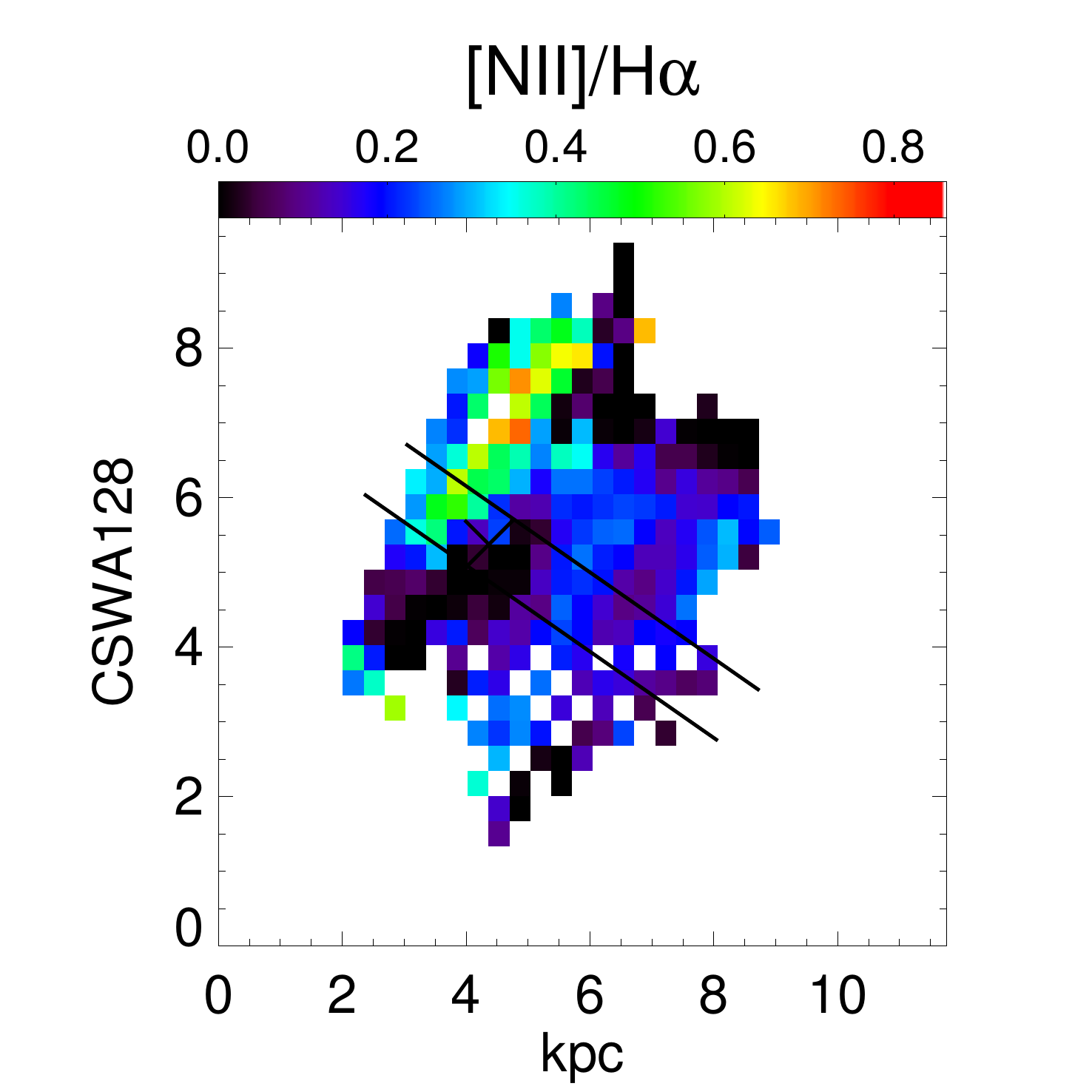}
	\includegraphics[width=5cm]{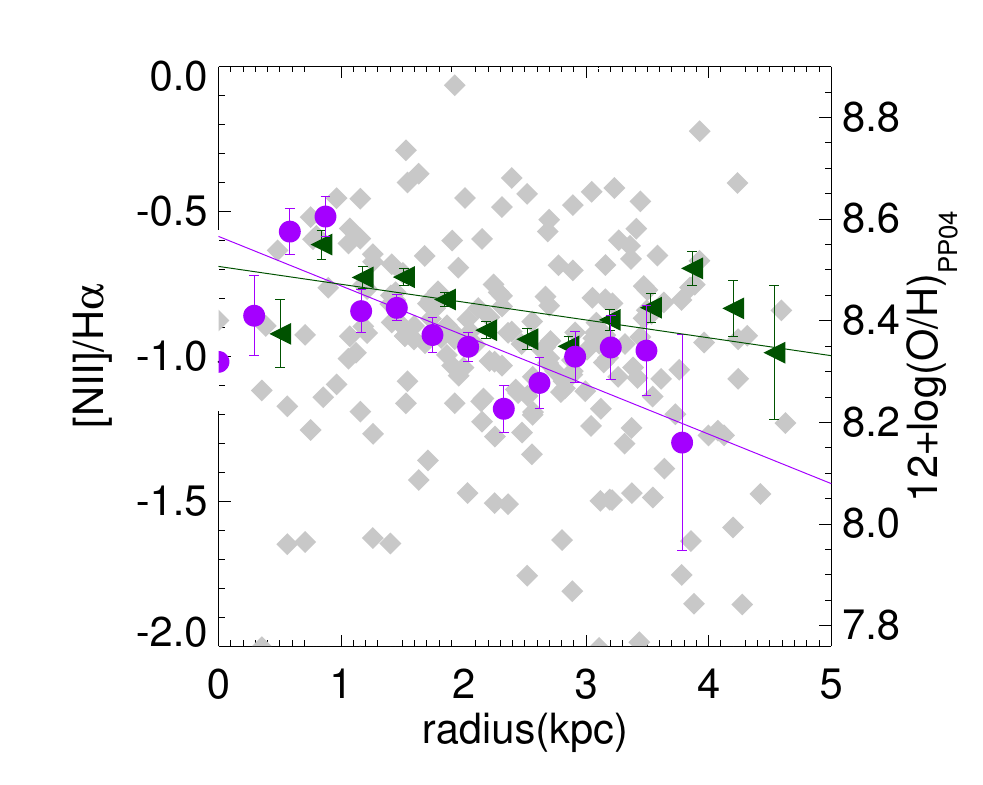}
	\includegraphics[width=5cm]{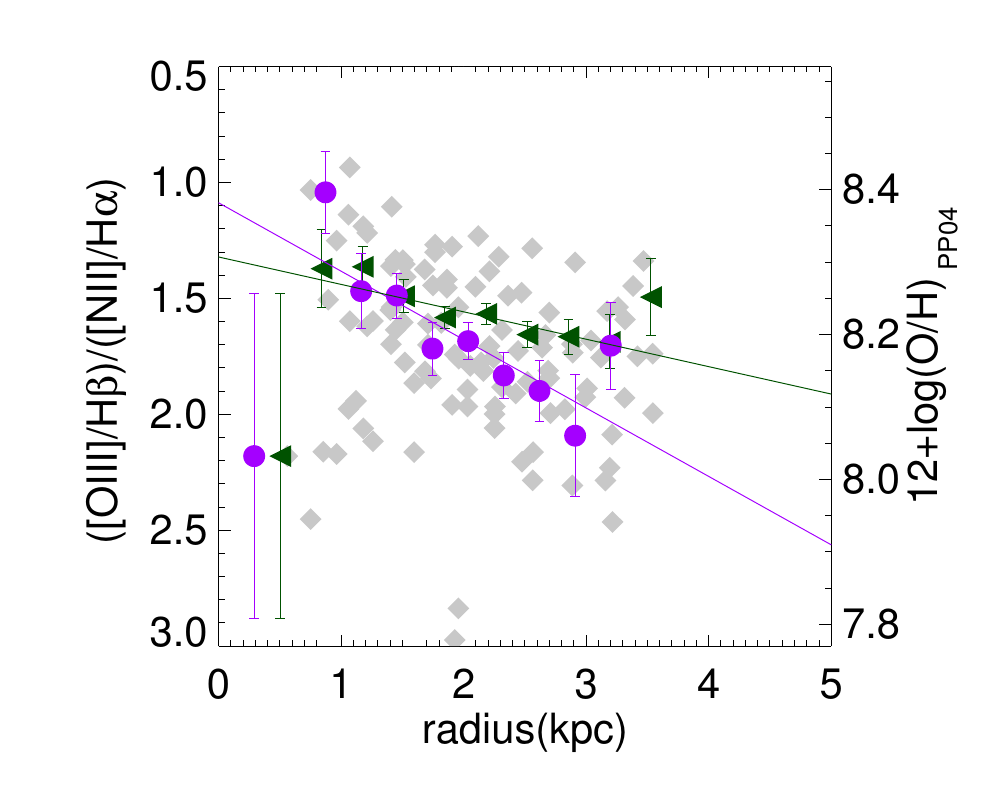}\\
	\includegraphics[width=4.3cm]{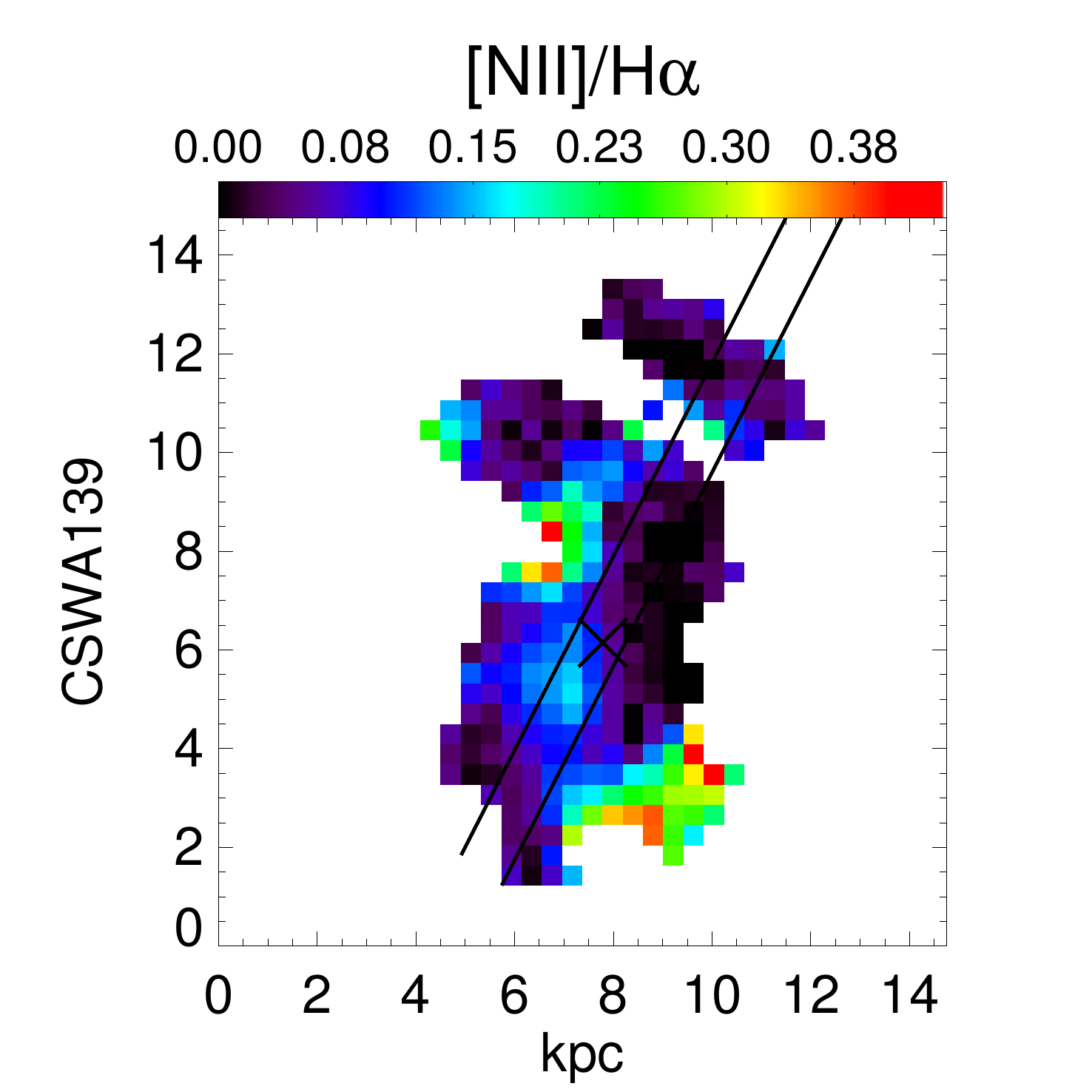}
	\includegraphics[width=5cm]{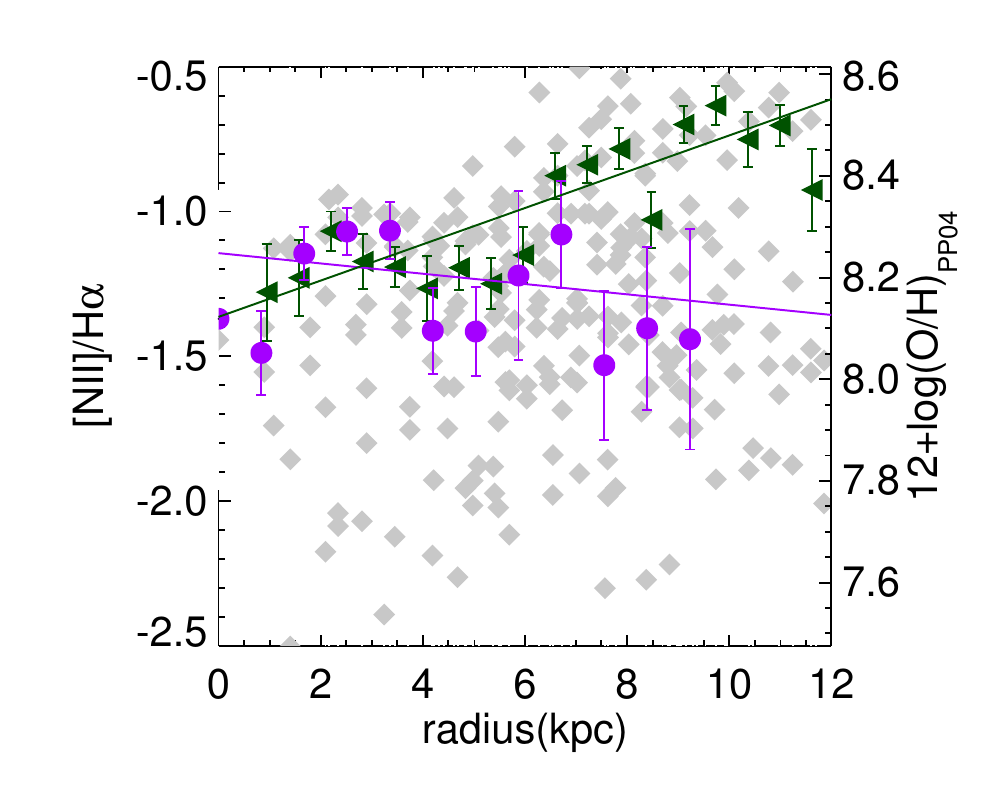}
	\includegraphics[width=5cm]{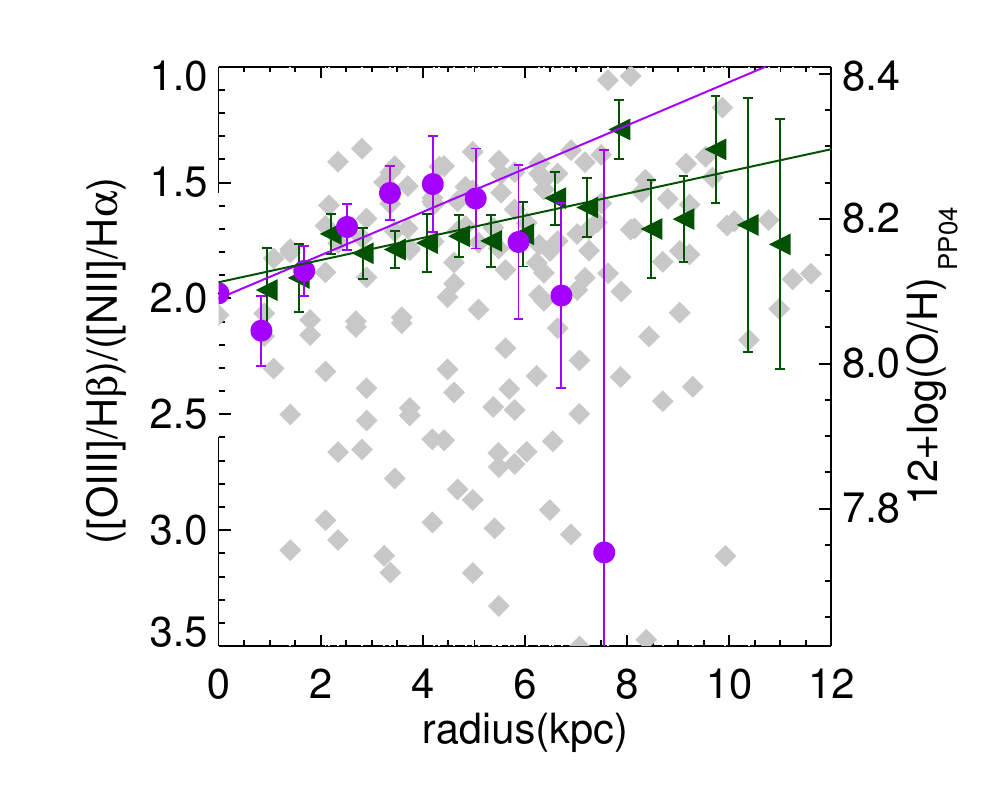}

\caption{ From left to right: source plane [N II]/H$\alpha$ map, radial metallicity gradient from N2 calibrators, and radial metallicity gradient from O3N2 calibrators for subsample of our targets. The gray dots are measurements from each source plane pixels. Green points and lines are measurements from radial binning. Purple points and lines are measurements along `major axis'.  \label{fig:NIIHa}}
\end{figure*}

\begin{figure*}
\ContinuedFloat
\centering
	\includegraphics[width=4.3cm]{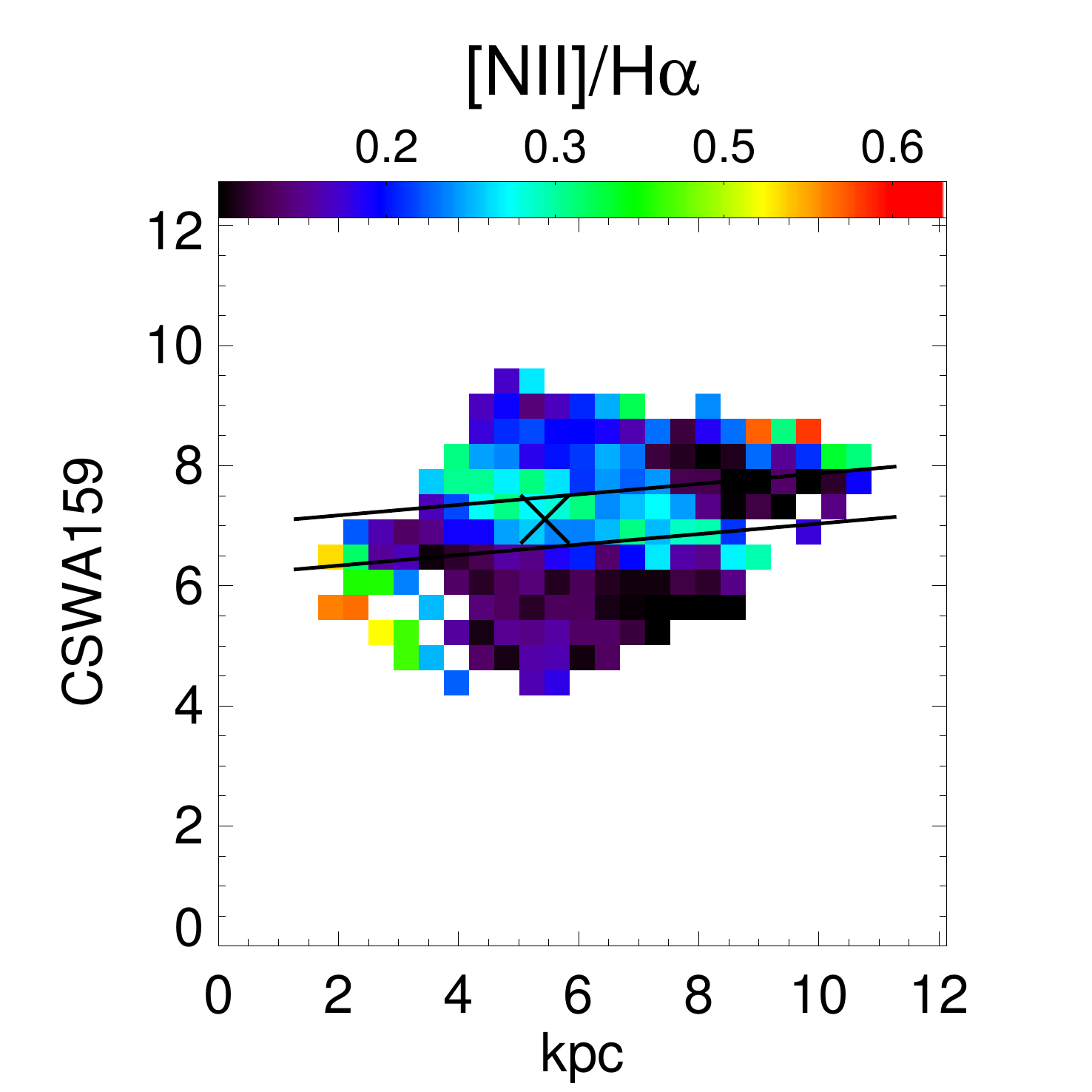}
	\includegraphics[width=5cm]{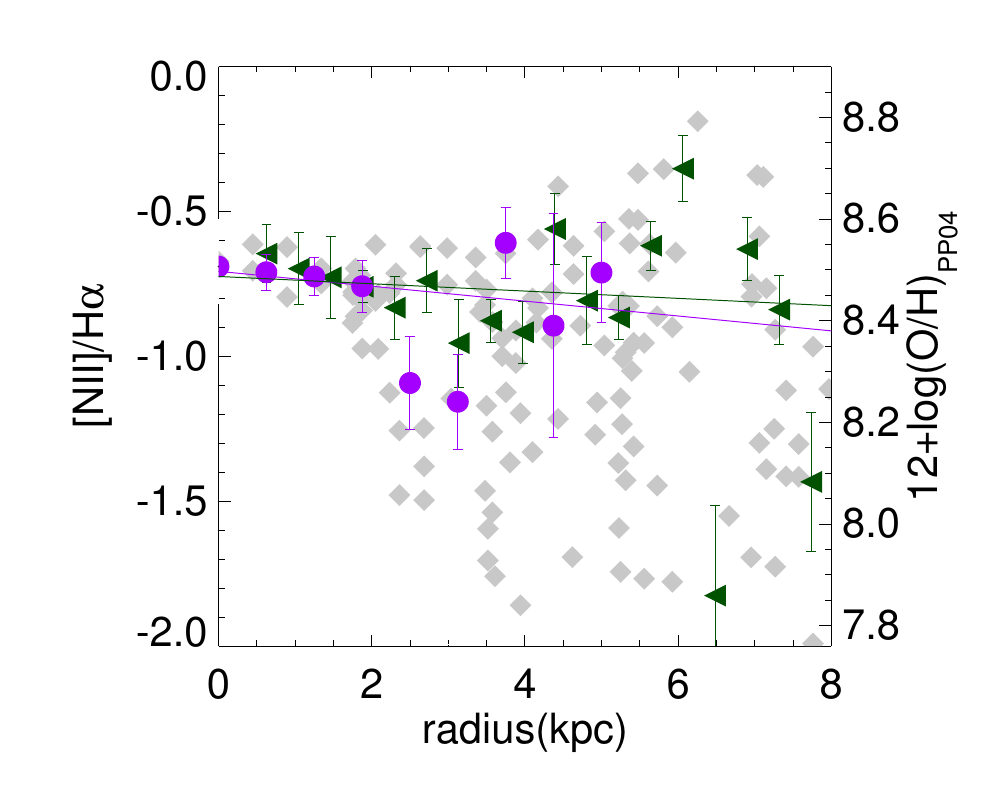}
	\includegraphics[width=5cm]{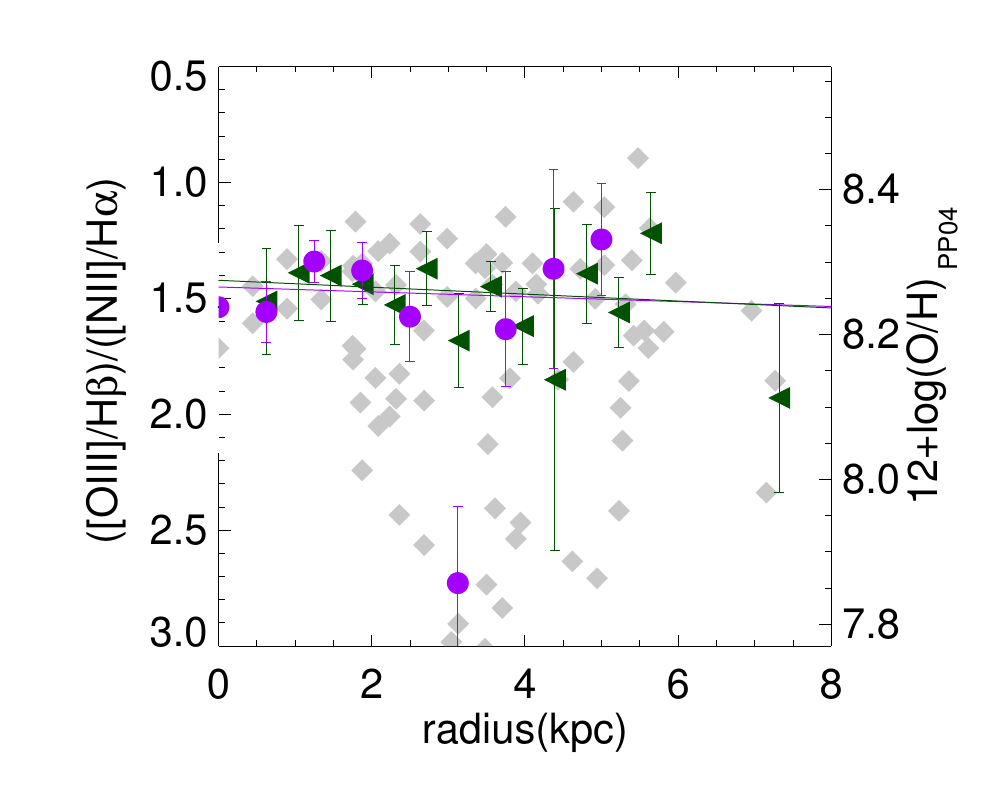}\\
	\includegraphics[width=4.3cm]{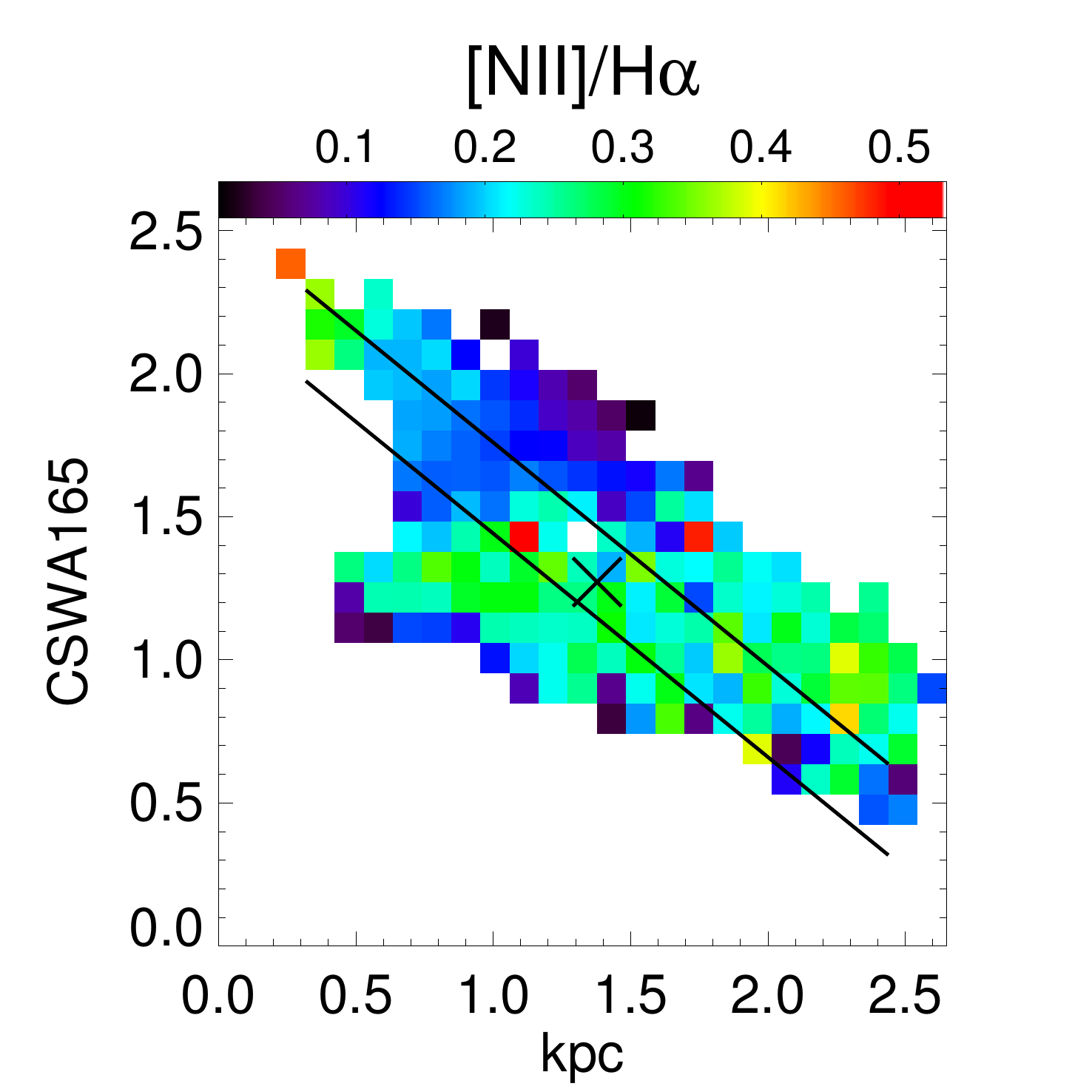}
	\includegraphics[width=5cm]{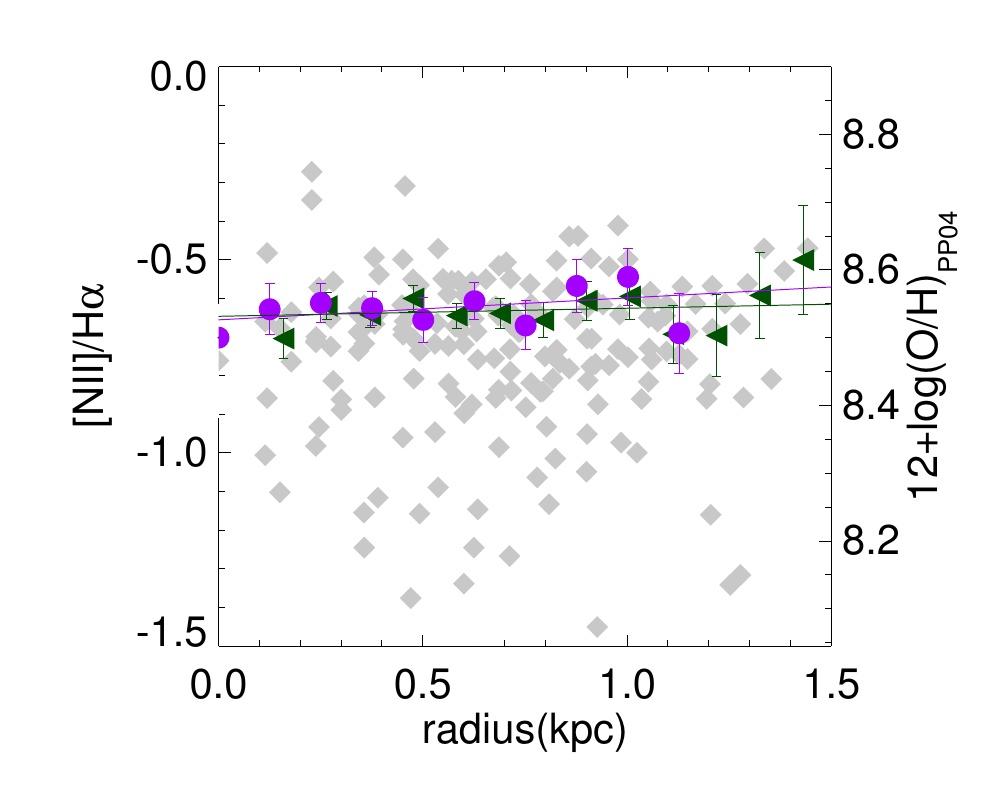}
	\includegraphics[width=5cm]{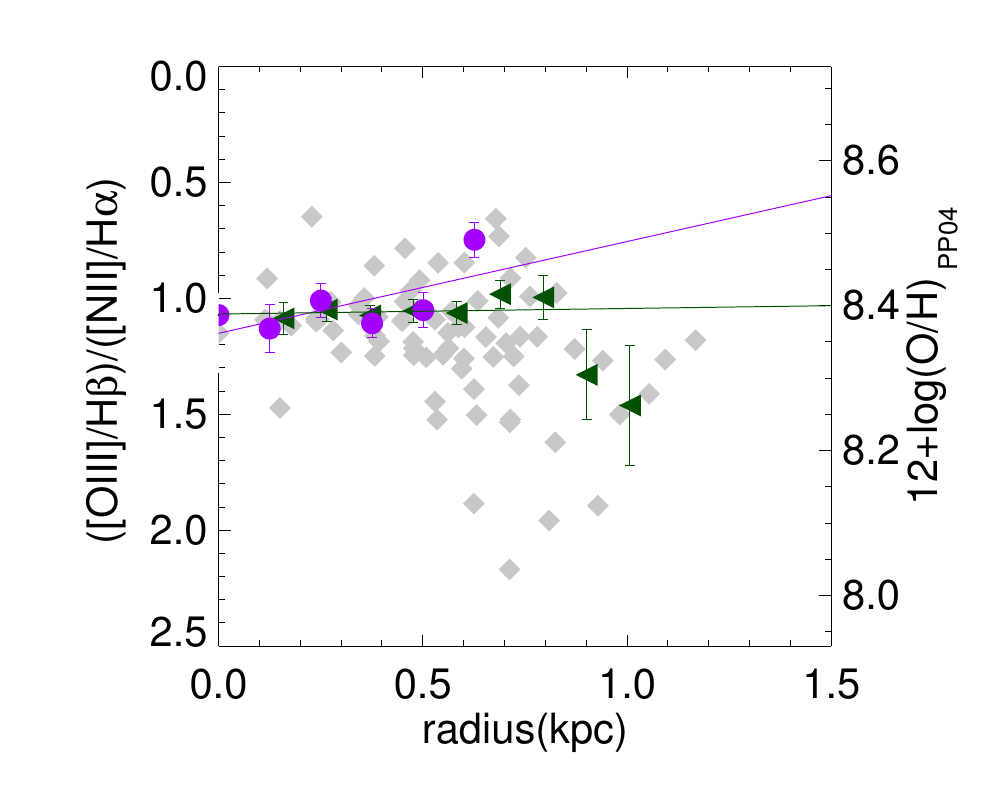}\\
	\includegraphics[width=4.3cm]{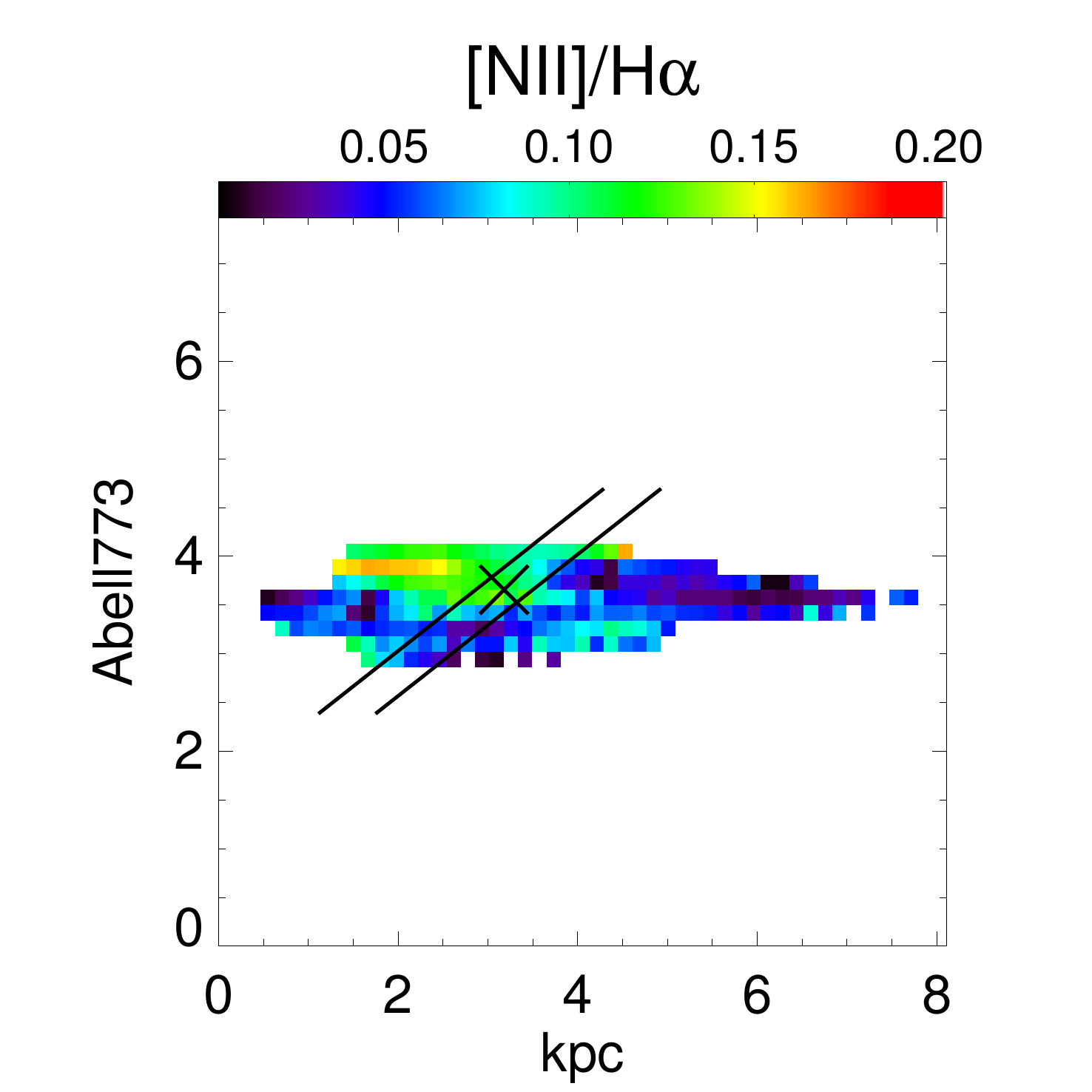}
	\includegraphics[width=5cm]{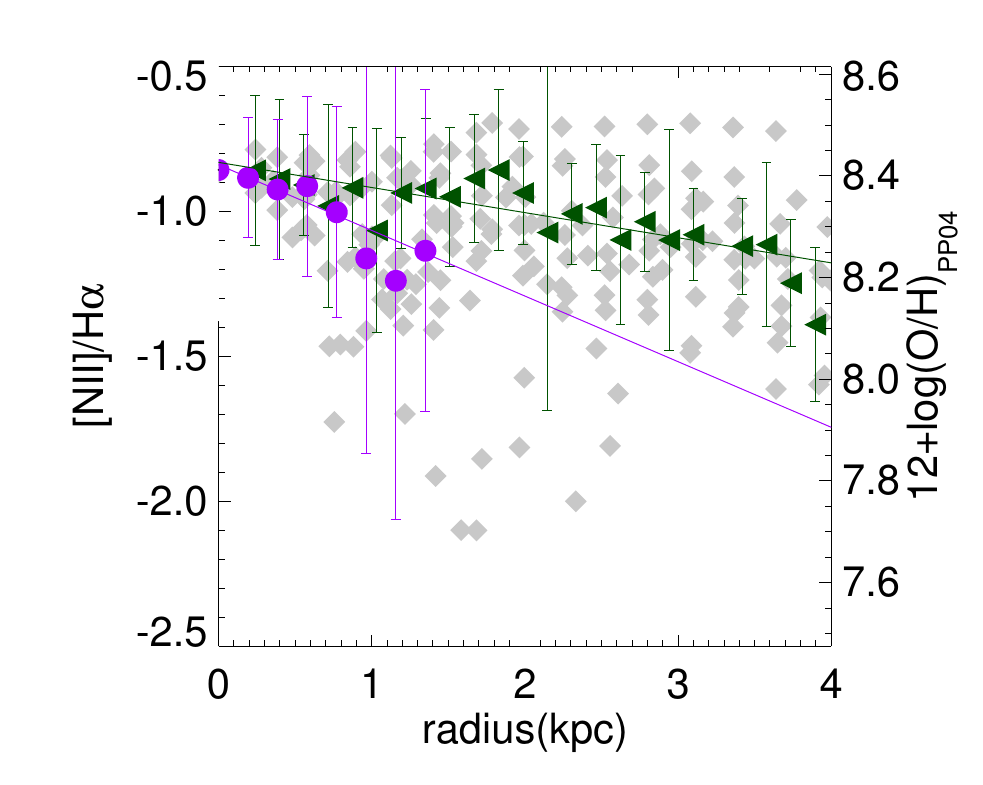}
	\includegraphics[width=5cm]{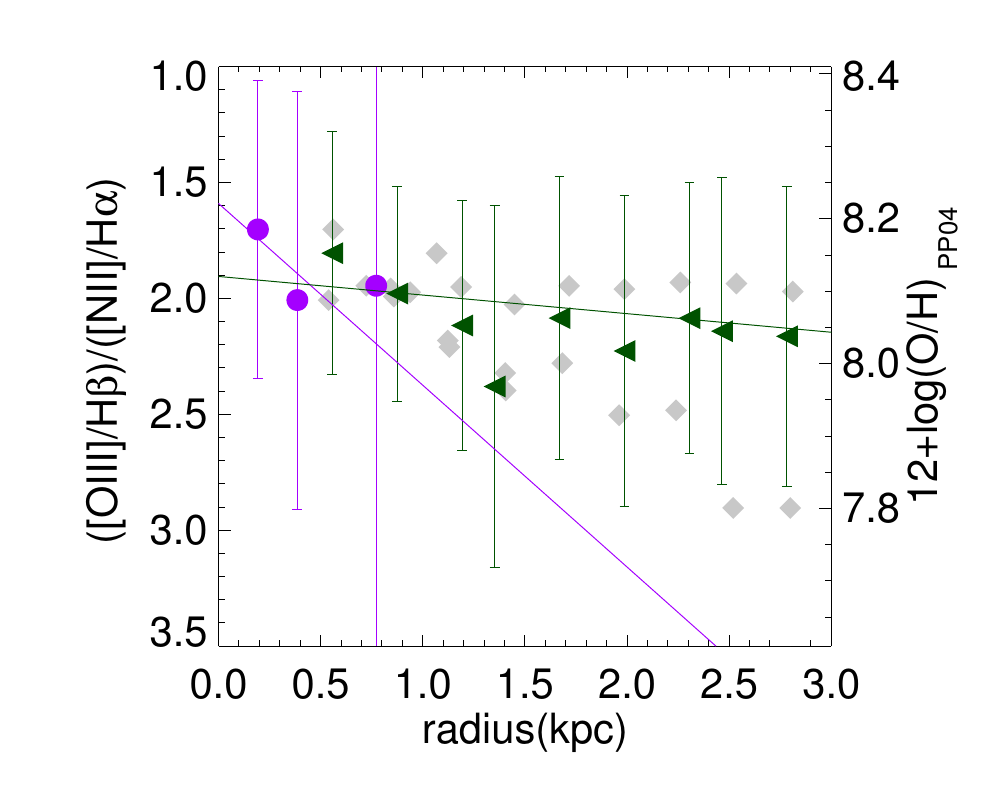}
	\caption{continued}
\end{figure*}

\begin{figure*}
\ContinuedFloat
\centering
	\includegraphics[width=4.3cm]{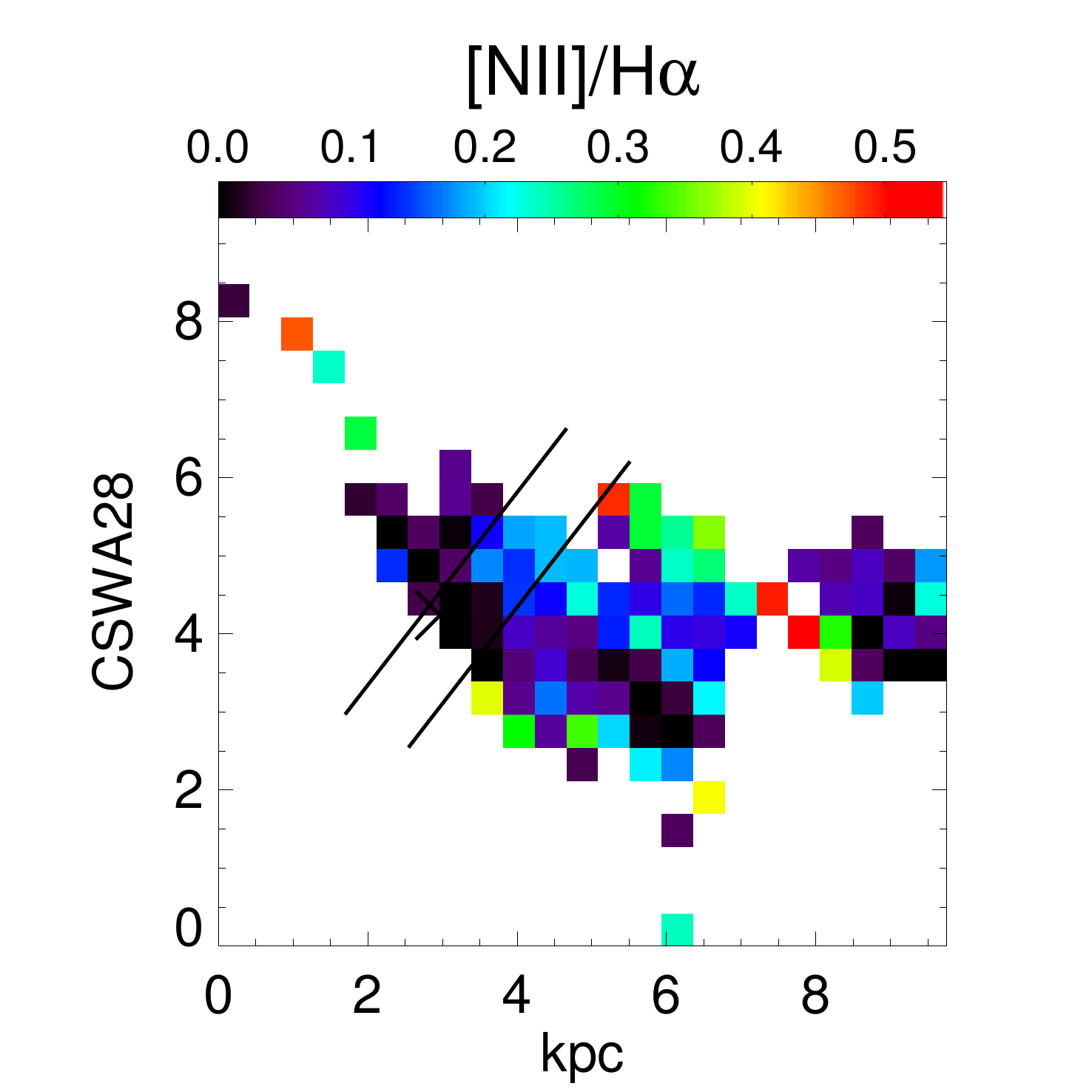}
	\includegraphics[width=5cm]{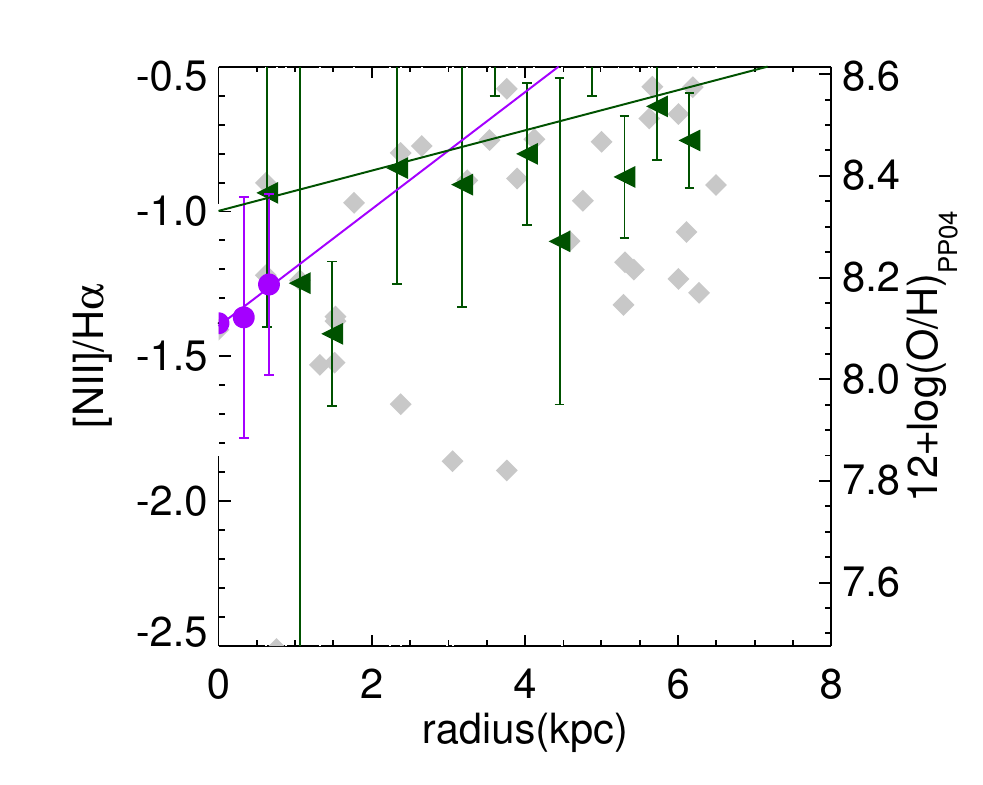}\\
	\includegraphics[width=4.3cm]{cswa28_N2map_n2}
	\includegraphics[width=5cm]{cswa28_N2radius}\\
	\includegraphics[width=4.3cm]{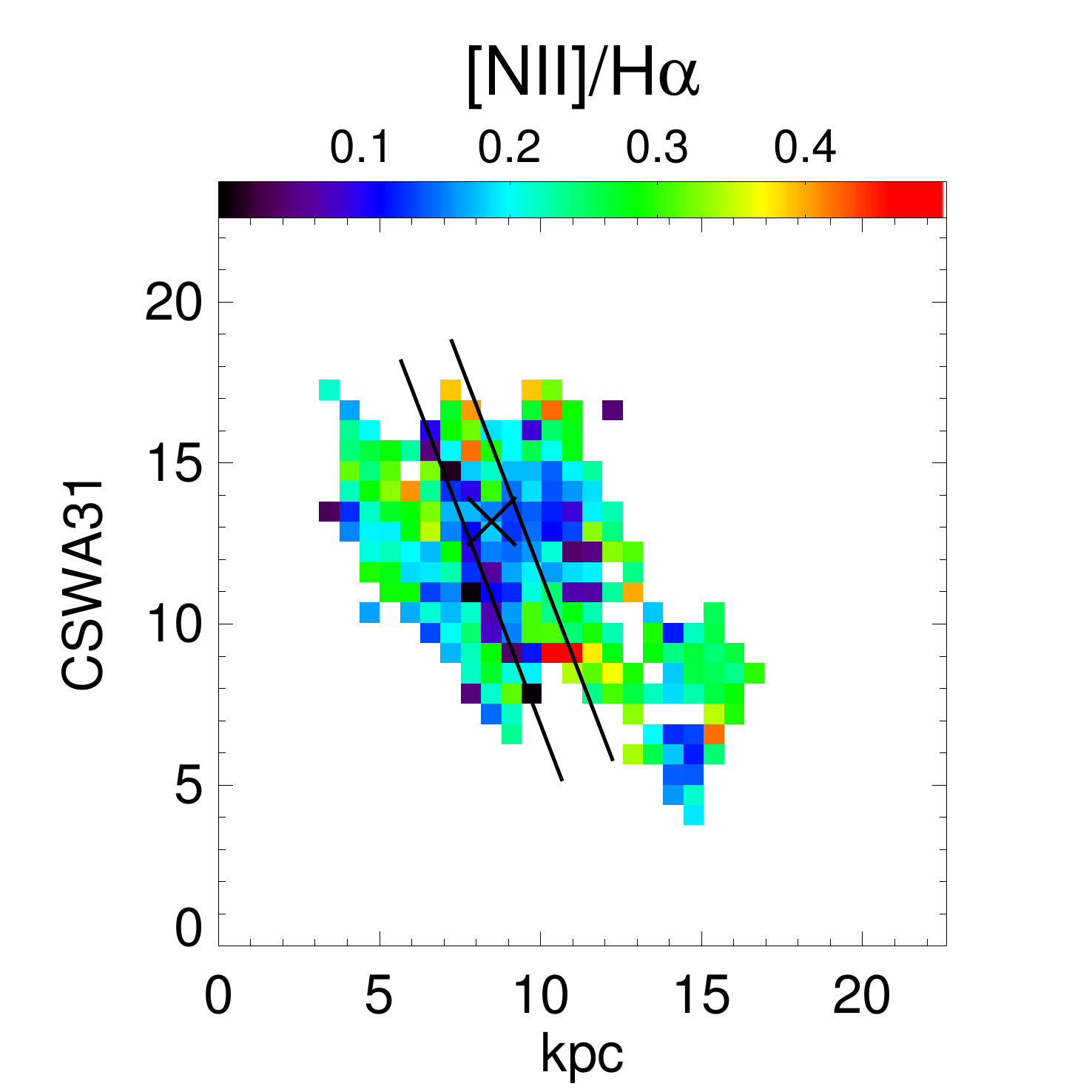}
	\includegraphics[width=5cm]{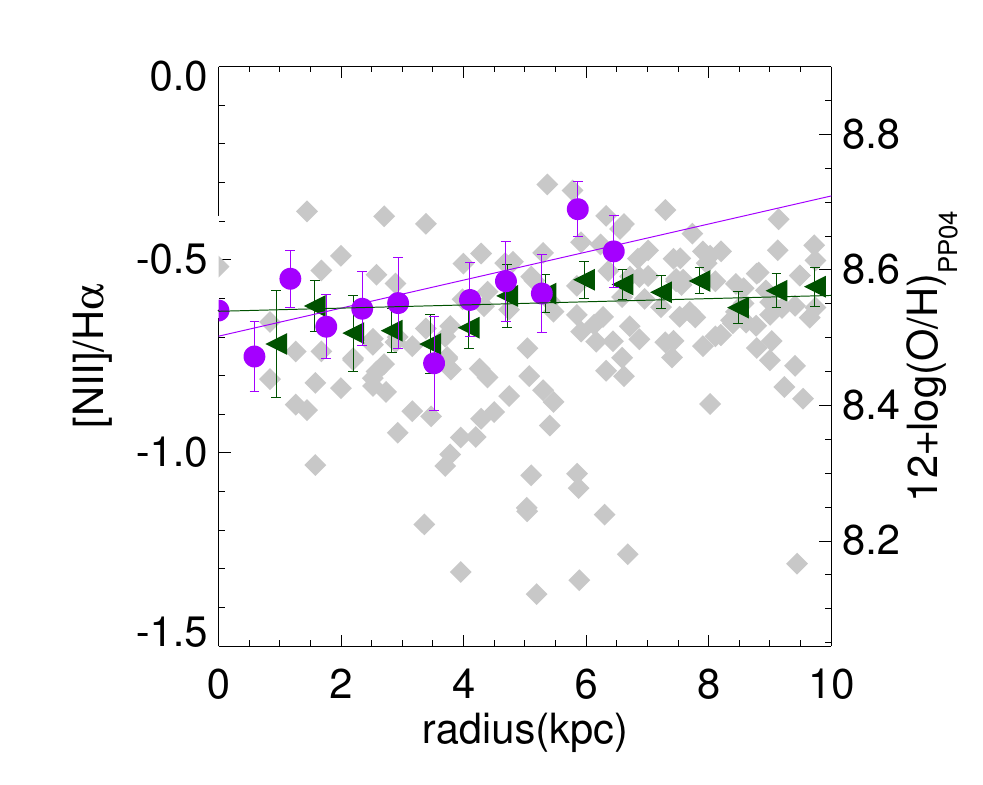}\\
	\caption{continued}
\end{figure*}

\acknowledgments

The authors recognize and acknowledge the very significant cultural role and reverence that the summit of Mauna Kea has always had within the indigenous Hawaiian community. We are most fortunate to have the opportunity to conduct observations from this mountain. We acknowledge useful discussions with Chuck Steidel, Phil Hopkins, Xiangcheng Ma, Drew Newman and Paul Torrey, and thank Paul Torrey for his {\it Illustris} metal gradient predictions tailored to the context of our observations. Support for AZ was provided by NASA through Hubble Fellowship grant \#HST-HF2-51334.001-A awarded by STScI. TAJ acknowledges support from NASA through Hubble Fellowship grant HST-HF2-51359.001-A awarded by the Space Telescope Science Institute, which is operated by the Association of Universities for Research in Astronomy, Inc., for NASA, under contract NAS 5-26555

\bibliography{Leethochawalit15}

\end{document}